\documentclass{aa}  

\usepackage{graphicx}
\usepackage{subcaption}
\usepackage{txfonts}
\usepackage{hyperref}
\usepackage{xcolor}
\usepackage{tikz,xcolor}
\usepackage{csquotes}
\usepackage{siunitx}

\usepackage{longtable}
\usepackage[para]{threeparttable}
\usepackage[para]{threeparttablex}
\usepackage{placeins}
\usepackage{multicol}
\usepackage{microtype}

\usepackage{scalerel}
\usepackage{tikz}
\usetikzlibrary{svg.path}

\definecolor{orcidlogocol}{HTML}{A6CE39}
\tikzset{
  orcidlogo/.pic={
    \fill[orcidlogocol] svg{M256,128c0,70.7-57.3,128-128,128C57.3,256,0,198.7,0,128C0,57.3,57.3,0,128,0C198.7,0,256,57.3,256,128z};
    \fill[white] svg{M86.3,186.2H70.9V79.1h15.4v48.4V186.2z}
                 svg{M108.9,79.1h41.6c39.6,0,57,28.3,57,53.6c0,27.5-21.5,53.6-56.8,53.6h-41.8V79.1z M124.3,172.4h24.5c34.9,0,42.9-26.5,42.9-39.7c0-21.5-13.7-39.7-43.7-39.7h-23.7V172.4z}electronically
                 svg{M88.7,56.8c0,5.5-4.5,10.1-10.1,10.1c-5.6,0-10.1-4.6-10.1-10.1c0-5.6,4.5-10.1,10.1-10.1C84.2,46.7,88.7,51.3,88.7,56.8z};
  }
}

\newcommand\orcidicon[1]{\href{https://orcid.org/#1}{\mbox{\scalerel*{
\begin{tikzpicture}[yscale=-1,transform shape]
\pic{orcidlogo};
\end{tikzpicture}
}{|}}}}

\newcommand{\review}[1]{\textcolor{black}{#1}} 
\newcommand{\reviewtwo}[1]{#1}

\begin{document}

   \title{Rapid jet ejection from PKS\,0215+015 coincident with a high-energy neutrino event}

   \subtitle{}
   \author{F.\,Eppel\inst{1,2,3}
          \and
          M.\,Kadler\inst{1}
          \and
          E.\,Ros\inst{2}
          \and
          P.\,Benke\inst{2,1,4}
          \and
          L.\,C.\,Debbrecht\inst{2}
          \and
          J.\,Eich\inst{1}
          \and
          P.\,G.\,Edwards\inst{5}
          \and
          M.\,Giroletti\inst{6}
          \and
          A.\,Gokus\inst{7}
          \and
          S.\,Hämmerich\inst{8}
          \and
          J.\,Heßdörfer\inst{1}
          \and
          M.\,Janssen\inst{9}
          \and
          S.\,Kim\inst{10}
          \and
          D.\,Kirchner\inst{1}
          \and
          Y.\,Y.\,Kovalev\inst{2}
          \and
          T.\,P.\,Krichbaum\inst{2}
          \and
          R.\,Ojha\inst{11}
          \and
          G.\,F.\,Paraschos\inst{\reviewtwo{12,13},2}
          \and
          F.\,Rösch\inst{1}
          \and
          W.\,Schulga\inst{1}
          \and
          J.\,Sinapius\inst{14}
          \and
          J.\,Stevens\inst{5}
          }

   \institute{Julius-Maximilians-Universität Würzburg, Institut für Theoretische Physik und Astrophysik, Lehrstuhl für Astronomie, Emil-Fischer-Straße 31, 97074 Würzburg, Germany\\
              \email{florian@eppel.space}
         \and
         Max-Planck-Institut für Radioastronomie, Auf dem Hügel 69, 53121 Bonn, Germany
         \and
         Joint Institute for VLBI ERIC, Oude Hoogeveensedijk 4, 7991 PD Dwingeloo, The Netherlands
         \and
         GFZ Helmholtz Centre for Geosciences, Telegrafenberg, 14476 Potsdam, Germany
         \and
         CSIRO Space and Astronomy, ATNF, PO Box 76, Epping NSW 1710, Australia
         \and
         INAF Istituto di Radioastronomia, via Gobetti 101, 40129 Bologna, Italy
         \and
         McDonnell Center for Space Science, Washington University, One Brookings Drive, St. Louis, MO 63130, USA
         \and
         Remeis Observatory and Erlangen Centre for Astroparticle Physics, Universtität Erlangen-Nürnberg, Sternwartstr. 7, 96049 Bamberg, Germany
         \and
         Department of Astrophysics, Institute for Mathematics, Astrophysics and Particle Physics (IMAPP), Radboud University, PO Box 9010, 6500 GL Nijmegen, The Netherlands
         \and
         Korea Astronomy and Space Science Institute, 776 Daedeok-daero, Yuseong-gu, Daejeon 34055, Republic of Korea
         \and
         NASA HQ, 300 E St SW, Washington, DC 20546-0002, USA
         \and
         \reviewtwo{Finnish Centre for Astronomy with ESO, University of Turku, 20014 Turku, Finland}
         \and
         \reviewtwo{Aalto University Metsähovi Radio Observatory, Metsähovintie 114, 02540 Kylmälä, Finland}
         \and
         Deutsches Elektronen-Synchrotron DESY, Platanenallee 6, 15738 Zeuthen, Germany
        }

   \date{Received Dec 16, 2025; accepted Feb 27, 2026}

 
  \abstract
  {}
  {We present a new neutrino-blazar multiwavelength-flare coincidence, observed in the blazar PKS\,0215+015 that showed a strong multiwavelength outburst in positional and temporal coincidence with the IceCube neutrino track alert IC220225A, similar to the case of TXS\,0506+056. We investigate the immediate response of the radio jet to the major flare and possible neutrino association.}
  {We performed target-of-opportunity observations of PKS\,0215+015 with the Very Long Baseline Array (VLBA) at 15\,GHz, 23\,GHz, and 43\,GHz in full polarization, for six epochs with monthly cadence following the neutrino event. We combine the VLBA observations with monitoring data from the Effelsberg 100-m telescope, the Australia Telescope Compact Array, and \textsl{Fermi}/LAT.}
  {Based on our VLBI kinematic analysis, we identified a new rapid jet component with an apparent speed of $\sim$60$-$80\,c, which was ejected around the arrival time of IC220225A. The fast component ejection is traced by a characteristic signature in polarization that suggests a shock-shock interaction with a quasi-stationary feature. By combining the VLBI results with radio variability data we estimated a bulk Lorentz factor of $\Gamma=105\pm56$, and a jet viewing angle $\vartheta=(1.47\pm0.31)^\circ$.}
  {We note that the properties of the rapid component exceed previously reported maximum apparent jet speeds and Lorentz factors reported by continuous VLBI monitoring programs. This is likely only possible because we are observing an exceptional flaring event at high redshift ($z=1.72$), with higher observing cadence than in typical monitoring programs. We suggest that neutrino production in PKS\,0215+015 can happen through $p\gamma$-interactions with protons possibly accelerated within the fast moving feature. The target photon field could be external to the jet or explained by a multi-layered jet. The latter scenario would be in agreement with the presence of quasi-stationary features revealed in our analysis.}

                          
\authorrunning{F.\,Eppel et al.}

\maketitle
\nolinenumbers 
%
\section{Introduction}

Following the discovery of TXS\,0506+056 as the first astrophysical high-energy neutrino-source in 2017 \citep{0506NeutrinoDiscovery}, the hunt for additional neutrino-emitting sources is ongoing. Meanwhile, the \cite{IceCubeNGC1068} have identified the active galaxy NGC\,1068, as well as the galactic plane \citep{IceCubeGalacticPlane} as additional neutrino emitters. Moreover\review{,} several works suggest a statistical correlation between bright active galactic nuclei (AGN), more specifically blazars, and high-energy neutrino events \citep[e.g.,][]{Plavin2020,Plavin2021,Plavin2023,Hovatta2021,Buson22,Kouch2024}. While these statistical associations are still under debate \citep[e.g.][]{Bellenghi2023,IceCubeCorrelation}, several individual blazars like PKS\,0735+178 \citep{0735ATelTelamon,Sahakyan2023_0735,Acharyya0735,Sanghyun2024_0735, Paraschos2025}, PKS\,0446+11 \citep{Telamon0446ATel,0446YuriATel,Kovalev0446} and PKS\,1725+123 \citep{Nanci2022} have raised the interest of the community, as they exhibited prominent flaring events close in time to IceCube neutrino events, similar to the case of TXS\,0506+056 \review{\citep[see also][]{Sumida2022,Jiang2024,Ji2024,Koemives2024,Perger2025}}.

We present the case of PKS\,0215+015 which was found to be in a strong multiwavelength outburst in coincidence with the IceCube neutrino event IC220225A \citep{0215Proceedings}. On February 25, 2022 the IceCube Collaboration reported the discovery of the neutrino event IC220225A\footnote{\url{https://gcn.gsfc.nasa.gov/gcn3/31650.gcn3}}, classified as \enquote{bronze} type, with a \enquote{signalness} (i.e., probability for the neutrino to be of astrophysical origin) of 38\,\% and an energy of $\sim150$\,TeV. PKS\,0215+015 exhibited a gamma-ray flux more than 7 times greater than the average reported in the 4FGL-DR3 catalog on the day preceding the neutrino event \citep{FermiAtel0215}. Additionally, the source showed strong optical \citep{Optical0215ATel} and radio flaring activity \citep{KadlerATel0215,Plavin0215ATel}. 

PKS\,0215+015 was initially classified as a BL Lac object by \cite{Gaskell1982}, and is located at a redshift of $z=1.72$ \citep{Redshift0215_first,Redshift0215}. In contrast to its initial classification, the detection of spectral lines by \cite{Redshift0215} suggests an FSRQ or changing-look AGN behavior. Moreover, \cite{Foschini2022} note that the BL Lac-type spectrum was observed during a faint state of the source, while the FSRQ spectrum was taken at a bright state, likely caused by a change of accretion rate.
This could be relevant in context of neutrino production since FSRQs are considered to be likely sources of high-energy neutrinos, due to their high density
of target photons for the $p\gamma$-channel \citep[e.g.,][]{Atoyan2001,Murase2014,Moretti2025}.

Triggered by the strong multiwavelength flare, coincident with IC220225A, we carried out a dense radio monitoring campaign of PKS\,0215+015 with the Effelsberg 100\,m-telescope, the Australia Telescope Compact Array (ATCA), and the Very Long Baseline Array (VLBA). Here, we present our results on the immediate response of the radio jet to the major multiwavelength outburst and possible neutrino association.
Throughout the paper, we assume a flat $\Lambda$CDM-model, with $H_0=67.4$\,km\,s$^{-1}$\,Mpc$^{-1}$, and $\Omega_\mathrm{m}=0.315$ \citep{Planck2020}. This puts PKS\,0215+015 at a luminosity distance of 13\,Gpc, with a linear scale of 8.7\,pc\,mas$^{-1}$, such that an apparent motion with 1\,mas\,yr$^{-1}$ corresponds to a speed of 77\,c.

\begin{figure}
    \centering
    \includegraphics[width=.95\columnwidth]{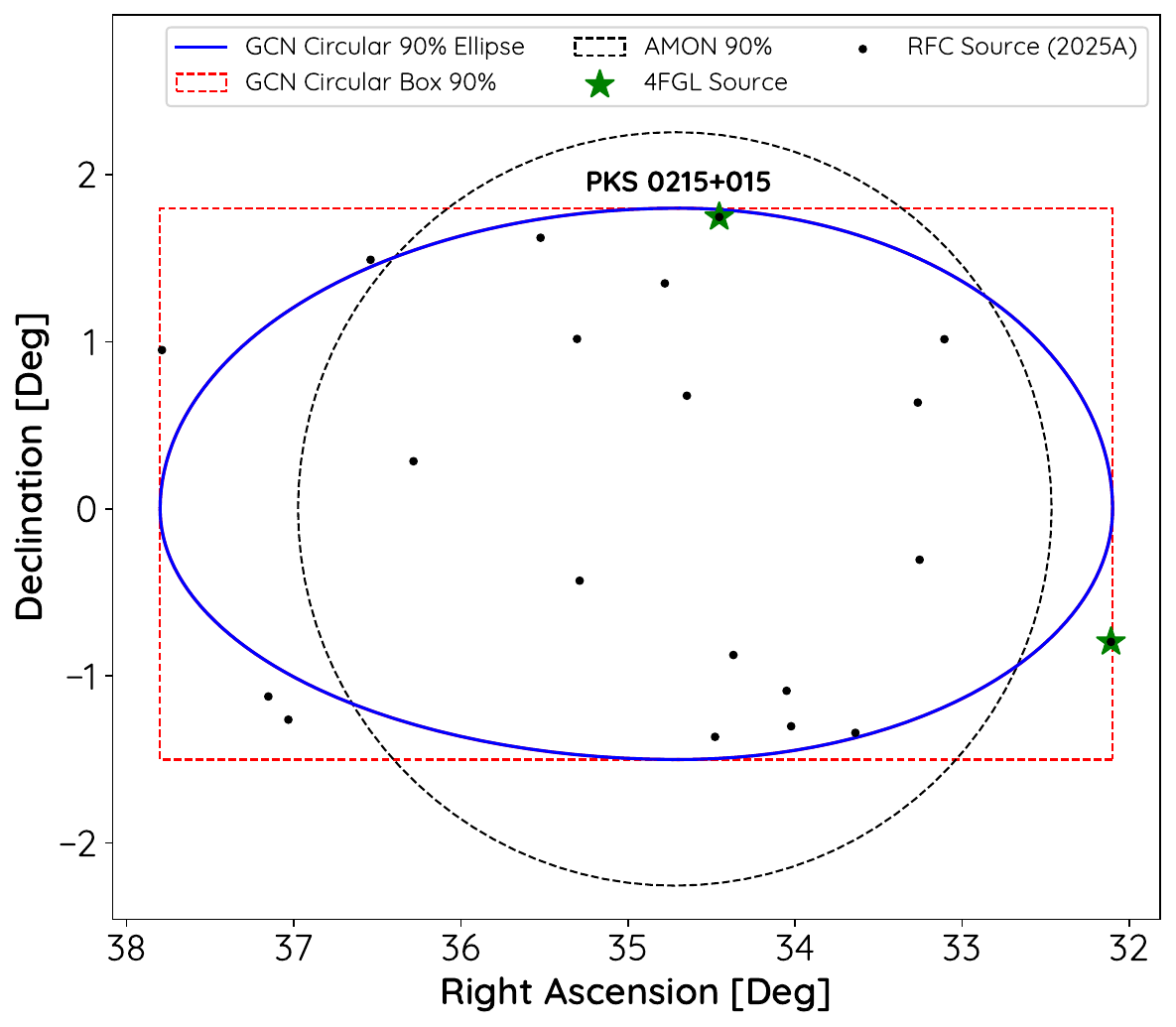}
    \caption{90\,\% uncertainty regions of the neutrino IC220225A, as reported by the IceCube Collaboration via AMON (black dashed line) and GCN circular (blue ellipse and red dashed box). RFC sources within the GCN box region are displayed as black dots, gamma-ray detected sources from the 4FGL catalog are highlighted as green stars. PKS\,0215+015 stands out as the only gamma-ray detected source within all three regions.}
    \label{fig:uncertainty_region}
    \vspace{-.5\baselineskip}
\end{figure}

\begin{figure*}
    \centering
    \includegraphics[width=.8\linewidth]{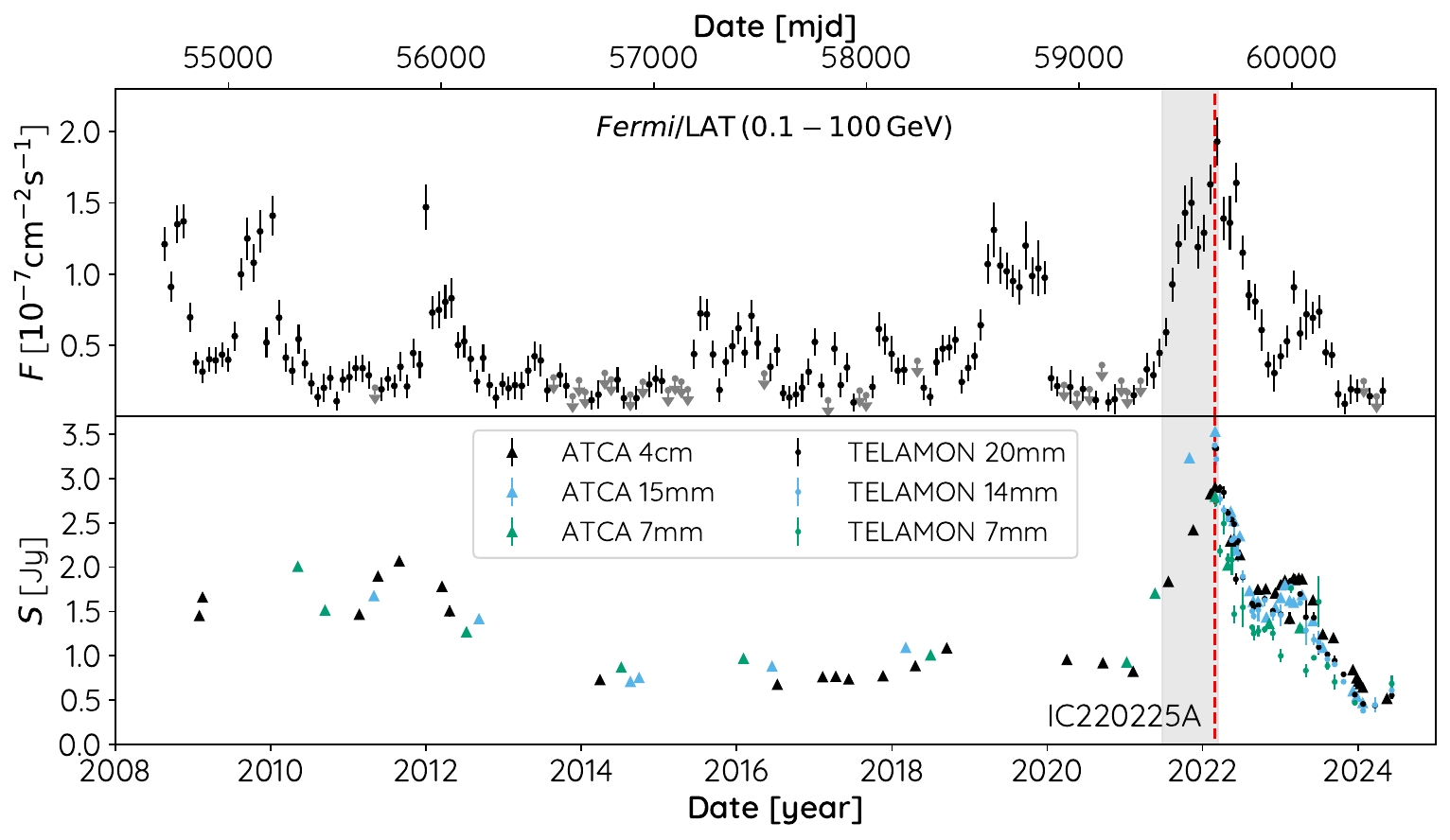}
    \caption{Gamma-ray (upper panel) and radio (lower panel) light curves of PKS\,0215+015. The source exhibited its historical high-state in both wavelengths coincident with the arrival time of IceCube neutrino event IC220224A (red dashed line). The new fast VLBI component 1 was ejected shortly before. Its possible ejection time range (1$\sigma$) is indicated by the gray shaded area.}
    \label{fig:lightcurves}
\end{figure*}

\section{Observations and analysis}
\label{sec:obs}

\subsection{Sources in the IC220225A field}

In order to find astrophysical sources coincident with the uncertainty region of IC220225A, we cross matched the neutrino region published by IceCube with the \textsl{Fermi}/LAT 4FGL-DR3 catalog \citep{4FGL_DR3}, as well as with the Radio Fundamental Catalog \citep[RFC 2025a,][]{RFC}. IceCube provides different stages of their high-energy track alerts, first, a real-time localization published via AMON\footnote{\url{https://gcn.gsfc.nasa.gov/notices_amon_g_b/136366_14203460.amon}} (Rev 0), followed by a refined localization (Rev 1), and usually a GCN circular\footnote{\url{https://gcn.nasa.gov/circulars/31650}}. While the AMON regions (Rev 1 and Rev 0) come with radial errors, the GCN circular contains asymmetric errors around a center position in right ascension and declination, usually resulting in a \enquote{box} shape. Based on what \cite{IceCat1} reported, these \enquote{GCN-boxes} can often be approximated with ellipses, however, the exact shape of the uncertainty region within the \enquote{GCN-box} is not made publicly available immediately. In accordance with \cite{IceCat1}, we therefore consider the full \enquote{GCN-box} as the best, and most conservative, approximation to the real uncertainty region. As an overview, in Fig.\,\ref{fig:uncertainty_region}, we show the latest AMON (Rev 1) 90\,\% uncertainty region of IC220225A, the \enquote{GCN-box}, as well as the interpolated GCN ellipse region, based on the \enquote{GCN-box}. Within the \enquote{GCN-box}, there are 20 sources from the RFC, with PKS\,0215+015 being the brightest one, based on the median X-band flux \review{density} listed in the RFC ($\sim$0.8\,Jy). Moreover, the region includes two sources from the 4FGL-DR3 catalog, 4FGL\,J0208.5-0046 and 4FGL\,J0217.8+0144. The latter is associated with PKS\,0215+015, and, due to its strong multiwavelength flare at the time of IC220225A, stands out as an intriguing counterpart to the neutrino event.

\subsection{VLBI observations}
\label{sec:vlba}

Following the neutrino association in February 2022, we obtained target-of-opportunity (ToO) observations of PKS\,0215+015 with the VLBA (project code: BE082). The observations were carried out at three different frequencies (15\,GHz/Ku-band, 23\,GHz/K-band \& 43\,GHz/Q-band), and covered six epochs with monthly cadence, starting on March 24, 2022. The data were calibrated using the \texttt{rPICARD} pipeline \citep{rpicard}, based on \texttt{CASA} \citep{CASA}. Imaging and self-calibration were carried out using standard methods in \texttt{DIFMAP} \citep{difmap}. Final Stokes I images were produced with an entropy-based cleaning method \citep{entropyclean}, using the final CLEAN windows obtained during self-calibration. In order to correct for polarization leakage (i.e., D-terms), we used the \texttt{PolSolve} self-calibration approach \citep{PolSolve}. Cleaning in Stokes Q and U between \texttt{PolSolve} iterations was performed using the entropy-based cleaning approach by \cite{entropyclean}, with the same CLEAN windows as used for Stokes I. The resulting \texttt{DIFMAP} \texttt{.mod} files were re-imported to \texttt{CASA} for every \texttt{PolSolve} iteration. We assumed the D-terms to have converged when their average variation between iterations was $<1\,\%$. The scripts for the polarization calibration are available on GitHub\footnote{\url{https://github.com/flep198/polcal}}.

After leakage calibration, we performed a correction for the absolute orientation of the electric vector positon angle $\chi$ (EVPA). As an EVPA calibrator, we used BL Lac (2200+420), which was included in the observations. Based on MOJAVE observations at 15\,GHz\footnote{\url{https://www.cv.nrao.edu/MOJAVE/sourcepages/2200+420.shtml}} \citep{MOJAVE}, and BEAM-ME observations at 43\,GHz\footnote{\url{https://www.bu.edu/blazars/VLBA_GLAST/bllac.html}} \citep{Weaver22}, the EVPA of BL Lac is well sampled and stable during the time of our observations. We interpolated the MOJAVE and BEAM-ME EVPAs for BL Lac in time, to get an EVPA value for every ToO observation, for 15\,GHz and 43\,GHz, respectively. EVPA values for 23\,GHz were interpolated between the 15\,GHz and 43\,GHz values, assuming a quadratic dependence of the EVPA with wavelength, as expected from Faraday rotation. For every frequency and epoch, we calculated $\Delta\chi=\chi_\textrm{cal}-\chi_\textrm{obs}$, the difference  between expected EVPA $\chi_\textrm{cal}$ and the measured EVPA $\chi_\textrm{obs}$ of BL Lac, and corrected the RL-phase difference (i.e., 2$\times\Delta\chi$) accordingly with the \texttt{AIPS}-based \texttt{RLCOR} task \citep{AIPS}, using \texttt{ParselTongue} \citep{ParselTongue}. 

\begin{table}[htbp]
\centering
\begin{threeparttable}

\caption{Scaling factors used to adjust the VLBI flux densities to single-dish measurements.}
\label{tab:scaling}

\begin{tabular}{cccc}
\hline\hline
Epoch & Ku-Band & K-Band & Q-Band \\
Date & 15\,GHz & 23\,GHz & 43\,GHz \\
\hline
2022-03-24 & 1.23 & 1.51 & 1.71 \\
2022-04-23 & 1.61 & 1.50 & 1.28 \\
2022-06-01 & 1.28 & 1.60 & 1.36 \\
2022-06-30 & 1.23 & 1.42 & 1.92 \\
2022-08-01 & 1.26 & 1.44 & 2.39 \\
2022-08-19 & 1.29 & 1.42 & 2.54 \\
\hline
Average & 1.32 & 1.48 & 1.87 \\
\hline
\end{tabular}
\begin{tablenotes}
\end{tablenotes}
\end{threeparttable}
\vspace{-.5\baselineskip}
\end{table}

\begin{figure*}
    \centering
    \includegraphics[width=.95\linewidth]{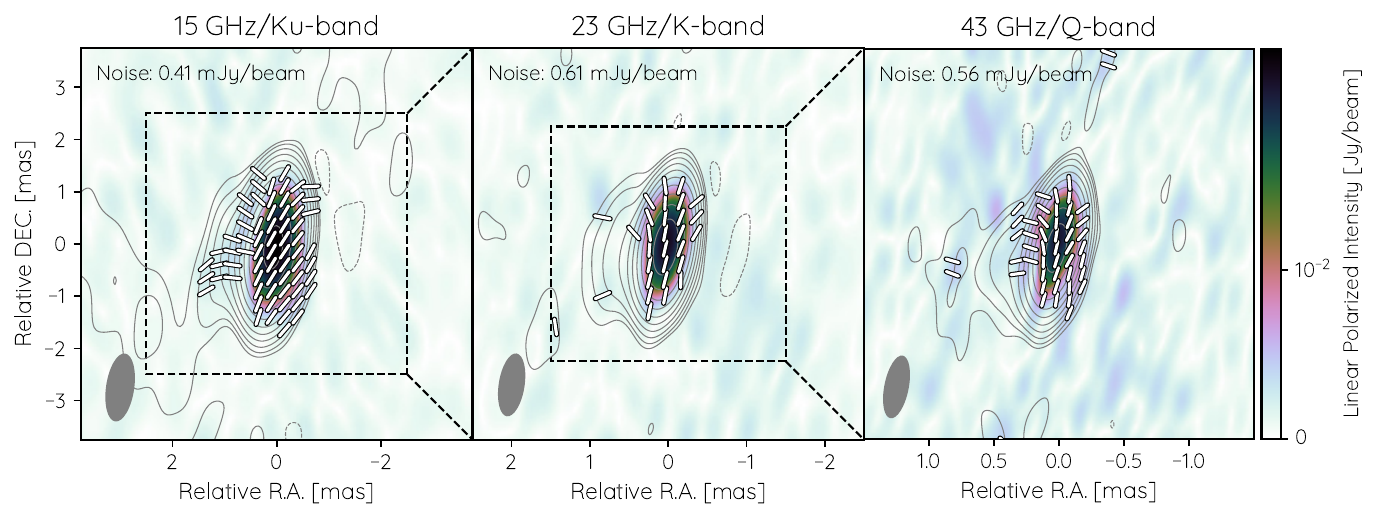}
    \caption{Stacked polarization images at 15\,GHz (left), 23\,GHz (center), and 43\,GHz (right). Before stacking the individual images were convolved with the median beam (indicated by the gray ellipse) at each frequency separately. Contours show the total intensity, starting at four times the noise level and increasing by factors of two. The colormap displays linear polarization, and the sticks indicate the direction of the EVPA.}
    \label{fig:stacked_pol_maps}
\end{figure*}

As the final calibration step, we performed an absolute flux \review{density} calibration of the VLBI data by using single-dish information from the TELAMON program (see Sect.\,\ref{sec:telamon}). Since the source has a very compact structure, we assumed a compactness factor of 1 (i.e. VLBI flux density equals single-dish flux density). This is justified since MOJAVE estimates of the 15\,GHz extended flux density for PKS\,0215+015 are less than 1\,mJy, based on a comparison between well calibrated parsec-scale total flux \review{density} and UMRAO \& OVRO values. The VLBI flux densities at 15\,GHz, 23\,GHz, and 43\,GHz (integrated CLEAN model) were scaled by using interpolated TELAMON 20\,mm, 14\,mm and 7\,mm flux densities of PKS\,0215+015, respectively. An overview of the resulting scaling factors is given in Tab.\,\ref{tab:scaling}. On average, there is $\sim$30\,\% of flux \review{density} missing at 15\,GHz, $\sim$50\,\% at 23\,GHz, and $\sim$90\,\% at 43\,GHz. The difference in the scaling factors between epochs and frequencies likely comes from pointing errors in the VLBI observations, which can be more significant at higher frequencies, due to smaller telescope beams. Additionally, the self-calibration process applied to the VLBI data can amplify the effect of initially low antenna amplitudes. This was especially severe in the second epoch at 15\,GHz, where several antennas had low amplitudes, resulting in a scaling factor of 1.61. Additionally, it has been discussed previously by \cite{Livingston2025} that the bandpass normalization used in the default \texttt{rPICARD} settings can result in lower amplitudes by $\sim10$\,\%. Since TELAMON single-dish data are free of self-calibration errors and pointing offsets are corrected during calibration (see Sect.\,\ref{sec:telamon}), their use as a reference for the absolute flux \review{density} calibration is justified. We assume a final flux \review{density} error of 5\,\%, which is similar to what is used in other programs \citep[e.g.,][]{MOJAVEI}.

We modeled the source structure with circular Gauss components for every \review{epoch} and frequency using the \texttt{modelfit} task in \texttt{DIFMAP}. \review{We used the \texttt{VCAT} package \citep{VCAT} to determine errors for every modelfit component with a signal-to-noise (S/N) based method \citep{Schinzel2012}, and their brightness temperatures, following \cite{Kovalev05} (see Tab.\,\ref{tab:components})}.

\subsection{TELAMON/Effelsberg}
\label{sec:telamon}

We started observing PKS\,0215+015 with the Effelsberg 100-m telescope as part of the TELAMON program after its neutrino association. The TELAMON program is using the Effelsberg 100-m telescope to monitor the radio spectra of active galactic nuclei (AGN) under scrutiny in astroparticle physics, namely TeV blazars and candidate neutrino-associated AGN \citep{TELAMON}. The first observation took place on Feb 27, 2022, only two days after the neutrino alert. The source was observed for a total of 28 epochs until June 7, 2024. We observed the source with the S20mm (i.e., 14$-$17\,GHz), S14mm (i.e., 19$-$25\,GHz) and S7mm (i.e., 36$-$44\,GHz) receivers in secondary focus, using multiple sub-bands per receiver. Data calibration and sub-band averaging was performed following \cite{TELAMON}. In addition to the total intensity data, we obtained linear polarized intensity data for 20\,mm and 7\,mm in 19 epochs. We followed the data processing and calibration for TELAMON polarization data described by \cite{TELAMONPolarization}. For sessions where not enough polarization calibrators were observed, an average Müller-matrix was used to correct the instrumental polarization.

\subsection{ATCA}

The ATCA has been monitoring PKS\,0215+015 since 2004 at multiple frequencies from 2.1\,GHz up to 50\,GHz as part of the AGN and calibrator monitoring program \citep{ATCAmonitoring}. The data were processed by a standard pipeline and are publicly available\footnote{\url{https://www.narrabri.atnf.csiro.au/calibrators/calibrator_database_viewcal.html?source=0215+015&detailed=true}}. We collected all data from the 4\,cm, 15\,mm and 7\,mm bands after carefully checking their quality based on the \enquote{defect} and \enquote{closure phase} parameters on the website. The ATCA light curve is shown together with the TELAMON data in Fig.\,\ref{fig:lightcurves}.

\subsection{Fermi/LAT}

Of the two gamma-ray sources in the \enquote{GCN-box} of IC220225A, only one, 4FGL\,J0217.8+0144 (PKS\,0215+015) has publicly available data in the \textsl{Fermi}/LAT light curve repository\footnote{\url{https://fermi.gsfc.nasa.gov/ssc/data/access/lat/LightCurveRepository/source.php?source_name=4FGL_J0217.8+0144}} \citep{FermiLCrepo}. We obtained its gamma-ray light curve in monthly binning, using a minimum detection significance of TS=4 (corresponds to 2$\sigma$). The gamma-ray lightcurve is shown on top of the radio light curve in Fig.\,\ref{fig:lightcurves} and reveals its historical maximum directly after the neutrino event.

\section{Results}
\label{sec:results}

\subsection{Jet structure}
\label{sec:jet_structure}

PKS\,0215+015 exhibits a very compact jet structure with a slightly extended jet to the East. We compiled stacked full polarization maps at 15\,GHz, 23\,GHz and 43\,GHz from the six observing epochs in 2022 in Fig.\,\ref{fig:stacked_pol_maps}. Before stacking, the individual images were centered on their core component, and convolved with the median beam for every frequency band (15\,GHz: $1.32\times0.54$\,mas, $-8.3^\circ$; 23\,GHz: $0.81\times0.33$\,mas, $-8.8^\circ$; 43\,GHz: $0.49 \times 0.19$\,mas, $-11.2^\circ$). The increased resolution at 23\,GHz and 43\,GHz, as compared to 15\,GHz, allows us to resolve a bit more of the faint extended jet structure. While earlier observations of this source \citep[e.g.,][]{MOJAVE} revealed extended structure up to $\sim5$\,mas away from the core, the source structure was more compact in 2022, close in time to the neutrino event. \review{We note that the brightness temperature of the core component reaches values of $\sim10^{13}$\,K around the peak of the flare (see Tab.\,\ref{tab:components}) and thus significantly exceeds the equipartition brightness temperature \citep[$T_\mathrm{b,eq}=5\times 10^{10}$\,K,][]{Readhead}.}

\begin{figure*}
\centering
    \includegraphics[width=\linewidth]{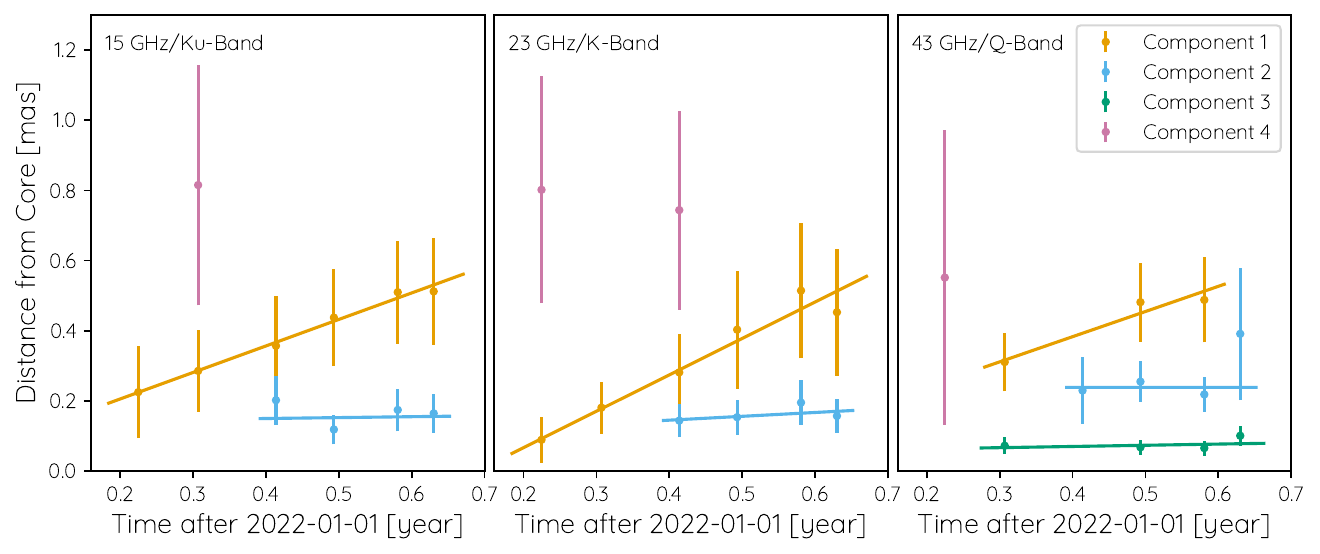}
    \caption{Kinematic modelling of \review{PKS}\,0215+015 at 15\,GHz (left), 23\,GHz (center) and 43\,GHz (right). Component 1 is moving outwards, while components 2 and 3 are quasi-stationary. Fit parameters for all components at all frequencies are shown in Tab.\,\ref{tab:kinematics}.}
    \label{fig:kinematics}
\end{figure*}

Following \cite{MOJAVE_opening}, we used all model components to calculate the apparent jet opening angle, $\phi_\mathrm{app}=$(36$\pm$18)$^\circ$, averaged over all observing epochs and frequencies. Compared to a larger sample of \textsl{Fermi}-detected AGN \citep{MOJAVEOpeningAngle2017}, this value is above the median of the population, but not unusually large. Moreover, we used the model components across all frequencies to fit a simple power law collimation profile \citep[e.g.,][]{Kovalev20collimation}, revealing a conical jet profile with power law index $k=1.17\pm0.30$ (see App.\,\ref{app:collimation}).

\subsection{Jet kinematics}
\label{sec:kinematics}

We cross-identified the modelfit components across epochs and frequencies to determine the multi-frequency kinematics of the jet (for an overview about the cross-identifications see App.\,\ref{app:modelfits}). In total, we identified six distinct components, component 0 is the core component, and components 1$-$5 are identified as jet components. In Fig.\,\ref{fig:kinematics}, we show the kinematic plots (i.e., core distance vs. time) for components 1$-$4 at 15\,GHz (left), 23\,GHz (center), and 43\,GHz (right). We note that the low-S/N component 5 is located $\sim$2$-$3\,mas away from the core and does not imply significant motion, and is thus not considered in the kinematic analysis. For all other components, we fitted their distance to the core vs. time linearly to determine an apparent component speed $\mu_\mathrm{app}$, and ejection epoch $t_0$ for moving components. The fit results for components 1$-$3 are shown in Tab.\,\ref{tab:kinematics}. Component 4 does not have enough detections to perform a sensible kinematic fit. In Fig.\,\ref{fig:kinematics} one can clearly see that component 1 is moving away from the core at all three frequencies. The fitted component speed $\mu_\mathrm{app}$ for component 1 varies slightly between the three frequencies but is consistent within the errorbars, suggesting fast motion of  $\sim$0.8$-$1.0\,mas/yr. Given the large redshift of PKS\,0215+015, this translates to an apparent superluminal speed $\beta_\mathrm{app}$ of $\sim$60$-$80\,c. The ejection times for component 1 suggest an ejection consistent with the arrival time of the neutrino, as indicated in Fig.\,\ref{fig:lightcurves}.
Components 2 and 3 show no significant motion, i.e., they are considered as stationary components \citep{Jorstad2001a}. Component 2 is only detected starting from the third epoch onward. This is likely the case since we cannot distinguish component 1 from component 2 in the first two epochs, since component 1 seems to move through stationary component 2. We decided to label this combined feature as component 1 to get the best constraints on the motion of component 1. Fitting component 1 only from epoch 3 onward results in comparable kinematics. Component 3 is located very close to the core ($<0.1$\,mas) and only detected at Q-band. It likely remains undetected at Ku- and K-band due to the limited resolution as compared to Q-band. As reported in Tab.\,\ref{tab:kinematics}, the speed of components 2 and 3 is consistent with zero.

\begin{table}[htbp]
\centering
\begin{threeparttable}
\caption{Kinematic fit results as shown in Fig.\,\ref{fig:kinematics}.}
\label{tab:kinematics}

\begin{tabular}{cccc}
\hline\hline
Frequency & $\mu_\mathrm{app}$ & $\beta_\mathrm{app}$ & $t_0$ \\
{[GHz]} & [mas/yr] & [c] & [yr] \\
\hline
\multicolumn{4}{c}{\textbf{Component 1 (moving)}} \\
15 & 0.76 $\pm$ 0.39  & 58 $\pm$ 30 & 2021.93 $\pm$ 0.27 \\
23 & 1.04 $\pm$ 0.34  & 80 $\pm$ 26 & 2022.136 $\pm$ 0.075 \\
43 & 0.72 $\pm$ 0.50  & 55 $\pm$ 38 & 2021.87 $\pm$ 0.39 \\
\hline
\multicolumn{4}{c}{\textbf{Component 2 (stationary)}} \\
15 & 0.03 $\pm$ 0.36  & 2 $\pm$ 28 & 2017 $\pm$ 82 \\
23 & 0.11 $\pm$ 0.30 & 8 $\pm$ 22 & 2021.1 $\pm$ 3.9 \\
43 & 0.00 $\pm$ 0.59  & 0 $\pm$ 45 & \review{-} \\
\hline
\multicolumn{4}{c}{\textbf{Component 3 (stationary)}} \\
43 & 0.03 $\pm$ 0.10 & 2.6 $\pm$ 7.8 & 2020.3 $\pm$ 6.4 \\
\hline
\hline
\end{tabular}
\begin{tablenotes}
\end{tablenotes}
\end{threeparttable}
\end{table}

\subsection{Radio variability properties}
\label{sec:variability}

\begin{figure}
    \includegraphics[width=\columnwidth]{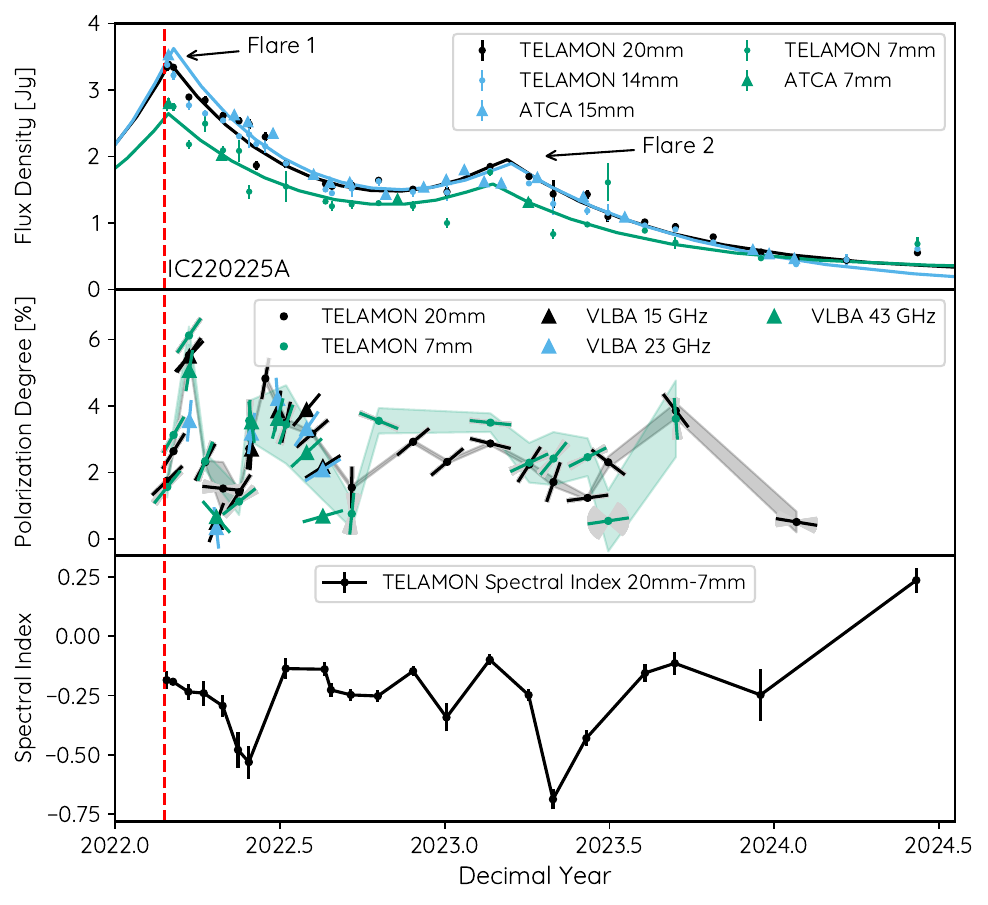}
    \caption{\textsl{Top:} Combined ATCA and TELAMON light curve with fitted flares. \textsl{Center:} TELAMON fractional polarization at 20\,mm and 7\,mm. The inclined lines indicate direction of the EVPA, errors are indicated by the shaded regions in EVPA and polarization. In addition to the TELAMON polarization values, we display the fractional polarization and EVPA of the VLBA data, for better visibility without uncertainties. \textsl{Bottom:} Spectral index evolution obtained from TELAMON observations between 20\,mm and 7\,mm. The red dashed line indicates the time of IC220225A.}
    \label{fig:variability}
\end{figure}

\begin{table*}[htbp]
\centering
\begin{threeparttable}

\caption{Flare fit results as shown in Fig.\,\ref{fig:variability}.}
\label{tab:variability}

\begin{tabular}{ccccccc}
\hline\hline
Frequency & Band & t$_0$ & S$_\mathrm{max}$ & $\tau$ & T$_\mathrm{b,var}$ & $\delta_\mathrm{var}$ \\
{[GHz]} & & [yr] & [Jy] & [days] & [$10^{14}$\,K] &  \\
\hline
\multicolumn{7}{c}{\textbf{Flare 1}} \\
15 & Ku & 2022.1650 $\pm$ 0.0043  & 3.088 $\pm$ 0.033 & 122.8 $\pm$ 2.5 & 8.64 $\pm$ 0.37 & 25.85 $\pm$ 0.37 \\
20 & K & 2022.1750 $\pm$ 0.0026  & 3.460 $\pm$ 0.032 & 121.9 $\pm$ 2.2 & 4.57 $\pm$ 0.17 & 20.90 $\pm$ 0.26 \\
40 & Q & 2022.1610 $\pm$ 0.0083  & 2.292 $\pm$ 0.054 & 135.3 $\pm$ 9.4 & 0.74 $\pm$ 0.10 & 11.41 $\pm$ 0.53 \\
\hline
\multicolumn{7}{c}{\textbf{Flare 2}} \\
15 & Ku & 2023.1826 $\pm$ 0.0055  & 1.451 $\pm$ 0.031 & 138.7 $\pm$ 8.4 & 3.18 $\pm$ 0.39 & 18.52 $\pm$ 0.76 \\
20 & K & 2023.1979 $\pm$ 0.0028  & 1.602 $\pm$ 0.041 & 181 $\pm$ 10 & 0.96 $\pm$ 0.11 & 12.44 $\pm$ 0.48 \\
40 & Q & 2023.1380 $\pm$ 0.0094  & 1.010 $\pm$ 0.068 & 123 $\pm$ 22 & 0.39 $\pm$ 0.14 & 9.2 $\pm$ 1.1 \\
\hline
\multicolumn{2}{c}{S$_\mathrm{0,Ku}=(0.227\pm0.025$)\,Jy} &
\multicolumn{3}{c}{S$_\mathrm{0,K}=(-0.022\pm0.043$)\,Jy} &
\multicolumn{2}{c}{S$_\mathrm{0,Q}=(0.298\pm0.058$)\,Jy} \\
\hline
\hline
\end{tabular}
\begin{tablenotes}
\end{tablenotes}
\end{threeparttable}
\end{table*}

We analyzed the flares in the radio light curve to estimate a variability time scale and associated variability Doppler factor of PKS\,0215+015, at the time of the neutrino event. We performed this analysis on the combined TELAMON and ATCA data, at three different frequency bands. We used TELAMON 20\,mm data (15\,GHz, Ku-band), combined TELAMON 14\,mm and ATCA 15\,mm (20\,GHz, K-band), and combined TELAMON and ATCA 7\,mm data (40\,GHz, Q-band). The combined light curves are shown in Fig.\,\ref{fig:variability} (top panel). We restricted the analysis to the period after the neutrino detection, since the rise of the first flare is not well enough sampled to perform a sensible fit. For every frequency we fitted two flares with exponential peaks of the form
\begin{equation}
    \Delta S(t)= \begin{cases}
\Delta S_\mathrm{max}\mathrm{e}^{(t-t_\mathrm{max})/\tau} & \text{if } t \leq t_\mathrm{max} \\
\Delta S_\mathrm{max}\mathrm{e}^{(t_\mathrm{max}-t)/1.3\tau} & \text{if } t > t_\mathrm{max}
\end{cases},
\end{equation}
following the approach by \cite{ValtaojaLaehteenmaki1999}, where $\Delta S_\mathrm{max}$ is the amplitude of the flare, $\tau$ the flare timescale, and $t_\mathrm{max}$ the time of the flare peak. After performing an initial CLEAN-like approach of fitting the two flares individually, we performed a common fit of both flares, using the previous fit values as starting values for the fit, and adding a fit parameter $S_0$ for the flux \review{density} baseline. The best fit parameters are shown in Tab.\,\ref{tab:variability}. From the obtained fit parameters, we estimated the variability brightness temperature (in the source frame),
\begin{equation}
T_\mathrm{b,var}=1.47 \times10^{13}\frac{D_L^2\Delta S_\mathrm{max}}{\nu^2\tau^2(1+z)},
\end{equation}
where $D_L$ is the luminosity distance in Mpc, $\nu$ the observing frequency in GHz,  and $z$ the redshift \citep{Hovatta2009,Liodakis}. Using $T_\mathrm{b,var}$ and assuming an equipartition brightness temperature $T_\mathrm{b,eq}=5\times10^{10}$\,K \citep{Readhead,Lahteenmaki_1999}, we calculated the variability Doppler factor
\begin{equation}
\label{eq:delta_var}
    \delta_\mathrm{var}=\sqrt[3]{\frac{T_\mathrm{b,var}}{T_\mathrm{b,eq}}},
\end{equation}
following \cite{Hovatta2009,Liodakis}. The obtained $T_\mathrm{b,var}$ and $\delta_\mathrm{var}$ values for every flare and frequency are presented in Tab.\,\ref{tab:variability}. For the first flare, whose peak times match the neutrino arrival time, we obtain Doppler factors between $\delta_\mathrm{var}=11.41\pm0.53$ at the highest frequencies (40\,GHz) up to $\delta_\mathrm{var}=25.85\pm0.37$ at 15\,GHz. The second flare, peaking in early 2023, exhibits slightly lower $\delta_\mathrm{var}$ values from $\delta_\mathrm{var}=9.2\pm1.1$ at 40\,GHz up to $\delta_\mathrm{var}=18.2\pm0.76$ at 15\,GHz. The difference between the $\delta_\mathrm{var}$ values across the different frequencies is likely due to the fact that we are \review{undersampling} the variability at the higher frequencies and probably missing the peak flux densities in our monitoring data. This could also affect the lower frequency $\delta_\mathrm{var}$ estimate, especially for flare 1, since we do not have any coverage of the source between late 2021 and February 2022, where the flux \review{density} peak likely occurred. This means that the obtained $\delta_\mathrm{var}$ values for flare 1, especially at the higher frequencies, are probably slightly underestimated and might be closer to the value obtained for 15\,GHz. \review{On the other hand, the intrinsic brightness temperature used in Eq.\,\eqref{eq:delta_var} can exceed the equipartition value and reach values of several times $10^{11}$\,K during flares \citep[e.g.,][]{Homan2006}. In such a scenario, the variability Doppler factors presented above could be systematically overestimated by a factor of 1.5$-$2 (see Sect.\,\ref{sec:lorentz} for further discussion).}

In addition to the flux density variability, we analyzed the variability of the radio spectral index in PKS\,0215+015. For the spectral index analysis, we used all TELAMON epochs where data across at least 19.5\,GHz up to 37\,GHz were available and fitted a power law of the form
\begin{equation}
    S(\nu) \propto \nu^\alpha,
\end{equation}
where $\alpha$ is the spectral index, to the data, as described by \cite{TELAMON}. The spectral index evolution is plotted in Fig.\,\ref{fig:variability} (bottom panel). We find a median spectral index of $\alpha=-0.24$. Immediately after the neutrino event, PKS\,0215+015 exhibits a rather flat spectrum in agreement with the median value. Shortly after the peak of the flare, the spectrum steepens during the decay of the radio flare and then becomes flat again. This behavior is consistent with the fast VLBI component 1 moving through the jet, as described in Sec.\,\ref{sec:kinematics}. The steeper spectral index suggests that optically thin emission dominates the jet, which is expected from an extended feature that is moving through the jet. Shortly after ($\sim 2022.5$) the source shows a flat spectrum again, consistent with the median spectral index which could indicate that the optically thick core component dominates the emission again (after $\sim2022.5$). After the second flare, we observe a similar spectral steepening, however, this part of the light curve is not as well sampled as the first flare and no dense VLBI data is available to probe for fast features associated with the spectral steepening.

\subsection{Polarization properties}

\begin{figure*}
    \centering
    \includegraphics[width=.73\linewidth]{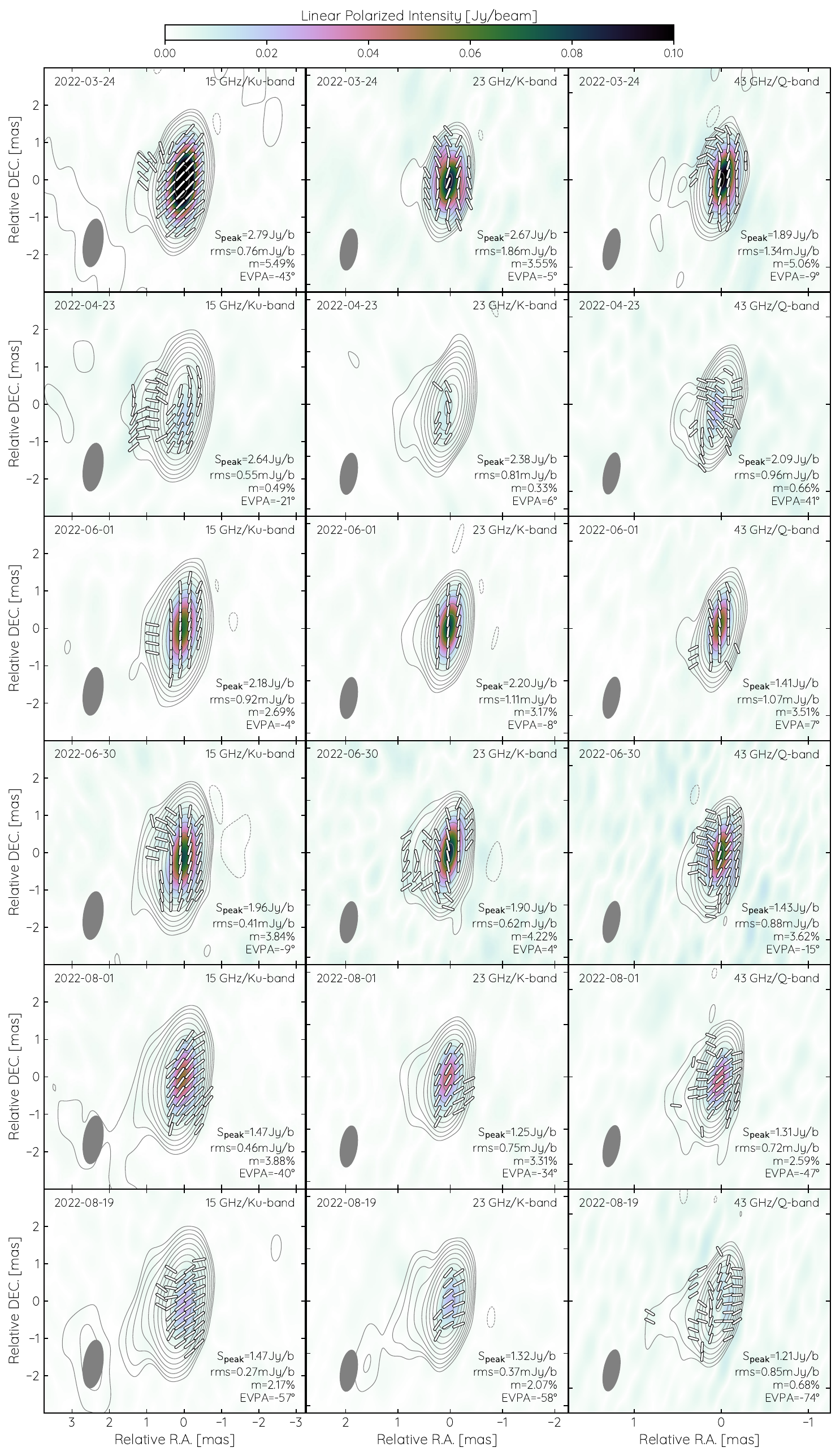}
    \caption{Full Polarization VLBI images of PKS\,0215+015 from the conducted ToO campaign, convolved with the median beam for each frequency. The colormap indicates linear polarized intensity, contours correspond to total intensity and start at five times the noise level, increasing by factors of two. The direction of the EVPA is indicated by the tilted lines.}
    \label{fig:all_pol_images}
\end{figure*}

Based on the VLBA and TELAMON data, we analyzed the radio polarization properties of PKS\,0215+015. In Fig.\,\ref{fig:variability} (center panel), we show the evolution of the fractional polarization with time, the position angles of the EVPA are indicated by the tilted lines plotted on top of the data points. Fig.\,\ref{fig:variability} displays both the TELAMON and VLBA polarization information that have completely independent calibration procedures, but show consistent results. The source shows strong variability in fractional polarization. Shortly after the neutrino event, we observed a flare of fractional polarization, reaching values close to $\sim6$\,\%, followed by a sudden drop of fractional polarization down to values $\lesssim1\,\%$, and an associated change in the EVPA direction. At the same time (i.e., $\sim2022.3$) the source is still in an elevated state in total intensity. After the fractional polarization dip, the source remains in the range of 1$-$4\,\% fractional polarization which is similar to previous activity, based on archival MOJAVE observations. In Fig.\,\ref{fig:all_pol_images}, we display all full polarization images from the VLBA ToO campaign at all frequencies. The images reveal that the polarization originates almost entirely from the region close to the core throughout the observing campaign. The strong polarization flare and dip in the second epoch is clearly visible in the images. Some epochs show hints of extended polarized emission with perpendicular EVPA as compared to the core, which becomes better visible in the stacked maps in Fig.\,\ref{fig:stacked_pol_maps}. The EVPA swing observed at the time of the polarization dip (consistent with VLBA epoch 2) is clearly visible in the VLBI images in Fig.\,\ref{fig:all_pol_images}. As discussed in Sect.\,\ref{sec:vlba}, this epoch was affected by low amplitudes at several stations. However, we have carefully checked the cross-hand signals to confirm that the low polarization is not due to instrumental reasons.

We have estimated the Faraday rotation measure, $\mathrm{RM} = (\chi_\mathrm{obs} - \chi_0)/\lambda^2$, where $\chi_\mathrm{obs}$ is the observed EVPA, $\chi_0$ the intrinsic EVPA, and $\lambda$ the observing wavelength. The median RM across all TELAMON epochs where 20\,mm and 7\,mm polarization data was available is 117\,rad/m$^2$. From the VLBI observations, we obtain a median RM value of $-154$\,rad/m$^2$, after fitting the RM across all three VLBI frequencies for every epoch. Typical uncertainties for the RM per epoch are on the order of 200$-$300\,rad/m$^2$ for TELAMON, and $\sim500$\,rad/m$^2$ for the VLBA data, i.e., both datasets suggest an RM consistent with zero. A clear outlier is seen in the second VLBI epoch, at the time of the polarization dip, where the fitted VLBI rotation measure is $-(3000\pm500)$\,rad/m$^2$.

\subsection{Doppler factor, Lorentz factor and viewing angle}
\label{sec:lorentz}

With knowledge of the apparent jet speed, $\beta_\mathrm{app}$, and the Doppler factor, $\delta$, one can estimate the bulk Lorentz factor,
\begin{equation}
\label{eq:gamma}
\Gamma=\frac{\beta_\mathrm{app}^2 + \delta^2+1}{2\delta},
\end{equation}
and the jet viewing angle, 
\begin{equation}
\vartheta=\mathrm{arctan}\frac{2\beta_\mathrm{app}}{\beta_\mathrm{app}^2 + \delta^2-1},
\end{equation}
following \cite{ValtaojaLaehteenmaki1999}. As the apparent speed $\beta_\mathrm{app}$, we consider the speed of the fast component 1 for every frequency individually, as determined by our kinematic analysis (see Tab.\,\ref{tab:kinematics}). For the Doppler factor, we use the variability Doppler factors of flare 1 as reported in Tab.\,\ref{tab:variability}, since it is likely associated with the jet ejection. We calculated the Lorentz factor and viewing angle for all frequencies independently. The Lorentz factor obtained at 15\,GHz is $\Gamma_\mathrm{Ku}=78\pm67$, consistent with the 23\,GHz value $\Gamma_\mathrm{K}=160\pm100$, while at 43\,GHz, the Lorentz factor cannot be constrained in a sensible way due to the large uncertainty of $\beta_\mathrm{app}$ in that band (formally $\Gamma_\mathrm{Q}=140\pm180$).
\review{Apart from $\beta_\mathrm{app}$, the main source of error here is the uncertainty of the Doppler factor used in Eq.\,\eqref{eq:gamma}. As discussed in Sect.\,\ref{sec:variability}, $\delta_\mathrm{var}$ can be \textbf{(i)} underestimated due to an undersampled light curve and \textbf{(ii)} overestimated due to an elevated intrinsic brightness temperature during the flare that exceeds the equipartition value. Independent of the variability properties, the Doppler factor can also be estimated from the VLBI core brightness temperature, $T_\mathrm{b,core}$, as $\delta_\mathrm{VLBI}=T_\mathrm{b,core}/T_\mathrm{b,int}$ \citep{Lahteenmaki_1999}. Indeed, making the same assumption as in Sect.\,\ref{sec:variability}, i.e., $T_\mathrm{b,int}=T_\mathrm{b,eq}$, and using the core $T_\mathrm{b}$ values from Tab.\,\ref{tab:components} results in significantly higher Doppler factors than obtained from the variability, i.e., $\delta_\mathrm{VLBI}\sim\mathcal{O}(10^2)$. This suggests that $T_\mathrm{b,int}$ likely exceeds the equipartition value, at least during some parts of the flare. Thus, the variability Doppler factors presented above would be overestimated. However, if $\delta\lesssim10$, the resulting Lorentz factor would become unrealistically high, i.e., $\Gamma\gtrsim200$. Therefore, we likely observe a combination of both effects, (i) and (ii), such that the $\delta_\mathrm{var}$ values presented in Sect.\,\ref{sec:variability} are still reasonable estimates. As a more conservative approach we have estimated a lower limit for $\Gamma$ by assuming the critical Doppler factor, $\delta_\mathrm{crit}=\sqrt{1+\beta_\mathrm{app}^2}$, i.e., the analytical minimum of Eq.\,\eqref{eq:gamma}. Following this approach, it becomes clear that} the Lorentz factor related to apparent speeds of 60$-$80\,c must be at least $\Gamma>60-80$, regardless \review{of the assumed Doppler factor}. For the following discussion, we adapt the weighted mean value between 15\,GHz and 23\,GHz, $\Gamma=105\pm56$, calculated with the $\delta_\mathrm{var}$ values, \review{consistent with the lower limit value}. The viewing angle between all frequency bands is consistent within its error, i.e., $\vartheta_\mathrm{Ku}=(1.65\pm0.57)^\circ$, $\vartheta_\mathrm{K}=(1.34\pm0.38)^\circ$, and $\vartheta_\mathrm{Q}=(2.0\pm1.3)^\circ$. This suggests a very small jet inclination angle, with a weighted mean of $\vartheta=(1.47\pm0.31)^\circ$ from all frequencies combined.

We note that already in previous studies, at times when PKS\,0215+015 did not show such exceptional flaring activity, \cite{Lister19} have reported superluminal jet speeds of $\beta_\mathrm{app}=(25.3\pm1.2)$\,c for this source. Based on these findings, \cite{Sebastian22} have estimated a Lorentz factor of $\Gamma=37.4$, and a viewing angle of $\theta=0.6^\circ$. A slight change of the viewing angle and an acceleration of the jet seems possible, given the major flaring event immediately before our observations.

\section{Discussion}
\label{sec:discussion}

\subsection{Fast component ejection \& shock-shock interaction}

\begin{figure}
    \centering
    \includegraphics[width=\linewidth]{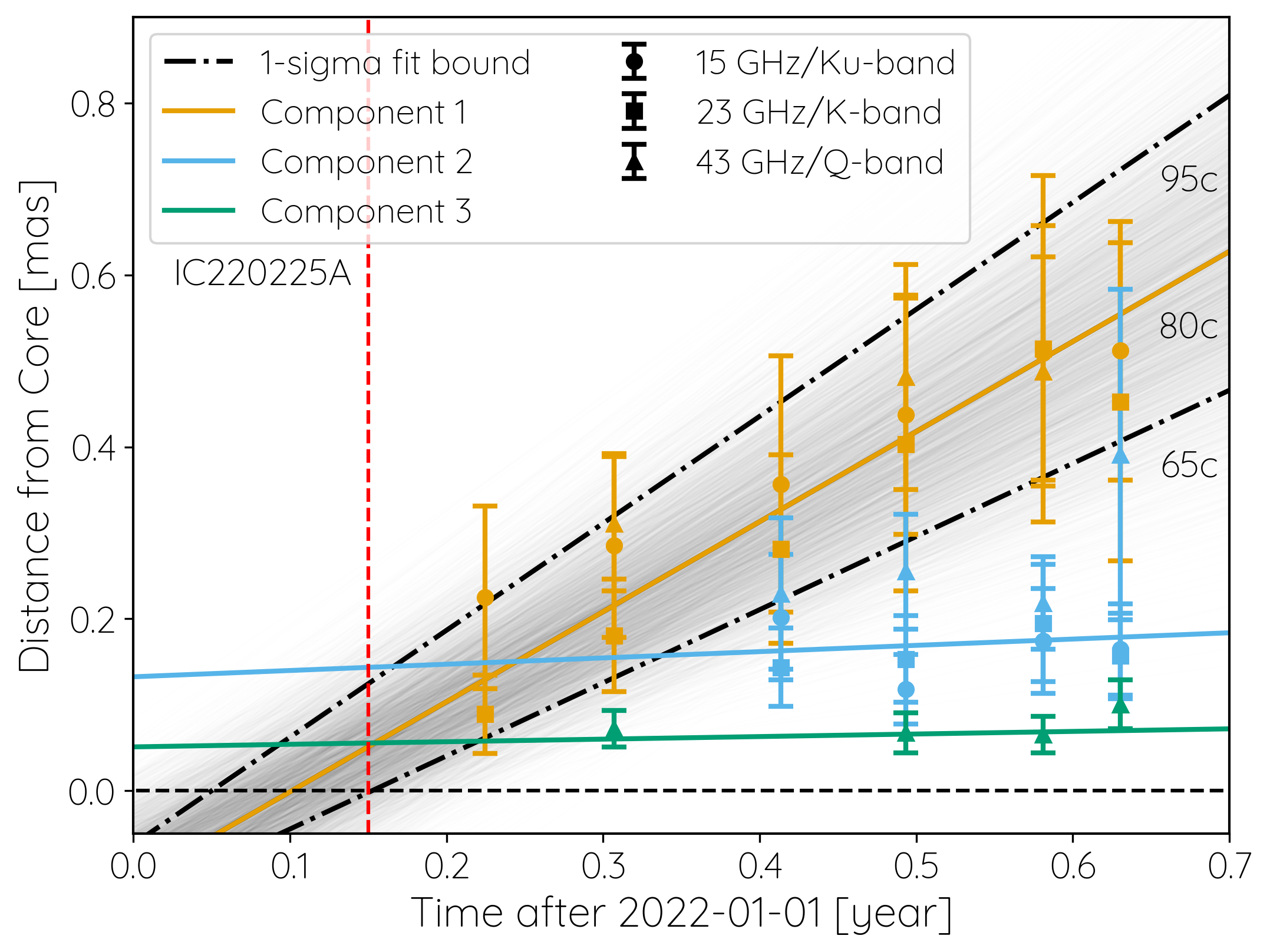}
    \caption{Combined kinematic fit from 15\,GHz, 23\,GHz, and 43\,GHz for component 1, 2 and 3. The solid line indicates the best fit. For component 1, the dash-dotted lines represent the 1$\sigma$ fit bounds, and the shaded area in the background reflects the confidence range of the fit. The time of arrival of the neutrino event IC220225A is indicated by the red dashed line and coincides with the passage of component 1 through component 3.}
    \label{fig:comps_combined}
    \vspace{-.5\baselineskip}
\end{figure}

We interpret the fast ejection of component 1, revealed in the kinematic analysis (see Sect.\,\ref{sec:kinematics}), as a new jet feature, i.e., a traveling shock wave, that was born in response to the major flare observed in PKS\,0215+015 in early 2022. The ejection of new jet features in response to major flares in AGN is well known \citep[e.g.,][]{Jorstad2001,Savolainen2002}, and their interpretation as traveling shock waves is in line with the shocked jet model \citep{MarscherGear}. The new feature appears to be very fast, with an apparent speed of $\beta_\mathrm{app}$ $\sim$ 60$-$80\,c, and resulting bulk Lorentz factor $\Gamma=105\pm56$. We note that both of these values exceed the maximum apparent speeds ($\beta_\mathrm{app}\lesssim 50$\,c) and Lorentz factors ($\Gamma\lesssim50$) reported for sources in the MOJAVE and VLBA-BU-BLAZAR AGN monitoring programs \citep{Lister21,Homan2021,Weaver22}. Several other works, however, suggest Lorentz factors $\Gamma>60$ in exceptional cases \citep[e.g.,][]{Fujisawa1999,Hovatta2009}. 

The case of PKS\,0215+015 is likely such an exceptional event, where we observe a source at high redshift (z=1.72), immediately after the peak of a major outburst. The MOJAVE program observes their sample sources with a cadence of several months or longer, which makes it difficult to detect features with $\beta_\mathrm{app}\gtrsim50$\,c reliably, especially at lower redshifts.
Revealing such (possibly short-lived) fast components is only possible with high-cadence (i.e., $\lesssim$ monthly) monitoring, and in response to major flaring events when new ejections are expected.

Due to its proximity to the core component, the uncertainty of the speed of component 1 is relatively large, when using the three observing frequencies separately as presented in Sect.\,\ref{sec:kinematics}. Under the assumption of a negligible core shift (see App.\,\ref{app:collimation}), it is possible to use the component positions across frequencies for a combined kinematic fit, which yields slightly lower uncertainties. This is shown in Fig.\,\ref{fig:comps_combined}, where we have performed a frequency-combined fit for jet components 1$-$3. While component 2 and 3 do not exhibit significant motion, the combined fit for component 1 suggests a speed of $\mu_\mathrm{app}=(80\pm15)$\,c, with an ejection epoch of \review{$2022.10\pm0.05$}, consistent with the results reported in Sect.\,\ref{sec:kinematics}. The simultaneous presence of stationary features suggests that this fast motion is not related to a change in synchrotron opacity \citep[e.g.,][]{PlavinCoreShift,Chamani} but represents the actual motion of the jet plasma. As shown in Fig.\,\ref{fig:comps_combined}, the ejection of component 1 from the core happened simultaneous to, or shortly before the arrival of the neutrino event IC220225A (see Fig.\,\ref{fig:lightcurves}). Shortly after its ejection, the fast component 1 passes through the stationary components 3 and 2, likely resulting in shock-shock interactions.

An additional trace of the fast motion of component 1 through the jet, and the shock-shock interactions, can be seen in the observed polarization light curve (see Fig.\,\ref{fig:variability}, center panel). If we assume that the component was ejected from the core around the time of the neutrino event, a sudden rise in fractional polarization is expected, due to the resulting opacity changes \citep[e.g.,][]{Lara2001}. Usually, the core component is expected to show relatively low fractional polarization due to synchrotron self-absorption \citep[e.g.,][]{Pollack2003}. Therefore, it is not surprising that at the peak of the total intensity flare, the fractional polarization is still in a relatively low state of $\sim$2\,\%. As the new component is ejected from the core, the emission becomes more optically thin, and therefore a rise in fractional polarization is observed. Alternatively, the initial low-state in fractional polarization could already be the state of the first shock-shock interaction between component 1 and 3. The subsequent dip of fractional polarization is likely explained by a shock-shock interaction \citep[e.g.,][]{Fromm2013Kinematic} of the fast component 1 with the stationary component 2, located between 0.1$-$0.2\,mas away from the core. This interaction is in agreement with the observed VLBI component propagation (see Fig.\,\ref{fig:kinematics}). If we assume that component 2 exhibits a different EVPA than component 1, the induced shock-shock interaction causes a net depolarization, which is observed in $\sim2022.3-2022.4$. This scenario is supported by the fact that in the stacked VLBI maps, we do see hints of a second polarized component with perpendicular EVPA to the core, especially in the stacked 15\,GHz map in Fig.\,\ref{fig:stacked_pol_maps} (left).
A shock-shock interaction can also explain the change in net EVPA direction \citep[e.g.,][]{Liodakis20,ParaschosOJ248}, observed around the time of the polarization dip, suggesting that component 2 slightly outshines component 1 and the core at that time. The net EVPA rotation could in principle also be caused by an external Faraday screen, however, since the RM in PKS\,0215+015 for almost all epochs is close to zero, this is an unlikely scenario. The large outlier RM value observed at the time of the suggested shock-shock interaction in the second VLBI epoch is likely caused by the fact that we are observing different stages of the interaction at the different frequencies, due to synchrotron opacity. While at Q-band, the polarization is likely already dominated by the stationary feature, at 15\,GHz, there is still a significant contribution from the core to the net EVPA visible, which causes an apparently large RM.

As component 1 moves away from component 2, its flux \review{density} decays and the core component dominates the emission again with typical fractional polarization values of $\sim$1$-$4\,\%. As described in Sect.\,\ref{sec:variability}, the spectral index behavior is in agreement with the fast component ejection, showing a steepening of the spectrum as the component moves downstream, and then returning to a flat spectrum (after $\sim2022.5$), when the core emission dominates again. After flare 2 (see Fig.\,\ref{fig:variability}), a similar behavior in polarization and spectral steepening is observed, however, no simultaneous, dense VLBI data is available at that time to probe for a second fast component ejection possibly related to the second flare. Another possible explanation of the second flare could be a shock-shock interaction between the fast feature and another standing shock further downstream, possibly component 4.

\subsection{Possibilities for neutrino production}

Neutrinos in relativistic jets can be produced through hadronic ($pp$) or photopion ($p\gamma$) interactions \citep[e.g.,][]{Mannheim1993,Mannheim1995,Atoyan2001}, with the latter being thought to be the dominant process \citep[e.g.,][]{Oikonomou22}. Most theoretical models assume a spherical emitting region (blob) with bulk Lorentz factor $\Gamma_\mathrm{blob}$, where protons are accelerated. The protons interact with a target photon field, resulting in the production of high-energy neutrinos through the $p\gamma$-channel if protons of sufficiently high energy are present. The target photon field can be internal to the jet (usually assumed for BL Lac objects) or external to the jet \citep[possible for FSRQs, e.g., photons from the accretion disk or broad line region,][]{Oikonomou19}. In the case of PKS\,0215+015, our results suggest that the fast blob where proton acceleration happens could be consistent with the newly ejected component 1 with bulk Lorentz factor $\Gamma=105\pm56$, but even the mildly relativistic quasi-stationary shocks close to the core are capable of accelerating protons to the required energies for neutrino production \citep[e.g.,][]{Plavin2021,Kalashev2023}. Additional evidence for the presence of $p\gamma$-interactions comes from the strong gamma-ray flare observed at the time of the neutrino event. Since the classification of PKS\,0215+015 between FSRQ and BL Lac is uncertain and the source previously showed a changing-look AGN behavior \citep[e.g.,][]{Foschini2022}, there are multiple possibilities for the target photon field. While external photons from the accretion disk or broad line region \review{are} a possible target photon field in FSRQ-type sources, the presence of fast moving shocks and stationary jet components at the same time possibly suggests a multi-layered jet configuration in PKS\,0215+015. The shock-shock interaction between component 1 and component 3, happening almost simultaneous to the time of neutrino arrival (see Fig.\,\ref{fig:comps_combined}) could present a scenario where a slower jet region acts as the target photon field for the relativistic protons that are accelerated in a faster shock region \citep[e.g.,][]{Ghisellini2005,Tavecchio2014,Tavecchio2015}. While we cannot resolve the detailed jet structure of PKS\,0215+015 due to its high redshift and small viewing angle, such a configuration would be consistent with our findings and has already been suggested for other neutrino-candidate blazars. \cite{Ansoldi18} were able to explain the neutrino emission in TXS\,0506+056 using a multi-layered spine-sheath model, but also for sources like PKS\,1424+240 and PKS\,0735+178 a multi-layered jet configuration seems possible \citep{Kovalev1424,Kim2025_0735,Paraschos2025}. In the latter case, however, studies of the multiwavelength spectral energy distribution prefer external target photon fields \citep{Acharyya0735,Sahakyan2023_0735,Prince24}. Additional insights on the nature of the target photon field in PKS\,0215+015 could come from similar modeling approaches in the future, if applied to the available multiwavelength data.


\subsection{Significance of the neutrino association}

\begin{figure*}
    \centering
    \includegraphics[width=\linewidth]{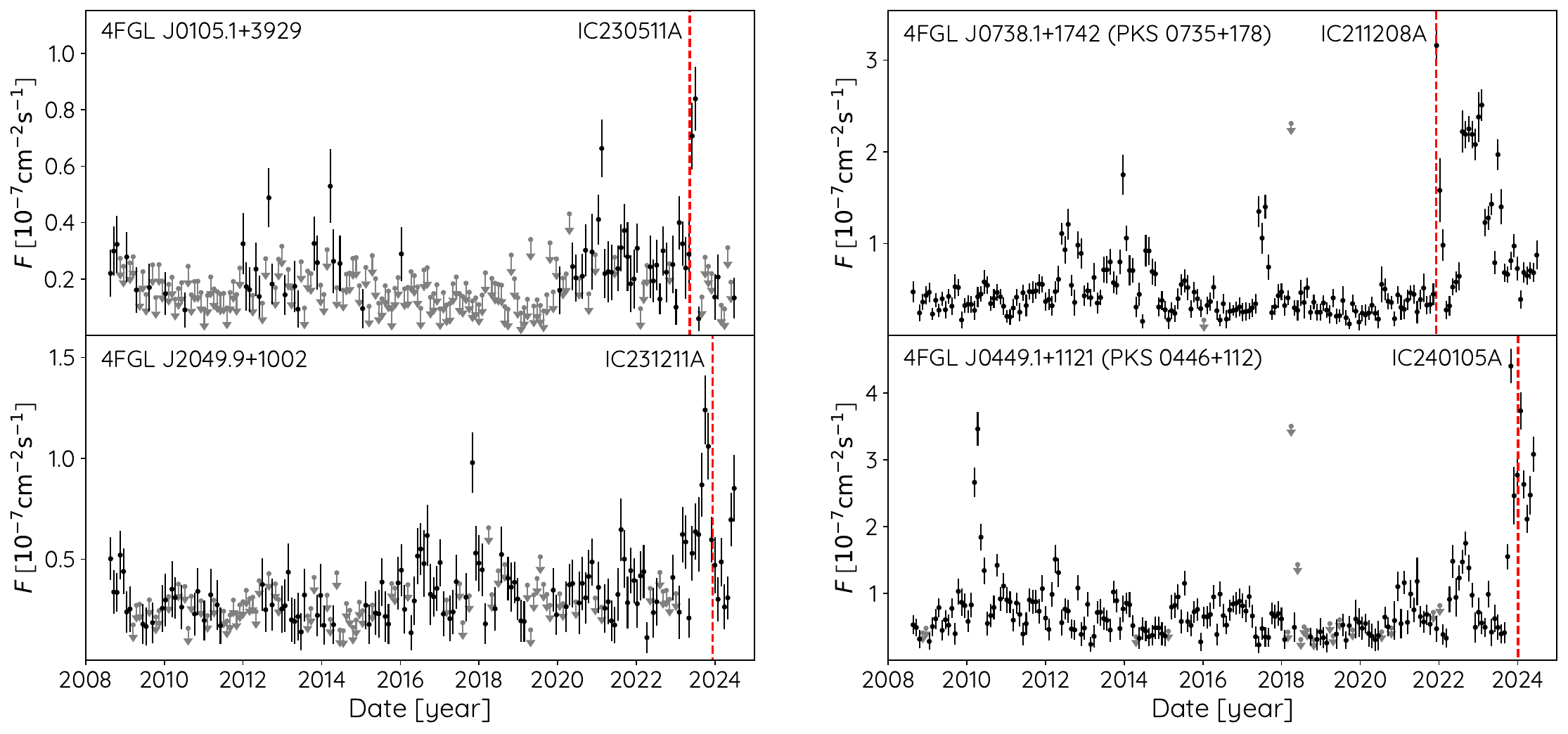}
    \caption{\reviewtwo{Monthly-binned} gamma-ray light curves of additional neutrino-candidate sources obtained from the \textsl{Fermi}/LAT light curve repository. All sources show gamma-ray flaring on historical levels in temporal ($\pm3$\,months) and spatial proximity with IceCube neutrino events. The two sources shown on the left are located inside the 90\,\% confidence regions of their respective neutrino events. The two sources shown on the right are located slightly outside of the listed neutrino events, but are discussed as likely associations in the literature.}
    \label{fig:fermi_flares}
\end{figure*}

The significance of the association between high-energy neutrino events and astrophysical sources can be addressed in different ways, and using different strategies. Several studies have previously focused on cross-matching various types of source catalogs with neutrino events \citep[e.g.,][]{Plavin2020,Plavin2021,Plavin2023,Buson22,FermiIceCube24,IceCubeFermi}, 
but also individual source flares were investigated \citep[e.g.,][]{Kadler2016,IceCubeTXS0506}. 
In order to estimate the significance of the correlation between the flare in PKS\,0215+015 and the neutrino event IC220225A we consider the arrival time of the neutrino in correlation to the monthly-binned \textsl{Fermi}/LAT light curve. The maximum of the \textsl{Fermi}/LAT light curve occurs immediately following the neutrino arrival time. Assuming the null hypothesis that the neutrino arrival time is uncorrelated with the light curve peak, the probability of such a coincidence occurring by chance is $p=1/194\approx0.52\,\%$, with 194 light curve bins. This corresponds to a nominal significance of $\sim2.6\sigma$ under Gaussian statistics. This significance needs to be corrected for the number of trials. In order to estimate the trial factor, we used all IceCube track alerts published since introduction of the real time alert system in June 2019 \citep{IceCubeAlerts} until June 2024. This comprises a total of 137 IceCube neutrinos. For each event, we consider the uncertainty boxes as published in the GCN circulars by IceCube, and cross match them with sources from the 4FGL-DR3 catalog. We find 61 neutrino events without a 4FGL source, 20 events with 1 source, 17 events with 2 sources, 10 events with 3 sources, and 29 events with more than 3 sources. Similar to the approach by \cite{FermiIceCube24}, in the following, we consider only well localized events, i.e., events with $\leq3$ 4FGL sources in the field. Inside these 108 events, we find a total number of 84 4FGL sources, of which only 20 have available light curves in the \textsl{Fermi}/LAT light curve repository \citep[i.e., their variability index is $>21.67$, see][]{FermiLCrepo}. This means the obtained pre-trial p-value of 0.0052 needs to be corrected by a trial factor of 20, resulting in a post-trial p-value of 10\,\%, equivalent to a significance of 1.27$\sigma$.
 
During the cross-matching of the \textsl{Fermi}/LAT light curves with the neutrino arrival times, we noticed two additional coincidences between historical gamma-ray flares and neutrino arrival times in the sources 4FGL\,J0105.1+3929 (IC230511A) and 4FGL\,J2049.9+1002 (IC231211A), shown in Fig.\,\ref{fig:fermi_flares}. For both sources, the historical gamma-ray flux maximum was observed $\pm3$ monthly bins away from the neutrino arrival time. Including PKS\,0215+015 this adds up to 3 out of 20 sources exhibiting their historical gamma-ray flux peak within 3\,months of a spatially coincident neutrino event. Similar to the single-source estimate for PKS\,0215+015, we estimate the chance coincidence that the arrival times of the high-energy neutrinos are uncorrelated to the flux-maxima within $\pm3$\, monthly bins. For a single light curve, the pre-trial p-value is p=6/194$\approx$3.1\,\% (3 bins before the neutrino + 3 bins after the neutrino). In order to test the null-hypothesis that three or more such detections happen \review{by} chance in 20 independent light curves, we describe the system as a binomial process
\begin{equation}
    P(k; n, p) = \binom{n}{k} p^k (1 - p)^{n - k},
\end{equation}
where $n$ is the number of light curves analyzed (here, $n=20$), $k$ is the number of historical flares within $\pm$3 monthly bins of the neutrino arrival time, and $p$ is the single light curve p-value (here, $p=3.1\,\%$). The probability of finding three or more such coincidences by chance can be calculated with
\begin{equation}
    P(k\geq3)=1-P(k=0)-P(k=1)-P(k=2)\approx2.27\,\%.
\end{equation}
This corresponds to a significance of 2$\sigma$, likely linking the historical flares to the neutrino arrival times. Additional evidence comes from the neutrino-candidate blazars PKS\,0735+178 \citep[IC211208A, e.g.,][]{FermiIceCube24} and PKS\,0446+112 \citep[IC240105A, e.g.,][]{0446FermiAtel}, shown in Fig.\,\ref{fig:fermi_flares}. They are formally located slightly outside of the GCN neutrino regions, but showed strong gamma-ray flaring in temporal coincidence with neutrino events, similar to PKS\,0215+015. Due to their location slightly outside of IceCube uncertainty regions, it is difficult to include them into our significance estimate. Moreover, a strong gamma-ray outburst was also observed in the first neutrino blazar, TXS\,0506+056 \citep{IceCubeTXS0506}, providing a strong link between gamma-ray flares in blazars and neutrino-emission.

Additionally, we emphasize that the reported physical properties of the jet (small inclination angle and rapid jet ejection with high Lorentz factor) provide strong evidence for the connection of PKS\,0215+015 with neutrino event IC220225A that cannot be expressed in terms of a statistical significance value.

\section{Summary and outlook} 
\label{sec:conclusions}

We observed the blazar PKS\,0215+015 (z=1.72) in response to a major multiwavelength outburst that happened simultaneous to the spatially coincident neutrino track-alert IC220225A. Our VLBA kinematic analysis revealed the ejection of a rapid new jet component with apparent speed $\beta_\mathrm{app}$ $\sim$60$-$80\,c. In addition to this fast component, we also observed two stationary features in the jet, located within $<0.3$\,mas ($<100$\,pc de-projected) from the core. Shortly after the rapid ejection, we found a characteristic signature in fractional polarization, which first showed a significant increase and then a sudden drop, associated with a rotation of the EVPA. We interpret this behavior as a shock-shock interaction of the new-born component with a stationary feature of perpendicular EVPA which results in a fast depolarization and net EVPA rotation as the fast component passes through it. We used radio monitoring data from the TELAMON and ATCA monitoring programs to estimate the variability Doppler factor related to the strong radio flare. Using this information, we estimated the bulk Lorentz factor of the fast component $\Gamma=105\pm56$, and the jet viewing angle $\vartheta=(1.47 \pm 0.31)^\circ$ of PKS\,0215+015. We note that the observed apparent speed, and as a result the Lorentz factor, is higher than the maximum values reported by continuous AGN VLBI monitoring programs \citep{Lister21,Weaver22}. The measurement of such high jet speeds is likely only possible in the aftermath of exceptional flares and with high-cadence (i.e., $\lesssim$\,monthly) observations. Additionally, the high redshift leads to a \enquote{slow-motion} effect, i.e., fast features with an apparent speed of $60-80$\,c are seen at angular speeds of only $\sim$0.8$-$1.0\,mas/yr. In the local universe (z$\sim$0.1), such fast features would appear about an order of magnitude faster which makes them challenging to trace with typical VLBI monitoring cadences of months to years. Dense polarization monitoring programs like TELAMON are therefore crucial as an additional tracer of the underlying jet dynamics. Even though the chance coincidence between the neutrino event and the historical gamma-ray flare is not negligible ($\sim10\,\%$ post-trial), the rapid jet ejection and shock-shock interaction close in time to the neutrino provide additional physical hints for the connection of jet activity and neutrino production that cannot be expressed as a significance value. Proton acceleration can in principle happen in the relativistic shocks close to the core \citep[cf.,][]{Plavin2021}, however, the target photon field for the $p\gamma$-process remains less constrained in our study. The presence of quasi-stationary and rapid jet components at the same time as revealed in our kinematic study could suggest a multi-layered jet configuration, where the slower jet regions provide the target photon field for the $p\gamma$-process \citep[e.g.,][]{Ghisellini2005,Tavecchio2014,Tavecchio2015}, but also target photon fields external to the jet are possible.
Our study shows that rapid, possibly short-lived components can be ejected in response to major flares in blazars, exceeding speeds usually reported by VLBI monitoring programs. Such a scenario would present a solution to the bulk Lorentz-factor crisis, also called \enquote{Doppler crisis} \citep{DopplerCrisis} for TeV-emitting blazars. The high Lorentz factors usually suggested by TeV observations could be matched by the high Lorentz factors associated with fast ejections in response to major flares. While PKS\,0215+015 is not yet detected at TeV-energies \citep[likely due to its high redshift, e.g.,][]{EBL_absorption}, and thus not affected by the Doppler crisis, triggered, dense VLBI follow-up observations of flaring Doppler crisis blazars should be able to reveal similarly fast features, if they exist in these sources.

\begin{acknowledgements}
\review{We thank the anonymous referee for valuable comments that helped us improve the manuscript.} 
This work is based on observations with the 100-m telescope of the MPIfR (Max-Planck-Institut für Radioastronomie) at Effelsberg. FE, MK, JH, and FR acknowledge support from the German Science Foundation (DFG, grants 447572188, 434448349, 465409577, 443220636 [FOR5195:
Relativistic Jets in Active Galaxies]) and the German Aerospace Center (DLR, grant 50OR2506). SH is partly supported by the German Science Foundation (DFG grant numbers WI 1860/14-1 and 434448349). We acknowledge the M2FINDERS project from the European Research Council (ERC) under the European Union's Horizon 2020 research and innovation programme (grant agreement No~101018682). YYK was supported by the MuSES project, which has received funding from the European Union (ERC grant agreement No~101142396). Views and opinions expressed are those of the authors only and do not necessarily reflect those of the European Union or ERCEA. Neither the European Union nor the granting authority can be held responsible for them. This research has made use of the Astrophysics Data System, funded by NASA under Cooperative Agreement 80NSSC21M00561. The VLBA is an instrument of the National Radio Astronomy Observatory. The National Radio Astronomy Observatory and Green Bank Observatory are facilities of the U.S. National Science Foundation operated under cooperative agreement by Associated Universities, Inc. This work made use of the Swinburne University of Technology software correlator, developed as part of the Australian Major National Research Facilities Programme and operated under licence \citep{DiFX}. The Australia Telescope Compact Array is part of the Australia Telescope National Facility (\href{https://ror.org/05qajvd42}{https://ror.org/05qajvd42}) which is funded by the Australian Government for operation as a National Facility managed by CSIRO. We acknowledge the Gomeroi people as the Traditional Owners of the Observatory site. This research has made use of data from the MOJAVE database that is maintained by the MOJAVE team \citep{MOJAVE}. This study makes use of VLBA data from the VLBA-BU Blazar Monitoring Program (BEAM-ME and VLBA-BU-BLAZAR; \href{http://www.bu.edu/blazars/BEAM-ME.html}{http://www.bu.edu/blazars/BEAM-ME.html}), funded by NASA through the Fermi Guest Investigator Program.
\end{acknowledgements}

\bibliographystyle{aa} 
\bibliography{bib} 

@ARTICLE{Planck2020,
       author = {{Planck Collaboration} and {Aghanim}, N. and {Akrami}, Y. and {Ashdown}, M. and {Aumont}, J. and {Baccigalupi}, C. and {Ballardini}, M. and {Banday}, A.~J. and {Barreiro}, R.~B. and {Bartolo}, N. and {Basak}, S. and {Battye}, R. and {Benabed}, K. and {Bernard}, J. -P. and {Bersanelli}, M. and {Bielewicz}, P. and {Bock}, J.~J. and {Bond}, J.~R. and {Borrill}, J. and {Bouchet}, F.~R. and {Boulanger}, F. and {Bucher}, M. and {Burigana}, C. and {Butler}, R.~C. and {Calabrese}, E. and {Cardoso}, J. -F. and {Carron}, J. and {Challinor}, A. and {Chiang}, H.~C. and {Chluba}, J. and {Colombo}, L.~P.~L. and {Combet}, C. and {Contreras}, D. and {Crill}, B.~P. and {Cuttaia}, F. and {de Bernardis}, P. and {de Zotti}, G. and {Delabrouille}, J. and {Delouis}, J. -M. and {Di Valentino}, E. and {Diego}, J.~M. and {Dor{\'e}}, O. and {Douspis}, M. and {Ducout}, A. and {Dupac}, X. and {Dusini}, S. and {Efstathiou}, G. and {Elsner}, F. and {En{\ss}lin}, T.~A. and {Eriksen}, H.~K. and {Fantaye}, Y. and {Farhang}, M. and {Fergusson}, J. and {Fernandez-Cobos}, R. and {Finelli}, F. and {Forastieri}, F. and {Frailis}, M. and {Fraisse}, A.~A. and {Franceschi}, E. and {Frolov}, A. and {Galeotta}, S. and {Galli}, S. and {Ganga}, K. and {G{\'e}nova-Santos}, R.~T. and {Gerbino}, M. and {Ghosh}, T. and {Gonz{\'a}lez-Nuevo}, J. and {G{\'o}rski}, K.~M. and {Gratton}, S. and {Gruppuso}, A. and {Gudmundsson}, J.~E. and {Hamann}, J. and {Handley}, W. and {Hansen}, F.~K. and {Herranz}, D. and {Hildebrandt}, S.~R. and {Hivon}, E. and {Huang}, Z. and {Jaffe}, A.~H. and {Jones}, W.~C. and {Karakci}, A. and {Keih{\"a}nen}, E. and {Keskitalo}, R. and {Kiiveri}, K. and {Kim}, J. and {Kisner}, T.~S. and {Knox}, L. and {Krachmalnicoff}, N. and {Kunz}, M. and {Kurki-Suonio}, H. and {Lagache}, G. and {Lamarre}, J. -M. and {Lasenby}, A. and {Lattanzi}, M. and {Lawrence}, C.~R. and {Le Jeune}, M. and {Lemos}, P. and {Lesgourgues}, J. and {Levrier}, F. and {Lewis}, A. and {Liguori}, M. and {Lilje}, P.~B. and {Lilley}, M. and {Lindholm}, V. and {L{\'o}pez-Caniego}, M. and {Lubin}, P.~M. and {Ma}, Y. -Z. and {Mac{\'\i}as-P{\'e}rez}, J.~F. and {Maggio}, G. and {Maino}, D. and {Mandolesi}, N. and {Mangilli}, A. and {Marcos-Caballero}, A. and {Maris}, M. and {Martin}, P.~G. and {Martinelli}, M. and {Mart{\'\i}nez-Gonz{\'a}lez}, E. and {Matarrese}, S. and {Mauri}, N. and {McEwen}, J.~D. and {Meinhold}, P.~R. and {Melchiorri}, A. and {Mennella}, A. and {Migliaccio}, M. and {Millea}, M. and {Mitra}, S. and {Miville-Desch{\^e}nes}, M. -A. and {Molinari}, D. and {Montier}, L. and {Morgante}, G. and {Moss}, A. and {Natoli}, P. and {N{\o}rgaard-Nielsen}, H.~U. and {Pagano}, L. and {Paoletti}, D. and {Partridge}, B. and {Patanchon}, G. and {Peiris}, H.~V. and {Perrotta}, F. and {Pettorino}, V. and {Piacentini}, F. and {Polastri}, L. and {Polenta}, G. and {Puget}, J. -L. and {Rachen}, J.~P. and {Reinecke}, M. and {Remazeilles}, M. and {Renzi}, A. and {Rocha}, G. and {Rosset}, C. and {Roudier}, G. and {Rubi{\~n}o-Mart{\'\i}n}, J.~A. and {Ruiz-Granados}, B. and {Salvati}, L. and {Sandri}, M. and {Savelainen}, M. and {Scott}, D. and {Shellard}, E.~P.~S. and {Sirignano}, C. and {Sirri}, G. and {Spencer}, L.~D. and {Sunyaev}, R. and {Suur-Uski}, A. -S. and {Tauber}, J.~A. and {Tavagnacco}, D. and {Tenti}, M. and {Toffolatti}, L. and {Tomasi}, M. and {Trombetti}, T. and {Valenziano}, L. and {Valiviita}, J. and {Van Tent}, B. and {Vibert}, L. and {Vielva}, P. and {Villa}, F. and {Vittorio}, N. and {Wandelt}, B.~D. and {Wehus}, I.~K. and {White}, M. and {White}, S.~D.~M. and {Zacchei}, A. and {Zonca}, A.},
        title = "{Planck 2018 results. VI. Cosmological parameters}",
      journal = {\aap},
         year = 2020,
        month = sep,
       volume = {641},
          eid = {A6},
        pages = {A6},
          doi = {10.1051/0004-6361/201833910},
archivePrefix = {arXiv},
       eprint = {1807.06209},
 primaryClass = {astro-ph.CO},
       adsurl = {https://ui.adsabs.harvard.edu/abs/2020A&A...641A...6P}
}

@ARTICLE{TELAMON,
       author = {{Eppel}, F. and {Kadler}, M. and {He{\ss}d{\"o}rfer}, J. and {Benke}, P. and {Debbrecht}, L. and {Eich}, J. and {Gokus}, A. and {H{\"a}mmerich}, S. and {Kirchner}, D. and {Paraschos}, G.~F. and {R{\"o}sch}, F. and {Schulga}, W. and {Sinapius}, J. and {Weber}, P. and {Bach}, U. and {Dorner}, D. and {Edwards}, P.~G. and {Giroletti}, M. and {Kraus}, A. and {Hervet}, O. and {Koyama}, S. and {Krichbaum}, T.~P. and {Mannheim}, K. and {Ros}, E. and {Zacharias}, M. and {Zensus}, J.~A.},
        title = "{TELAMON: Effelsberg monitoring of AGN jets with very-high-energy astroparticle emission. I. Program description and sample characterization}",
      journal = {\aap},
         year = 2024,
        month = apr,
       volume = {684},
          eid = {A11},
        pages = {A11},
          doi = {10.1051/0004-6361/202348262},
archivePrefix = {arXiv},
       eprint = {2401.06296},
 primaryClass = {astro-ph.HE},
       adsurl = {https://ui.adsabs.harvard.edu/abs/2024A&A...684A..11E}
}

@ARTICLE{MOJAVE,
       author = {{Lister}, M.~L. and {Aller}, M.~F. and {Aller}, H.~D. and {Hodge}, M.~A. and {Homan}, D.~C. and {Kovalev}, Y.~Y. and {Pushkarev}, A.~B. and {Savolainen}, T.},
        title = "{MOJAVE. XV. VLBA 15 GHz Total Intensity and Polarization Maps of 437 Parsec-scale AGN Jets from 1996 to 2017}",
      journal = {\apjs},
         year = 2018,
        month = jan,
       volume = {234},
       number = {1},
          eid = {12},
        pages = {12},
          doi = {10.3847/1538-4365/aa9c44},
archivePrefix = {arXiv},
       eprint = {1711.07802},
 primaryClass = {astro-ph.GA},
       adsurl = {https://ui.adsabs.harvard.edu/abs/2018ApJS..234...12L}
}

@INPROCEEDINGS{ParselTongue,
       author = {{Kettenis}, M. and {van Langevelde}, H.~J. and {Reynolds}, C. and {Cotton}, B.},
        title = "{ParselTongue: AIPS Talking Python}",
    booktitle = {Astronomical Data Analysis Software and Systems XV},
         year = 2006,
       editor = {{Gabriel}, C. and {Arviset}, C. and {Ponz}, D. and {Enrique}, S.},
       series = {Astronomical Society of the Pacific Conference Series},
       volume = {351},
        month = jul,
        pages = {497},
       adsurl = {https://ui.adsabs.harvard.edu/abs/2006ASPC..351..497K}
}

@INPROCEEDINGS{AIPS,
       author = {{Greisen}, E.~W.},
        title = "{The Astronomical Image Processing System.}",
    booktitle = {Acquisition, Processing and Archiving of Astronomical Images},
         year = 1990,
       editor = {{Longo}, Guiseppe and {Sedmak}, Giorgio},
        month = jan,
        pages = {125-142},
       adsurl = {https://ui.adsabs.harvard.edu/abs/1990apaa.conf..125G}
}

@ARTICLE{CASA,
       author = {{CASA Team} and {Bean}, Ben and {Bhatnagar}, Sanjay and {Castro}, Sandra and {Donovan Meyer}, Jennifer and {Emonts}, Bjorn and {Garcia}, Enrique and {Garwood}, Robert and {Golap}, Kumar and {Gonzalez Villalba}, Justo and {Harris}, Pamela and {Hayashi}, Yohei and {Hoskins}, Josh and {Hsieh}, Mingyu and {Jagannathan}, Preshanth and {Kawasaki}, Wataru and {Keimpema}, Aard and {Kettenis}, Mark and {Lopez}, Jorge and {Marvil}, Joshua and {Masters}, Joseph and {McNichols}, Andrew and {Mehringer}, David and {Miel}, Renaud and {Moellenbrock}, George and {Montesino}, Federico and {Nakazato}, Takeshi and {Ott}, Juergen and {Petry}, Dirk and {Pokorny}, Martin and {Raba}, Ryan and {Rau}, Urvashi and {Schiebel}, Darrell and {Schweighart}, Neal and {Sekhar}, Srikrishna and {Shimada}, Kazuhiko and {Small}, Des and {Steeb}, Jan-Willem and {Sugimoto}, Kanako and {Suoranta}, Ville and {Tsutsumi}, Takahiro and {van Bemmel}, Ilse M. and {Verkouter}, Marjolein and {Wells}, Akeem and {Xiong}, Wei and {Szomoru}, Arpad and {Griffith}, Morgan and {Glendenning}, Brian and {Kern}, Jeff},
        title = "{CASA, the Common Astronomy Software Applications for Radio Astronomy}",
      journal = {\pasp},
         year = 2022,
        month = nov,
       volume = {134},
       number = {1041},
          eid = {114501},
        pages = {114501},
          doi = {10.1088/1538-3873/ac9642},
archivePrefix = {arXiv},
       eprint = {2210.02276},
 primaryClass = {astro-ph.IM},
       adsurl = {https://ui.adsabs.harvard.edu/abs/2022PASP..134k4501C}
}

@ARTICLE{PolSolve,
       author = {{Mart{\'\i}-Vidal}, I. and {Mus}, A. and {Janssen}, M. and {de Vicente}, P. and {Gonz{\'a}lez}, J.},
        title = "{Polarization calibration techniques for the new-generation VLBI}",
      journal = {\aap},
         year = 2021,
        month = feb,
       volume = {646},
          eid = {A52},
        pages = {A52},
          doi = {10.1051/0004-6361/202039527},
archivePrefix = {arXiv},
       eprint = {2012.05581},
 primaryClass = {astro-ph.IM},
       adsurl = {https://ui.adsabs.harvard.edu/abs/2021A&A...646A..52M}
}

@ARTICLE{entropyclean,
       author = {{Homan}, D.~C. and {Roth}, J.~S. and {Pushkarev}, A.~B.},
        title = "{Residual Entropy as a Diagnostic and Stopping Metric for CLEAN}",
      journal = {\aj},
         year = 2024,
        month = jan,
       volume = {167},
       number = {1},
          eid = {11},
        pages = {11},
          doi = {10.3847/1538-3881/ad0beb},
archivePrefix = {arXiv},
       eprint = {2311.05427},
 primaryClass = {astro-ph.IM},
       adsurl = {https://ui.adsabs.harvard.edu/abs/2024AJ....167...11H}
}

@INPROCEEDINGS{difmap,
       author = {{Shepherd}, M.~C.},
        title = "{Difmap: an Interactive Program for Synthesis Imaging}",
    booktitle = {Astronomical Data Analysis Software and Systems VI},
         year = 1997,
       editor = {{Hunt}, Gareth and {Payne}, Harry},
       series = {Astronomical Society of the Pacific Conference Series},
       volume = {125},
        month = jan,
        pages = {77},
       adsurl = {https://ui.adsabs.harvard.edu/abs/1997ASPC..125...77S}
}

@ARTICLE{rpicard,
       author = {{Janssen}, M. and {Goddi}, C. and {van Bemmel}, I.~M. and {Kettenis}, M. and {Small}, D. and {Liuzzo}, E. and {Rygl}, K. and {Mart{\'\i}-Vidal}, I. and {Blackburn}, L. and {Wielgus}, M. and {Falcke}, H.},
        title = "{rPICARD: A CASA-based calibration pipeline for VLBI data. Calibration and imaging of 7 mm VLBA observations of the AGN jet in M 87}",
      journal = {\aap},
         year = 2019,
        month = jun,
       volume = {626},
          eid = {A75},
        pages = {A75},
          doi = {10.1051/0004-6361/201935181},
archivePrefix = {arXiv},
       eprint = {1905.01905},
 primaryClass = {astro-ph.IM},
       adsurl = {https://ui.adsabs.harvard.edu/abs/2019A&A...626A..75J}
}

@ARTICLE{MOJAVEOpeningAngle2017,
       author = {{Pushkarev}, A.~B. and {Kovalev}, Y.~Y. and {Lister}, M.~L. and {Savolainen}, T.},
        title = "{MOJAVE - XIV. Shapes and opening angles of AGN jets}",
      journal = {\mnras},
         year = 2017,
        month = jul,
       volume = {468},
       number = {4},
        pages = {4992-5003},
          doi = {10.1093/mnras/stx854},
archivePrefix = {arXiv},
       eprint = {1705.02888},
 primaryClass = {astro-ph.HE},
       adsurl = {https://ui.adsabs.harvard.edu/abs/2017MNRAS.468.4992P}
}

@ARTICLE{ATCAmonitoring,
       author = {{Stevens}, J. and {Edwards}, P.~G. and {Ojha}, R. and {Kadler}, M. and {Hungwe}, F. and {Dutka}, M. and {Tingay}, S. and {Macquart}, J.~P. and {Moin}, A. and {Lovell}, J. and {Blanchard}, J.},
        title = "{ATCA monitoring of gamma-ray loud AGN}",
      journal = {arXiv e-prints},
         year = 2012,
        month = may,
          eid = {arXiv:1205.2403},
        pages = {arXiv:1205.2403},
          doi = {10.48550/arXiv.1205.2403},
archivePrefix = {arXiv},
       eprint = {1205.2403},
 primaryClass = {astro-ph.HE},
       adsurl = {https://ui.adsabs.harvard.edu/abs/2012arXiv1205.2403S}
}

@article{TELAMONpolarization,
  author = "Hessdoerfer, Jonas  and  Kadler, Matthias  and  Benke, Petra  and  Debbrecht, Lena  and  Eich, Julia  and  Eppel, Florian  and  Gokus, Andrea  and  Hämmerich, Steven  and  Kirchner, Dana  and  Paraschos, Georgios  and  Rösch, Florian  and  Schulga, Wladislaw  and  Sinapius, Jonas L.  and  Weber, Philip  and  Bach, Uwe  and  Berge, David  and  Buson, Sara  and  Dorner, Daniela  and  Edwards, Philip G.  and  Fromm, Christian M.  and  Giroletti, Marcello  and  Hervet, Olivier  and  Kappes, Alexander  and  Koyama, Shoko  and  Kraus, Alex  and  Krichbaum, Thomas  and  Lindfors, Elina  and  Mannheim, Karl  and  de Menezes, Raniere  and  Ojha, Roopesh  and  Pueschel, Elisa  and  Ros, Eduardo  and  Schleicher, Bernd  and  Sitarek, Julian  and  Wilms, Jörn  and  Zacharias, Michael  and  Zensus, J. Anton",
  title = "{TELAMON: Effelsberg Monitoring of AGN Jets with Very-High-Energy Astroparticle Emissions - Polarization properties}",
  doi = "10.22323/1.444.1545",
  journal = "PoS",
  year = 2023,
  volume = "ICRC2023",
  pages = "1545"
}

@ARTICLE{Tavecchio2014,
       author = {{Tavecchio}, Fabrizio and {Ghisellini}, Gabriele and {Guetta}, Dafne},
        title = "{Structured Jets in BL Lac Objects: Efficient PeV Neutrino Factories?}",
      journal = {\apjl},
         year = 2014,
        month = sep,
       volume = {793},
       number = {1},
          eid = {L18},
        pages = {L18},
          doi = {10.1088/2041-8205/793/1/L18},
archivePrefix = {arXiv},
       eprint = {1407.0907},
 primaryClass = {astro-ph.HE},
       adsurl = {https://ui.adsabs.harvard.edu/abs/2014ApJ...793L..18T}
}

@ARTICLE{Plavin2021,
       author = {{Plavin}, A.~V. and {Kovalev}, Y.~Y. and {Kovalev}, Yu. A. and {Troitsky}, S.~V.},
        title = "{Directional Association of TeV to PeV Astrophysical Neutrinos with Radio Blazars}",
      journal = {\apj},
         year = 2021,
        month = feb,
       volume = {908},
       number = {2},
          eid = {157},
        pages = {157},
          doi = {10.3847/1538-4357/abceb8},
archivePrefix = {arXiv},
       eprint = {2009.08914},
 primaryClass = {astro-ph.HE},
       adsurl = {https://ui.adsabs.harvard.edu/abs/2021ApJ...908..157P}
}

@ARTICLE{Plavin2020,
       author = {{Plavin}, Alexander and {Kovalev}, Yuri Y. and {Kovalev}, Yuri A. and {Troitsky}, Sergey},
        title = "{Observational Evidence for the Origin of High-energy Neutrinos in Parsec-scale Nuclei of Radio-bright Active Galaxies}",
      journal = {\apj},
         year = 2020,
        month = may,
       volume = {894},
       number = {2},
          eid = {101},
        pages = {101},
          doi = {10.3847/1538-4357/ab86bd},
archivePrefix = {arXiv},
       eprint = {2001.00930},
 primaryClass = {astro-ph.HE},
       adsurl = {https://ui.adsabs.harvard.edu/abs/2020ApJ...894..101P}
}

@ARTICLE{DopplerCrisis,
       author = {{Henri}, Gilles and {Saug{\'e}}, Ludovic},
        title = "{The Bulk Lorentz Factor Crisis of TeV Blazars: Evidence for an Inhomogeneous Pileup Energy Distribution?}",
      journal = {\apj},
         year = 2006,
        month = mar,
       volume = {640},
       number = {1},
        pages = {185-195},
          doi = {10.1086/500039},
archivePrefix = {arXiv},
       eprint = {astro-ph/0511610},
 primaryClass = {astro-ph},
       adsurl = {https://ui.adsabs.harvard.edu/abs/2006ApJ...640..185H}
}

@ARTICLE{IceCat1,
       author = {{Abbasi}, R. and {Ackermann}, M. and {Adams}, J. and {Agarwalla}, S.~K. and {Aguilar}, J.~A. and {Ahlers}, M. and {Alameddine}, J.~M. and {Amin}, N.~M. and {Andeen}, K. and {Anton}, G. and {Arg{\"u}elles}, C. and {Ashida}, Y. and {Athanasiadou}, S. and {Axani}, S.~N. and {Bai}, X. and {Balagopal}, V.~A. and {Baricevic}, M. and {Barwick}, S.~W. and {Basu}, V. and {Bay}, R. and {Beatty}, J.~J. and {Becker}, K. -H. and {Becker Tjus}, J. and {Beise}, J. and {Bellenghi}, C. and {BenZvi}, S. and {Berley}, D. and {Bernardini}, E. and {Besson}, D.~Z. and {Binder}, G. and {Bindig}, D. and {Blaufuss}, E. and {Blot}, S. and {Bontempo}, F. and {Book}, J.~Y. and {Boscolo Meneguolo}, C. and {B{\"o}ser}, S. and {Botner}, O. and {B{\"o}ttcher}, J. and {Bourbeau}, E. and {Braun}, J. and {Brinson}, B. and {Brostean-Kaiser}, J. and {Burley}, R.~T. and {Busse}, R.~S. and {Butterfield}, D. and {Campana}, M.~A. and {Carloni}, K. and {Carnie-Bronca}, E.~G. and {Chattopadhyay}, S. and {Chau}, N. and {Chen}, C. and {Chen}, Z. and {Chirkin}, D. and {Choi}, S. and {Clark}, B.~A. and {Classen}, L. and {Coleman}, A. and {Collin}, G.~H. and {Connolly}, A. and {Conrad}, J.~M. and {Coppin}, P. and {Correa}, P. and {Countryman}, S. and {Cowen}, D.~F. and {Dave}, P. and {De Clercq}, C. and {DeLaunay}, J.~J. and {Delgado}, D. and {Dembinski}, H. and {Deng}, S. and {Deoskar}, K. and {Desai}, A. and {Desiati}, P. and {de Vries}, K.~D. and {de Wasseige}, G. and {DeYoung}, T. and {Diaz}, A. and {D{\'\i}az-V{\'e}lez}, J.~C. and {Dittmer}, M. and {Domi}, A. and {Dujmovic}, H. and {DuVernois}, M.~A. and {Ehrhardt}, T. and {Eller}, P. and {Engel}, R. and {Erpenbeck}, H. and {Evans}, J. and {Evenson}, P.~A. and {Fan}, K.~L. and {Fang}, K. and {Farrag}, K. and {Fazely}, A.~R. and {Fedynitch}, A. and {Feigl}, N. and {Fiedlschuster}, S. and {Finley}, C. and {Fischer}, L. and {Fox}, D. and {Franckowiak}, A. and {Friedman}, E. and {Fritz}, A. and {F{\"u}rst}, P. and {Gaisser}, T.~K. and {Gallagher}, J. and {Ganster}, E. and {Garcia}, A. and {Gerhardt}, L. and {Ghadimi}, A. and {Glaser}, C. and {Glauch}, T. and {Gl{\"u}senkamp}, T. and {Goehlke}, N. and {Gonzalez}, J.~G. and {Goswami}, S. and {Grant}, D. and {Gray}, S.~J. and {Griffin}, S. and {Griswold}, S. and {G{\"u}nther}, C. and {Gutjahr}, P. and {Haack}, C. and {Hallgren}, A. and {Halliday}, R. and {Halve}, L. and {Halzen}, F. and {Hamdaoui}, H. and {Ha Minh}, M. and {Hanson}, K. and {Hardin}, J. and {Harnisch}, A.~A. and {Hatch}, P. and {Haungs}, A. and {Helbing}, K. and {Hellrung}, J. and {Henningsen}, F. and {Heuermann}, L. and {Heyer}, N. and {Hickford}, S. and {Hidvegi}, A. and {Hill}, C. and {Hill}, G.~C. and {Hoffman}, K.~D. and {Hoshina}, K. and {Hou}, W. and {Huber}, T. and {Hultqvist}, K. and {H{\"u}nnefeld}, M. and {Hussain}, R. and {Hymon}, K. and {In}, S. and {Ishihara}, A. and {Jacquart}, M. and {Janik}, O. and {Jansson}, M. and {Japaridze}, G.~S. and {Jayakumar}, K. and {Jeong}, M. and {Jin}, M. and {Jones}, B.~J.~P. and {Kang}, D. and {Kang}, W. and {Kang}, X. and {Kappes}, A. and {Kappesser}, D. and {Kardum}, L. and {Karg}, T. and {Karl}, M. and {Karle}, A. and {Katz}, U. and {Kauer}, M. and {Kelley}, J.~L. and {Zathul}, A. Khatee and {Kheirandish}, A. and {Kiryluk}, J. and {Klein}, S.~R. and {Kochocki}, A. and {Koirala}, R. and {Kolanoski}, H. and {Kontrimas}, T. and {K{\"o}pke}, L. and {Kopper}, C. and {Koskinen}, D.~J. and {Koundal}, P. and {Kovacevich}, M. and {Kowalski}, M. and {Kozynets}, T. and {Kruiswijk}, K. and {Krupczak}, E. and {Kumar}, A. and {Kun}, E. and {Kurahashi}, N. and {Lad}, N. and {Lagunas Gualda}, C. and {Lamoureux}, M. and {Larson}, M.~J. and {Lauber}, F. and {Lazar}, J.~P. and {Lee}, J.~W. and {Leonard DeHolton}, K.},
        title = "{IceCat-1: The IceCube Event Catalog of Alert Tracks}",
      journal = {\apjs},
         year = 2023,
        month = nov,
       volume = {269},
       number = {1},
          eid = {25},
        pages = {25},
          doi = {10.3847/1538-4365/acfa95},
archivePrefix = {arXiv},
       eprint = {2304.01174},
 primaryClass = {astro-ph.HE},
       adsurl = {https://ui.adsabs.harvard.edu/abs/2023ApJS..269...25A}
}

@ARTICLE{IceCubeTXS0506,
       author = {{IceCube Collaboration} and {Aartsen}, M.~G. and {Ackermann}, M. and {Adams}, J. and {Aguilar}, J.~A. and {Ahlers}, M. and {Ahrens}, M. and {Al Samarai}, I. and {Altmann}, D. and {Andeen}, K. and {Anderson}, T. and {Ansseau}, I. and {Anton}, G. and {Arg{\"u}elles}, C. and {Auffenberg}, J. and {Axani}, S. and {Bagherpour}, H. and {Bai}, X. and {Barron}, J.~P. and {Barwick}, S.~W. and {Baum}, V. and {Bay}, R. and {Beatty}, J.~J. and {Becker Tjus}, J. and {Becker}, K. -H. and {BenZvi}, S. and {Berley}, D. and {Bernardini}, E. and {Besson}, D.~Z. and {Binder}, G. and {Bindig}, D. and {Blaufuss}, E. and {Blot}, S. and {Bohm}, C. and {B{\"o}rner}, M. and {Bos}, F. and {B{\"o}ser}, S. and {Botner}, O. and {Bourbeau}, E. and {Bourbeau}, J. and {Bradascio}, F. and {Braun}, J. and {Brenzke}, M. and {Bretz}, H. -P. and {Bron}, S. and {Brostean-Kaiser}, J. and {Burgman}, A. and {Busse}, R.~S. and {Carver}, T. and {Cheung}, E. and {Chirkin}, D. and {Christov}, A. and {Clark}, K. and {Classen}, L. and {Coenders}, S. and {Collin}, G.~H. and {Conrad}, J.~M. and {Coppin}, P. and {Correa}, P. and {Cowen}, D.~F. and {Cross}, R. and {Dave}, P. and {Day}, M. and {de Andr{\'e}}, J.~P.~A.~M. and {De Clercq}, C. and {DeLaunay}, J.~J. and {Dembinski}, H. and {De Ridder}, S. and {Desiati}, P. and {de Vries}, K.~D. and {de Wasseige}, G. and {de With}, M. and {DeYoung}, T. and {D{\'\i}az-V{\'e}lez}, J.~C. and {di Lorenzo}, V. and {Dujmovic}, H. and {Dumm}, J.~P. and {Dunkman}, M. and {Dvorak}, E. and {Eberhardt}, B. and {Ehrhardt}, T. and {Eichmann}, B. and {Eller}, P. and {Evenson}, P.~A. and {Fahey}, S. and {Fazely}, A.~R. and {Felde}, J. and {Filimonov}, K. and {Finley}, C. and {Flis}, S. and {Franckowiak}, A. and {Friedman}, E. and {Fritz}, A. and {Gaisser}, T.~K. and {Gallagher}, J. and {Gerhardt}, L. and {Ghorbani}, K. and {Glauch}, T. and {Gl{\"u}senkamp}, T. and {Goldschmidt}, A. and {Gonzalez}, J.~G. and {Grant}, D. and {Griffith}, Z. and {Haack}, C. and {Hallgren}, A. and {Halzen}, F. and {Hanson}, K. and {Hebecker}, D. and {Heereman}, D. and {Helbing}, K. and {Hellauer}, R. and {Hickford}, S. and {Hignight}, J. and {Hill}, G.~C. and {Hoffman}, K.~D. and {Hoffmann}, R. and {Hoinka}, T. and {Hokanson-Fasig}, B. and {Hoshina}, K. and {Huang}, F. and {Huber}, M. and {Hultqvist}, K. and {H{\"u}nnefeld}, M. and {Hussain}, R. and {In}, S. and {Iovine}, N. and {Ishihara}, A. and {Jacobi}, E. and {Japaridze}, G.~S. and {Jeong}, M. and {Jero}, K. and {Jones}, B.~J.~P. and {Kalaczynski}, P. and {Kang}, W. and {Kappes}, A. and {Kappesser}, D. and {Karg}, T. and {Karle}, A. and {Katz}, U. and {Kauer}, M. and {Keivani}, A. and {Kelley}, J.~L. and {Kheirandish}, A. and {Kim}, J. and {Kim}, M. and {Kintscher}, T. and {Kiryluk}, J. and {Kittler}, T. and {Klein}, S.~R. and {Koirala}, R. and {Kolanoski}, H. and {K{\"o}pke}, L. and {Kopper}, C. and {Kopper}, S. and {Koschinsky}, J.~P. and {Koskinen}, D.~J. and {Kowalski}, M. and {Krings}, K. and {Kroll}, M. and {Kr{\"u}ckl}, G. and {Kunwar}, S. and {Kurahashi}, N. and {Kuwabara}, T. and {Kyriacou}, A. and {Labare}, M. and {Lanfranchi}, J.~L. and {Larson}, M.~J. and {Lauber}, F. and {Leonard}, K. and {Lesiak-Bzdak}, M. and {Leuermann}, M. and {Liu}, Q.~R. and {Lozano Mariscal}, C.~J. and {Lu}, L. and {L{\"u}nemann}, J. and {Luszczak}, W. and {Madsen}, J. and {Maggi}, G. and {Mahn}, K.~B.~M. and {Mancina}, S. and {Maruyama}, R. and {Mase}, K. and {Maunu}, R. and {Meagher}, K. and {Medici}, M. and {Meier}, M. and {Menne}, T. and {Merino}, G. and {Meures}, T. and {Miarecki}, S. and {Micallef}, J. and {Moment{\'e}}, G. and {Montaruli}, T. and {Moore}, R.~W. and {Morse}, R. and {Moulai}, M. and {Nahnhauer}, R. and {Nakarmi}, P. and {Naumann}, U. and {Neer}, G.},
        title = "{Multimessenger observations of a flaring blazar coincident with high-energy neutrino IceCube-170922A}",
      journal = {Science},
         year = 2018,
        month = jul,
       volume = {361},
       number = {6398},
          eid = {eaat1378},
        pages = {eaat1378},
          doi = {10.1126/science.aat1378},
archivePrefix = {arXiv},
       eprint = {1807.08816},
 primaryClass = {astro-ph.HE},
       adsurl = {https://ui.adsabs.harvard.edu/abs/2018Sci...361.1378I}
}

@ARTICLE{Kadler2016,
       author = {{Kadler}, M. and {Krau{\ss}}, F. and {Mannheim}, K. and {Ojha}, R. and {M{\"u}ller}, C. and {Schulz}, R. and {Anton}, G. and {Baumgartner}, W. and {Beuchert}, T. and {Buson}, S. and {Carpenter}, B. and {Eberl}, T. and {Edwards}, P.~G. and {Eisenacher Glawion}, D. and {Els{\"a}sser}, D. and {Gehrels}, N. and {Gr{\"a}fe}, C. and {Gulyaev}, S. and {Hase}, H. and {Horiuchi}, S. and {James}, C.~W. and {Kappes}, A. and {Kappes}, A. and {Katz}, U. and {Kreikenbohm}, A. and {Kreter}, M. and {Kreykenbohm}, I. and {Langejahn}, M. and {Leiter}, K. and {Litzinger}, E. and {Longo}, F. and {Lovell}, J.~E.~J. and {McEnery}, J. and {Natusch}, T. and {Phillips}, C. and {Pl{\"o}tz}, C. and {Quick}, J. and {Ros}, E. and {Stecker}, F.~W. and {Steinbring}, T. and {Stevens}, J. and {Thompson}, D.~J. and {Tr{\"u}stedt}, J. and {Tzioumis}, A.~K. and {Weston}, S. and {Wilms}, J. and {Zensus}, J.~A.},
        title = "{Coincidence of a high-fluence blazar outburst with a PeV-energy neutrino event}",
      journal = {Nature Physics},
         year = 2016,
        month = aug,
       volume = {12},
       number = {8},
        pages = {807-814},
          doi = {10.1038/nphys3715},
archivePrefix = {arXiv},
       eprint = {1602.02012},
 primaryClass = {astro-ph.HE},
       adsurl = {https://ui.adsabs.harvard.edu/abs/2016NatPh..12..807K}
}

@article{Mannheim1995,
title = {High-energy neutrinos from extragalactic jets},
journal = {Astroparticle Physics},
volume = {3},
number = {3},
pages = {295-302},
year = {1995},
issn = {0927-6505},
doi = {https://doi.org/10.1016/0927-6505(94)00044-4},
url = {https://www.sciencedirect.com/science/article/pii/0927650594000444},
author = {Karl Mannheim}
}

@ARTICLE{Mannheim1993,
       author = {{Mannheim}, K.},
        title = "{The proton blazar.}",
      journal = {\aap},
         year = 1993,
        month = mar,
       volume = {269},
        pages = {67-76},
archivePrefix = {arXiv},
       eprint = {astro-ph/9302006},
 primaryClass = {astro-ph},
       adsurl = {https://ui.adsabs.harvard.edu/abs/1993A&A...269...67M}
}

@ARTICLE{Hovatta2009,
       author = {{Hovatta}, T. and {Valtaoja}, E. and {Tornikoski}, M. and {L{\"a}hteenm{\"a}ki}, A.},
        title = "{Doppler factors, Lorentz factors and viewing angles for quasars, BL Lacertae objects and radio galaxies}",
      journal = {\aap},
         year = 2009,
        month = feb,
       volume = {494},
       number = {2},
        pages = {527-537},
          doi = {10.1051/0004-6361:200811150},
archivePrefix = {arXiv},
       eprint = {0811.4278},
 primaryClass = {astro-ph},
       adsurl = {https://ui.adsabs.harvard.edu/abs/2009A&A...494..527H}
}

@ARTICLE{Homan2021,
       author = {{Homan}, D.~C. and {Cohen}, M.~H. and {Hovatta}, T. and {Kellermann}, K.~I. and {Kovalev}, Y.~Y. and {Lister}, M.~L. and {Popkov}, A.~V. and {Pushkarev}, A.~B. and {Ros}, E. and {Savolainen}, T.},
        title = "{MOJAVE. XIX. Brightness Temperatures and Intrinsic Properties of Blazar Jets}",
      journal = {\apj},
         year = 2021,
        month = dec,
       volume = {923},
       number = {1},
          eid = {67},
        pages = {67},
          doi = {10.3847/1538-4357/ac27af},
archivePrefix = {arXiv},
       eprint = {2109.04977},
 primaryClass = {astro-ph.HE},
       adsurl = {https://ui.adsabs.harvard.edu/abs/2021ApJ...923...67H}
}

@ARTICLE{Lister19,
       author = {{Lister}, M.~L. and {Homan}, D.~C. and {Hovatta}, T. and {Kellermann}, K.~I. and {Kiehlmann}, S. and {Kovalev}, Y.~Y. and {Max-Moerbeck}, W. and {Pushkarev}, A.~B. and {Readhead}, A.~C.~S. and {Ros}, E. and {Savolainen}, T.},
        title = "{MOJAVE. XVII. Jet Kinematics and Parent Population Properties of Relativistically Beamed Radio-loud Blazars}",
      journal = {\apj},
         year = 2019,
        month = mar,
       volume = {874},
       number = {1},
          eid = {43},
        pages = {43},
          doi = {10.3847/1538-4357/ab08ee},
archivePrefix = {arXiv},
       eprint = {1902.09591},
 primaryClass = {astro-ph.GA},
       adsurl = {https://ui.adsabs.harvard.edu/abs/2019ApJ...874...43L}
}

@ARTICLE{Lara2001,
       author = {{Lara}, L. and {Alberdi}, A. and {Marcaide}, J.~M.},
        title = "{Variability and polarization in the inner jet of 3C 395}",
      journal = {\aap},
         year = 2001,
        month = mar,
       volume = {368},
        pages = {817-823},
          doi = {10.1051/0004-6361:20010059},
archivePrefix = {arXiv},
       eprint = {astro-ph/0101325},
 primaryClass = {astro-ph},
       adsurl = {https://ui.adsabs.harvard.edu/abs/2001A&A...368..817L}
}

@ARTICLE{Pollack2003,
       author = {{Pollack}, L.~K. and {Taylor}, G.~B. and {Zavala}, R.~T.},
        title = "{VLBI Polarimetry of 177 Sources from the Caltech-Jodrell Bank Flat-Spectrum Survey}",
      journal = {\apj},
         year = 2003,
        month = jun,
       volume = {589},
       number = {2},
        pages = {733-751},
          doi = {10.1086/374712},
archivePrefix = {arXiv},
       eprint = {astro-ph/0302211},
 primaryClass = {astro-ph},
       adsurl = {https://ui.adsabs.harvard.edu/abs/2003ApJ...589..733P}
}

@ARTICLE{Prince24,
       author = {{Prince}, Raj and {Das}, Saikat and {Gupta}, Nayantara and {Majumdar}, Pratik and {Czerny}, Bo{\.z}ena},
        title = "{Dissecting the broad-band emission from {\ensuremath{\gamma}}-ray blazar PKS 0735+178 in search of neutrinos}",
      journal = {\mnras},
         year = 2024,
        month = jan,
       volume = {527},
       number = {3},
        pages = {8746-8754},
          doi = {10.1093/mnras/stad3804},
archivePrefix = {arXiv},
       eprint = {2301.06565},
 primaryClass = {astro-ph.HE},
       adsurl = {https://ui.adsabs.harvard.edu/abs/2024MNRAS.527.8746P}
}

@ARTICLE{DiFX,
       author = {{Deller}, A.~T. and {Brisken}, W.~F. and {Phillips}, C.~J. and {Morgan}, J. and {Alef}, W. and {Cappallo}, R. and {Middelberg}, E. and {Romney}, J. and {Rottmann}, H. and {Tingay}, S.~J. and {Wayth}, R.},
        title = "{DiFX-2: A More Flexible, Efficient, Robust, and Powerful Software Correlator}",
      journal = {\pasp},
         year = 2011,
        month = mar,
       volume = {123},
       number = {901},
        pages = {275},
          doi = {10.1086/658907},
archivePrefix = {arXiv},
       eprint = {1101.0885},
 primaryClass = {astro-ph.IM},
       adsurl = {https://ui.adsabs.harvard.edu/abs/2011PASP..123..275D}
}

@ARTICLE{Jorstad2001,
       author = {{Jorstad}, Svetlana G. and {Marscher}, Alan P. and {Mattox}, John R. and {Aller}, Margo F. and {Aller}, Hugh D. and {Wehrle}, Ann E. and {Bloom}, Steven D.},
        title = "{Multiepoch Very Long Baseline Array Observations of EGRET-detected Quasars and BL Lacertae Objects: Connection between Superluminal Ejections and Gamma-Ray Flares in Blazars}",
      journal = {\apj},
         year = 2001,
        month = aug,
       volume = {556},
       number = {2},
        pages = {738-748},
          doi = {10.1086/321605},
archivePrefix = {arXiv},
       eprint = {astro-ph/0102012},
 primaryClass = {astro-ph},
       adsurl = {https://ui.adsabs.harvard.edu/abs/2001ApJ...556..738J}
}

@ARTICLE{Liodakis20,
       author = {{Liodakis}, I. and {Blinov}, D. and {Jorstad}, S.~G. and {Arkharov}, A.~A. and {Di Paola}, A. and {Efimova}, N.~V. and {Grishina}, T.~S. and {Kiehlmann}, S. and {Kopatskaya}, E.~N. and {Larionov}, V.~M. and {Larionova}, L.~V. and {Larionova}, E.~G. and {Marscher}, A.~P. and {Morozova}, D.~A. and {Nikiforova}, A.~A. and {Pavlidou}, V. and {Traianou}, E. and {Troitskaya}, Yu. V. and {Troitsky}, I.~S. and {Uemura}, M. and {Weaver}, Z.~R.},
        title = "{Two Flares with One Shock: The Interesting Case of 3C 454.3}",
      journal = {\apj},
         year = 2020,
        month = oct,
       volume = {902},
       number = {1},
          eid = {61},
        pages = {61},
          doi = {10.3847/1538-4357/abb1b8},
archivePrefix = {arXiv},
       eprint = {2008.08603},
 primaryClass = {astro-ph.HE},
       adsurl = {https://ui.adsabs.harvard.edu/abs/2020ApJ...902...61L}
}

@ARTICLE{ParaschosOJ248,
       author = {{Paraschos}, G.~F.},
        title = "{A shocking outcome: Jet dynamics and polarimetric signatures of the multi-band flare in blazar OJ 248}",
      journal = {\aap},
         year = 2025,
        month = mar,
       volume = {695},
          eid = {L3},
        pages = {L3},
          doi = {10.1051/0004-6361/202553689},
archivePrefix = {arXiv},
       eprint = {2502.12232},
 primaryClass = {astro-ph.HE},
       adsurl = {https://ui.adsabs.harvard.edu/abs/2025A&A...695L...3P}
}

@ARTICLE{Kovalev1424,
       author = {{Kovalev}, Y.~Y. and {Pushkarev}, A.~B. and {G{\'o}mez}, J.~L. and {Homan}, D.~C. and {Lister}, M.~L. and {Livingston}, J.~D. and {Pashchenko}, I.~N. and {Plavin}, A.~V. and {Savolainen}, T. and {Troitsky}, S.~V.},
        title = "{Looking into the jet cone of the neutrino-associated very high-energy blazar PKS 1424+240}",
      journal = {\aap},
         year = 2025,
        month = aug,
       volume = {700},
          eid = {L12},
        pages = {L12},
          doi = {10.1051/0004-6361/202555400},
archivePrefix = {arXiv},
       eprint = {2504.09287},
 primaryClass = {astro-ph.HE},
       adsurl = {https://ui.adsabs.harvard.edu/abs/2025A&A...700L..12K}
}

@ARTICLE{Tavecchio2015,
       author = {{Tavecchio}, F. and {Ghisellini}, G.},
        title = "{High-energy cosmic neutrinos from spine-sheath BL Lac jets}",
      journal = {\mnras},
         year = 2015,
        month = aug,
       volume = {451},
       number = {2},
        pages = {1502-1510},
          doi = {10.1093/mnras/stv1023},
archivePrefix = {arXiv},
       eprint = {1411.2783},
 primaryClass = {astro-ph.HE},
       adsurl = {https://ui.adsabs.harvard.edu/abs/2015MNRAS.451.1502T}
}

@ARTICLE{Ghisellini2005,
       author = {{Ghisellini}, G. and {Tavecchio}, F. and {Chiaberge}, M.},
        title = "{Structured jets in TeV BL Lac objects and radiogalaxies.  Implications for the observed properties}",
      journal = {\aap},
         year = 2005,
        month = mar,
       volume = {432},
       number = {2},
        pages = {401-410},
          doi = {10.1051/0004-6361:20041404},
archivePrefix = {arXiv},
       eprint = {astro-ph/0406093},
 primaryClass = {astro-ph},
       adsurl = {https://ui.adsabs.harvard.edu/abs/2005A&A...432..401G}
}

@ARTICLE{Oikonomou19,
       author = {{Oikonomou}, Foteini and {Murase}, Kohta and {Padovani}, Paolo and {Resconi}, Elisa and {M{\'e}sz{\'a}ros}, Peter},
        title = "{High-energy neutrino flux from individual blazar flares}",
      journal = {\mnras},
         year = 2019,
        month = nov,
       volume = {489},
       number = {3},
        pages = {4347-4366},
          doi = {10.1093/mnras/stz2246},
archivePrefix = {arXiv},
       eprint = {1906.05302},
 primaryClass = {astro-ph.HE},
       adsurl = {https://ui.adsabs.harvard.edu/abs/2019MNRAS.489.4347O}
}

@ARTICLE{Ansoldi18,
       author = {{Ansoldi}, S. and {Antonelli}, L.~A. and {Arcaro}, C. and {Baack}, D. and {Babi{\'c}}, A. and {Banerjee}, B. and {Bangale}, P. and {Barres de Almeida}, U. and {Barrio}, J.~A. and {Becerra Gonz{\'a}lez}, J. and {Bednarek}, W. and {Bernardini}, E. and {Berse}, R. Ch. and {Berti}, A. and {Besenrieder}, J. and {Bhattacharyya}, W. and {Bigongiari}, C. and {Biland}, A. and {Blanch}, O. and {Bonnoli}, G. and {Carosi}, R. and {Ceribella}, G. and {Chatterjee}, A. and {Colak}, S.~M. and {Colin}, P. and {Colombo}, E. and {Contreras}, J.~L. and {Cortina}, J. and {Covino}, S. and {Cumani}, P. and {D'Elia}, V. and {Da Vela}, P. and {Dazzi}, F. and {De Angelis}, A. and {De Lotto}, B. and {Delfino}, M. and {Delgado}, J. and {Di Pierro}, F. and {Dom{\'\i}nguez}, A. and {Dominis Prester}, D. and {Dorner}, D. and {Doro}, M. and {Einecke}, S. and {Elsaesser}, D. and {Fallah Ramazani}, V. and {Fattorini}, A. and {Fern{\'a}ndez-Barral}, A. and {Ferrara}, G. and {Fidalgo}, D. and {Foffano}, L. and {Fonseca}, M.~V. and {Font}, L. and {Fruck}, C. and {Gallozzi}, S. and {Garc{\'\i}a L{\'o}pez}, R.~J. and {Garczarczyk}, M. and {Gaug}, M. and {Giammaria}, P. and {Godinovi{\'c}}, N. and {Guberman}, D. and {Hadasch}, D. and {Hahn}, A. and {Hassan}, T. and {Hayashida}, M. and {Herrera}, J. and {Hoang}, J. and {Hrupec}, D. and {Inoue}, S. and {Ishio}, K. and {Iwamura}, Y. and {Konno}, Y. and {Kubo}, H. and {Kushida}, J. and {Lamastra}, A. and {Lelas}, D. and {Leone}, F. and {Lindfors}, E. and {Lombardi}, S. and {Longo}, F. and {L{\'o}pez}, M. and {Maggio}, C. and {Majumdar}, P. and {Makariev}, M. and {Maneva}, G. and {Manganaro}, M. and {Mannheim}, K. and {Maraschi}, L. and {Mariotti}, M. and {Mart{\'\i}nez}, M. and {Masuda}, S. and {Mazin}, D. and {Mielke}, K. and {Minev}, M. and {Miranda}, J.~M. and {Mirzoyan}, R. and {Moralejo}, A. and {Moreno}, V. and {Moretti}, E. and {Neustroev}, V. and {Niedzwiecki}, A. and {Nievas Rosillo}, M. and {Nigro}, C. and {Nilsson}, K. and {Ninci}, D. and {Nishijima}, K. and {Noda}, K. and {Nogu{\'e}s}, L. and {Paiano}, S. and {Palacio}, J. and {Paneque}, D. and {Paoletti}, R. and {Paredes}, J.~M. and {Pedaletti}, G. and {Pe{\~n}il}, P. and {Peresano}, M. and {Persic}, M. and {Pfrang}, K. and {Prada Moroni}, P.~G. and {Prandini}, E. and {Puljak}, I. and {Garcia}, J.~R. and {Rhode}, W. and {Rib{\'o}}, M. and {Rico}, J. and {Righi}, C. and {Rugliancich}, A. and {Saha}, L. and {Saito}, T. and {Satalecka}, K. and {Schweizer}, T. and {Sitarek}, J. and {{\v{S}}nidari{\'c}}, I. and {Sobczynska}, D. and {Stamerra}, A. and {Strzys}, M. and {Suri{\'c}}, T. and {Tavecchio}, F. and {Temnikov}, P. and {Terzi{\'c}}, T. and {Teshima}, M. and {Torres-Alb{\'a}}, N. and {Tsujimoto}, S. and {Vanzo}, G. and {Vazquez Acosta}, M. and {Vovk}, I. and {Ward}, J.~E. and {Will}, M. and {Zari{\'c}}, D. and {Cerruti}, Matteo},
        title = "{The Blazar TXS 0506+056 Associated with a High-energy Neutrino: Insights into Extragalactic Jets and Cosmic-Ray Acceleration}",
      journal = {\apjl},
         year = 2018,
        month = aug,
       volume = {863},
       number = {1},
          eid = {L10},
        pages = {L10},
          doi = {10.3847/2041-8213/aad083},
archivePrefix = {arXiv},
       eprint = {1807.04300},
 primaryClass = {astro-ph.HE},
       adsurl = {https://ui.adsabs.harvard.edu/abs/2018ApJ...863L..10A}
}

@INPROCEEDINGS{Oikonomou22,
       author = {{Oikonomou}, F.},
        title = "{High-energy neutrino emission from blazars}",
    booktitle = {37th International Cosmic Ray Conference},
         year = 2022,
        month = mar,
          eid = {30},
        pages = {30},
          doi = {10.22323/1.395.0030},
archivePrefix = {arXiv},
       eprint = {2201.05623},
 primaryClass = {astro-ph.HE},
       adsurl = {https://ui.adsabs.harvard.edu/abs/2022icrc.confE..30O}
}

@ARTICLE{Atoyan2001,
       author = {{Atoyan}, Armen and {Dermer}, Charles D.},
        title = "{High-Energy Neutrinos from Photomeson Processes in Blazars}",
      journal = {\prl},
         year = 2001,
        month = nov,
       volume = {87},
       number = {22},
          eid = {221102},
        pages = {221102},
          doi = {10.1103/PhysRevLett.87.221102},
archivePrefix = {arXiv},
       eprint = {astro-ph/0108053},
 primaryClass = {astro-ph},
       adsurl = {https://ui.adsabs.harvard.edu/abs/2001PhRvL..87v1102A}
}

@ARTICLE{Weaver22,
       author = {{Weaver}, Zachary R. and {Jorstad}, Svetlana G. and {Marscher}, Alan P. and {Morozova}, Daria A. and {Troitsky}, Ivan S. and {Agudo}, Iv{\'a}n and {G{\'o}mez}, Jos{\'e} L. and {L{\"a}hteenm{\"a}ki}, Anne and {Tammi}, Joni and {Tornikoski}, Merja},
        title = "{Kinematics of Parsec-scale Jets of Gamma-Ray Blazars at 43 GHz during 10 yr of the VLBA-BU-BLAZAR Program}",
      journal = {\apjs},
         year = 2022,
        month = may,
       volume = {260},
       number = {1},
          eid = {12},
        pages = {12},
          doi = {10.3847/1538-4365/ac589c},
archivePrefix = {arXiv},
       eprint = {2202.12290},
 primaryClass = {astro-ph.HE},
       adsurl = {https://ui.adsabs.harvard.edu/abs/2022ApJS..260...12W}
}

@ARTICLE{Lister21,
       author = {{Lister}, M.~L. and {Homan}, D.~C. and {Kellermann}, K.~I. and {Kovalev}, Y.~Y. and {Pushkarev}, A.~B. and {Ros}, E. and {Savolainen}, T.},
        title = "{Monitoring Of Jets in Active Galactic Nuclei with VLBA Experiments. XVIII. Kinematics and Inner Jet Evolution of Bright Radio-loud Active Galaxies}",
      journal = {\apj},
         year = 2021,
        month = dec,
       volume = {923},
       number = {1},
          eid = {30},
        pages = {30},
          doi = {10.3847/1538-4357/ac230f},
archivePrefix = {arXiv},
       eprint = {2108.13358},
 primaryClass = {astro-ph.HE},
       adsurl = {https://ui.adsabs.harvard.edu/abs/2021ApJ...923...30L}
}

@ARTICLE{MarscherGear,
       author = {{Marscher}, A.~P. and {Gear}, W.~K.},
        title = "{Models for high-frequency radio outbursts in extragalactic sources, with application to the early 1983 millimeter-to-infrared flare of 3C 273.}",
      journal = {\apj},
         year = 1985,
        month = nov,
       volume = {298},
        pages = {114-127},
          doi = {10.1086/163592},
       adsurl = {https://ui.adsabs.harvard.edu/abs/1985ApJ...298..114M}
}

@ARTICLE{IceCubeFermi,
       author = {{Hori}, Sam and {Desai}, Abhishek and {Vandenbroucke}, Justin},
        title = "{IceCube population constraints on neutrino emission by Fermi-LAT detected active galactic nuclei}",
      journal = {arXiv e-prints},
         year = 2025,
        month = jul,
          eid = {arXiv:2507.07098},
        pages = {arXiv:2507.07098},
          doi = {10.48550/arXiv.2507.07098},
archivePrefix = {arXiv},
       eprint = {2507.07098},
 primaryClass = {astro-ph.HE},
       adsurl = {https://ui.adsabs.harvard.edu/abs/2025arXiv250707098H}
}

@ARTICLE{Moretti2025,
       author = {{Moretti}, A. and {Caccianiga}, A.},
        title = "{Flat-spectrum radio quasars as high-energy neutrino sources}",
      journal = {\aap},
         year = 2025,
        month = dec,
       volume = {704},
          eid = {A184},
        pages = {A184},
          doi = {10.1051/0004-6361/202554527},
archivePrefix = {arXiv},
       eprint = {2507.17908},
 primaryClass = {astro-ph.HE},
       adsurl = {https://ui.adsabs.harvard.edu/abs/2025A&A...704A.184M}
}

@ARTICLE{Kovalev0446,
       author = {{Kovalev}, Y.~Y. and {Aller}, M.~F. and {Erkenov}, A.~K. and {G{\'o}mez}, J.~L. and {Homan}, D.~C. and {Kivokurtseva}, P.~I. and {Kovalev}, Yu. A. and {Lister}, M.~L. and {de la Parra}, P.~V. and {Plavin}, A.~V. and {Popkov}, A.~V. and {Pushkarev}, A.~B. and {Readhead}, A.~C.~S. and {Shablovinskaia}, E. and {Sotnikova}, Yu. V. and {Spiridonova}, O.~I. and {Troitsky}, S.~V. and {Vlasyuk}, V.~V.},
        title = "{Multi-messenger flare in the quasar PKS 0446+11}",
      journal = {arXiv e-prints},
         year = 2025,
        month = nov,
          eid = {arXiv:2511.07535},
        pages = {arXiv:2511.07535},
archivePrefix = {arXiv},
       eprint = {2511.07535},
 primaryClass = {astro-ph.HE},
       adsurl = {https://ui.adsabs.harvard.edu/abs/2025arXiv251107535K}
}

@ARTICLE{Paraschos2025,
       author = {{Paraschos}, G.~F. and {Traianou}, E. and {Debbrecht}, L.~C. and {Liodakis}, I. and {Ros}, E.},
        title = "{Polarization as a Probe of Neutrino Emission from Blazars}",
      journal = {\apj},
         year = 2025,
        month = aug,
       volume = {989},
       number = {2},
          eid = {208},
        pages = {208},
          doi = {10.3847/1538-4357/adf110},
archivePrefix = {arXiv},
       eprint = {2507.16929},
 primaryClass = {astro-ph.HE},
       adsurl = {https://ui.adsabs.harvard.edu/abs/2025ApJ...989..208P}
}

@ARTICLE{Fujisawa1999,
       author = {{Fujisawa}, Kenta and {Kobayashi}, Hideyuki and {Wajima}, Kiyoaki and {Hirabayashi}, Hisashi and {Kameno}, Seiji and {Inoue}, Makoto},
        title = "{Discovery of Large Doppler Factors in Radio-Loud Active Galactic Nuclei}",
      journal = {\pasj},
         year = 1999,
        month = aug,
       volume = {51},
        pages = {537-545},
          doi = {10.1093/pasj/51.4.537},
       adsurl = {https://ui.adsabs.harvard.edu/abs/1999PASJ...51..537F}
}

@ARTICLE{ValtaojaLaehteenmaki1999,
       author = {{L{\"a}hteenm{\"a}ki}, A. and {Valtaoja}, E.},
        title = "{Total Flux Density Variations in Extragalactic Radio Sources. III. Doppler Boosting Factors, Lorentz Factors, and Viewing Angles for Active Galactic Nuclei}",
      journal = {\apj},
         year = 1999,
        month = aug,
       volume = {521},
       number = {2},
        pages = {493-501},
          doi = {10.1086/307587},
       adsurl = {https://ui.adsabs.harvard.edu/abs/1999ApJ...521..493L}
}

@ARTICLE{Sebastian22,
       author = {{Sebastian}, Biny and {Kharb}, Preeti and {Lister}, Matthew L. and {Marshall}, Herman L. and {O'Dea}, Christopher P. and {Baum}, Stefi A.},
        title = "{Investigating the Origin of X-Ray Jets: A Case Study of Four Hybrid Morphology MOJAVE Blazars}",
      journal = {\apj},
         year = 2022,
        month = aug,
       volume = {935},
       number = {1},
          eid = {59},
        pages = {59},
          doi = {10.3847/1538-4357/ac7c10},
archivePrefix = {arXiv},
       eprint = {2206.13665},
 primaryClass = {astro-ph.GA},
       adsurl = {https://ui.adsabs.harvard.edu/abs/2022ApJ...935...59S}
}

@ARTICLE{Jorstad2001a,
       author = {{Jorstad}, Svetlana G. and {Marscher}, Alan P. and {Mattox}, John R. and {Wehrle}, Ann E. and {Bloom}, Steven D. and {Yurchenko}, Alexei V.},
        title = "{Multiepoch Very Long Baseline Array Observations of EGRET-detected Quasars and BL Lacertae Objects: Superluminal Motion of Gamma-Ray Bright Blazars}",
      journal = {\apjs},
         year = 2001,
        month = jun,
       volume = {134},
       number = {2},
        pages = {181-240},
          doi = {10.1086/320858},
archivePrefix = {arXiv},
       eprint = {astro-ph/0101570},
 primaryClass = {astro-ph},
       adsurl = {https://ui.adsabs.harvard.edu/abs/2001ApJS..134..181J}
}

@ARTICLE{EBL_absorption,
       author = {{Aharonian}, F. and {Akhperjanian}, A.~G. and {Bazer-Bachi}, A.~R. and {Beilicke}, M. and {Benbow}, W. and {Berge}, D. and {Bernl{\"o}hr}, K. and {Boisson}, C. and {Bolz}, O. and {Borrel}, V. and {Braun}, I. and {Breitling}, F. and {Brown}, A.~M. and {Chadwick}, P.~M. and {Chounet}, L.-M. and {Cornils}, R. and {Costamante}, L. and {Degrange}, B. and {Dickinson}, H.~J. and {Djannati-Ata{\"\i}}, A. and {Drury}, L. O'c. and {Dubus}, G. and {Emmanoulopoulos}, D. and {Espigat}, P. and {Feinstein}, F. and {Fontaine}, G. and {Fuchs}, Y. and {Funk}, S. and {Gallant}, Y.~A. and {Giebels}, B. and {Gillessen}, S. and {Glicenstein}, J.~F. and {Goret}, P. and {Hadjichristidis}, C. and {Hauser}, D. and {Hauser}, M. and {Heinzelmann}, G. and {Henri}, G. and {Hermann}, G. and {Hinton}, J.~A. and {Hofmann}, W. and {Holleran}, M. and {Horns}, D. and {Jacholkowska}, A. and {de Jager}, O.~C. and {Kh{\'e}lifi}, B. and {Klages}, S. and {Komin}, Nu. and {Konopelko}, A. and {Latham}, I.~J. and {Le Gallou}, R. and {Lemi{\`e}re}, A. and {Lemoine-Goumard}, M. and {Leroy}, N. and {Lohse}, T. and {Martin}, J.~M. and {Martineau-Huynh}, O. and {Marcowith}, A. and {Masterson}, C. and {McComb}, T.~J.~L. and {de Naurois}, M. and {Nolan}, S.~J. and {Noutsos}, A. and {Orford}, K.~J. and {Osborne}, J.~L. and {Ouchrif}, M. and {Panter}, M. and {Pelletier}, G. and {Pita}, S. and {P{\"u}hlhofer}, G. and {Punch}, M. and {Raubenheimer}, B.~C. and {Raue}, M. and {Raux}, J. and {Rayner}, S.~M. and {Reimer}, A. and {Reimer}, O. and {Ripken}, J. and {Rob}, L. and {Rolland}, L. and {Rowell}, G. and {Sahakian}, V. and {Saug{\'e}}, L. and {Schlenker}, S. and {Schlickeiser}, R. and {Schuster}, C. and {Schwanke}, U. and {Siewert}, M. and {Sol}, H. and {Spangler}, D. and {Steenkamp}, R. and {Stegmann}, C. and {Tavernet}, J.-P. and {Terrier}, R. and {Th{\'e}oret}, C.~G. and {Tluczykont}, M. and {van Eldik}, C. and {Vasileiadis}, G. and {Venter}, C. and {Vincent}, P. and {V{\"o}lk}, H.~J. and {Wagner}, S.~J.},
        title = "{A low level of extragalactic background light as revealed by {\ensuremath{\gamma}}-rays from blazars}",
      journal = {\nat},
         year = 2006,
        month = apr,
       volume = {440},
       number = {7087},
        pages = {1018-1021},
          doi = {10.1038/nature04680},
archivePrefix = {arXiv},
       eprint = {astro-ph/0508073},
 primaryClass = {astro-ph},
       adsurl = {https://ui.adsabs.harvard.edu/abs/2006Natur.440.1018A}
}

@ARTICLE{Savolainen2002,
       author = {{Savolainen}, T. and {Wiik}, K. and {Valtaoja}, E. and {Jorstad}, S.~G. and {Marscher}, A.~P.},
        title = "{Connections between millimetre continuum variations and VLBI structure in 27 AGN}",
      journal = {\aap},
         year = 2002,
        month = nov,
       volume = {394},
        pages = {851-861},
          doi = {10.1051/0004-6361:20021236},
       adsurl = {https://ui.adsabs.harvard.edu/abs/2002A&A...394..851S}
}

@article{Lahteenmaki_1999,
	doi = {10.1086/306649},
	url = {https://doi.org/10.1086/306649},
	year = 1999,
	month = {jan},
	publisher = {American Astronomical Society},
	volume = {511},
	number = {1},
	pages = {112--117},
	author = {A. Lahteenmaki and E. Valtaoja and K. Wiik},
	title = {Total Flux Density Variations in Extragalactic Radio Sources. {II}. Determining the Limiting Brightness Temperature for Synchrotron Sources},
	journal = {\apj}
}

@ARTICLE{Readhead,
       author = {{Readhead}, Anthony C.~S.},
        title = "{Equipartition Brightness Temperature and the Inverse Compton Catastrophe}",
      journal = {\apj},
         year = 1994,
        month = may,
       volume = {426},
        pages = {51},
          doi = {10.1086/174038},
       adsurl = {https://ui.adsabs.harvard.edu/abs/1994ApJ...426...51R}
}

@ARTICLE{Liodakis,
       author = {{Liodakis}, I. and {Marchili}, N. and {Angelakis}, E. and {Fuhrmann}, L. and {Nestoras}, I. and {Myserlis}, I. and {Karamanavis}, V. and {Krichbaum}, T.~P. and {Sievers}, A. and {Ungerechts}, H. and {Zensus}, J.~A.},
        title = "{F-GAMMA: variability Doppler factors of blazars from multiwavelength monitoring}",
      journal = {\mnras},
         year = 2017,
        month = apr,
       volume = {466},
       number = {4},
        pages = {4625-4632},
          doi = {10.1093/mnras/stx002},
archivePrefix = {arXiv},
       eprint = {1701.01452},
 primaryClass = {astro-ph.HE},
       adsurl = {https://ui.adsabs.harvard.edu/abs/2017MNRAS.466.4625L}
}

@ARTICLE{Livingston2025,
       author = {{Livingston}, J.~D. and {Nikonov}, A.~S. and {Dzib}, S.~A. and {Debbrecht}, L.~C. and {Kovalev}, Y.~Y. and {Lisakov}, M.~M. and {MacDonald}, N.~R. and {Paraschos}, G.~F. and {R{\"o}der}, J. and {Wielgus}, M.},
        title = "{A helical magnetic field in quasar NRAO150 revealed by Faraday rotation (Corrigendum)}",
      journal = {\aap},
         year = 2025,
        month = aug,
       volume = {700},
          eid = {C4},
        pages = {C4},
          doi = {10.1051/0004-6361/202556359e},
       adsurl = {https://ui.adsabs.harvard.edu/abs/2025A&A...700C...4L}
}

@ARTICLE{Chamani,
       author = {{Chamani}, Wara and {Savolainen}, Tuomas and {Ros}, Eduardo and {Kovalev}, Yuri Y. and {Wiik}, Kaj and {L{\"a}hteenm{\"a}ki}, Anne and {Tornikoski}, Merja and {Tammi}, Joni},
        title = "{Time variability of the core-shift effect in the blazar 3C 454.3}",
      journal = {\aap},
         year = 2023,
        month = apr,
       volume = {672},
          eid = {A130},
        pages = {A130},
          doi = {10.1051/0004-6361/202243435},
archivePrefix = {arXiv},
       eprint = {2209.13301},
 primaryClass = {astro-ph.HE},
       adsurl = {https://ui.adsabs.harvard.edu/abs/2023A&A...672A.130C}
}

@misc{VCAT,
  author       = {Baczko, Anne-Kathrin and
                  Eppel, Florian and
                  Bartolini, Vieri and
                  Ricci, Luca and
                  P\"otzl, Felix Marc and
                  Roesch, Florian},
  title        = {{mpifr-vlbi/VCAT}: v0.1.0: First public release},
  year         = 2025,
  month        = oct,
  publisher    = {Zenodo},
  version      = {v0.1.0},
  doi          = {10.5281/zenodo.17404907},
  url          = {https://doi.org/10.5281/zenodo.17404907},
  note         = {{doi:10.5281/zenodo.17404907}},
}

@ARTICLE{Perger2025,
       author = {{Perger}, K. and {Frey}, S. and {Gab{\'a}nyi}, K. {\'E}. and {Kun}, E.},
        title = "{Zooming into the neutrino-associated blazar candidate J1718+4239}",
      journal = {\aap},
         year = 2025,
        month = jun,
       volume = {698},
          eid = {L2},
        pages = {L2},
          doi = {10.1051/0004-6361/202555177},
archivePrefix = {arXiv},
       eprint = {2505.07400},
 primaryClass = {astro-ph.HE},
       adsurl = {https://ui.adsabs.harvard.edu/abs/2025A&A...698L...2P}
}

@ARTICLE{Koemives2024,
       author = {{K{\H{o}}m{\'\i}ves}, Janka and {Gab{\'a}nyi}, Krisztina {\'E}va and {Frey}, S{\'a}ndor and {Kun}, Emma},
        title = "{VLBI Analysis of a Potential High-Energy Neutrino Emitter Blazar}",
      journal = {Universe},
         year = 2024,
        month = feb,
       volume = {10},
       number = {2},
          eid = {78},
        pages = {78},
          doi = {10.3390/universe10020078},
archivePrefix = {arXiv},
       eprint = {2402.04011},
 primaryClass = {astro-ph.HE},
       adsurl = {https://ui.adsabs.harvard.edu/abs/2024Univ...10...78K}
}

@ARTICLE{Ji2024,
       author = {{Ji}, Shunhao and {Wang}, Zhongxiang},
        title = "{PKS 2254+074: A Blazar in Likely Association with the Neutrino Event IceCube-190619A}",
      journal = {\apjl},
         year = 2024,
        month = nov,
       volume = {975},
       number = {2},
          eid = {L30},
        pages = {L30},
          doi = {10.3847/2041-8213/ad87ea},
archivePrefix = {arXiv},
       eprint = {2410.14079},
 primaryClass = {astro-ph.HE},
       adsurl = {https://ui.adsabs.harvard.edu/abs/2024ApJ...975L..30J}
}

@ARTICLE{Jiang2024,
       author = {{Jiang}, Xiong and {Liao}, Neng-Hui and {Wang}, Yi-Bo and {Xue}, Rui and {Jiang}, Ning and {Wang}, Ting-Gui},
        title = "{The Awakening of a Blazar at Redshift 2.7 Temporally Coincident with the Arrival of Cospatial Neutrino Event IceCube-201221A}",
      journal = {\apjl},
         year = 2024,
        month = apr,
       volume = {965},
       number = {1},
          eid = {L2},
        pages = {L2},
          doi = {10.3847/2041-8213/ad36b9},
archivePrefix = {arXiv},
       eprint = {2401.12122},
 primaryClass = {astro-ph.HE},
       adsurl = {https://ui.adsabs.harvard.edu/abs/2024ApJ...965L...2J}
}

@ARTICLE{Homan2006,
       author = {{Homan}, D.~C. and {Kovalev}, Y.~Y. and {Lister}, M.~L. and {Ros}, E. and {Kellermann}, K.~I. and {Cohen}, M.~H. and {Vermeulen}, R.~C. and {Zensus}, J.~A. and {Kadler}, M.},
        title = "{Intrinsic Brightness Temperatures of AGN Jets}",
      journal = {\apjl},
         year = 2006,
        month = may,
       volume = {642},
       number = {2},
        pages = {L115-L118},
          doi = {10.1086/504715},
archivePrefix = {arXiv},
       eprint = {astro-ph/0603837},
 primaryClass = {astro-ph},
       adsurl = {https://ui.adsabs.harvard.edu/abs/2006ApJ...642L.115H}
}

@ARTICLE{Sumida2022,
       author = {{Sumida}, Viktor Y.~D. and {Schutzer}, A. de A. and {Caproni}, A. and {Abraham}, Z.},
        title = "{The relativistic parsec-scale jets of the blazars TXS 0506+056 and PKS 0502+049 and their possible association with gamma-ray flares and neutrino production}",
      journal = {\mnras},
         year = 2022,
        month = jan,
       volume = {509},
       number = {2},
        pages = {1646-1663},
          doi = {10.1093/mnras/stab3022},
archivePrefix = {arXiv},
       eprint = {2112.08858},
 primaryClass = {astro-ph.HE},
       adsurl = {https://ui.adsabs.harvard.edu/abs/2022MNRAS.509.1646S}
}

@ARTICLE{PlavinCoreShift,
       author = {{Plavin}, A.~V. and {Kovalev}, Y.~Y. and {Pushkarev}, A.~B. and {Lobanov}, A.~P.},
        title = "{Significant core shift variability in parsec-scale jets of active galactic nuclei}",
      journal = {\mnras},
         year = 2019,
        month = may,
       volume = {485},
       number = {2},
        pages = {1822-1842},
          doi = {10.1093/mnras/stz504},
archivePrefix = {arXiv},
       eprint = {1811.02544},
 primaryClass = {astro-ph.GA},
       adsurl = {https://ui.adsabs.harvard.edu/abs/2019MNRAS.485.1822P}
}

@ARTICLE{4FGL_DR3,
       author = {{Abdollahi}, S. and {Acero}, F. and {Baldini}, L. and {Ballet}, J. and {Bastieri}, D. and {Bellazzini}, R. and {Berenji}, B. and {Berretta}, A. and {Bissaldi}, E. and {Blandford}, R.~D. and {Bloom}, E. and {Bonino}, R. and {Brill}, A. and {Britto}, R.~J. and {Bruel}, P. and {Burnett}, T.~H. and {Buson}, S. and {Cameron}, R.~A. and {Caputo}, R. and {Caraveo}, P.~A. and {Castro}, D. and {Chaty}, S. and {Cheung}, C.~C. and {Chiaro}, G. and {Cibrario}, N. and {Ciprini}, S. and {Coronado-Bl{\'a}zquez}, J. and {Crnogorcevic}, M. and {Cutini}, S. and {D'Ammando}, F. and {De Gaetano}, S. and {Digel}, S.~W. and {Di Lalla}, N. and {Dirirsa}, F. and {Di Venere}, L. and {Dom{\'\i}nguez}, A. and {Fallah Ramazani}, V. and {Fegan}, S.~J. and {Ferrara}, E.~C. and {Fiori}, A. and {Fleischhack}, H. and {Franckowiak}, A. and {Fukazawa}, Y. and {Funk}, S. and {Fusco}, P. and {Galanti}, G. and {Gammaldi}, V. and {Gargano}, F. and {Garrappa}, S. and {Gasparrini}, D. and {Giacchino}, F. and {Giglietto}, N. and {Giordano}, F. and {Giroletti}, M. and {Glanzman}, T. and {Green}, D. and {Grenier}, I.~A. and {Grondin}, M. -H. and {Guillemot}, L. and {Guiriec}, S. and {Gustafsson}, M. and {Harding}, A.~K. and {Hays}, E. and {Hewitt}, J.~W. and {Horan}, D. and {Hou}, X. and {J{\'o}hannesson}, G. and {Karwin}, C. and {Kayanoki}, T. and {Kerr}, M. and {Kuss}, M. and {Landriu}, D. and {Larsson}, S. and {Latronico}, L. and {Lemoine-Goumard}, M. and {Li}, J. and {Liodakis}, I. and {Longo}, F. and {Loparco}, F. and {Lott}, B. and {Lubrano}, P. and {Maldera}, S. and {Malyshev}, D. and {Manfreda}, A. and {Mart{\'\i}-Devesa}, G. and {Mazziotta}, M.~N. and {Mereu}, I. and {Meyer}, M. and {Michelson}, P.~F. and {Mirabal}, N. and {Mitthumsiri}, W. and {Mizuno}, T. and {Moiseev}, A.~A. and {Monzani}, M.~E. and {Morselli}, A. and {Moskalenko}, I.~V. and {Negro}, M. and {Nuss}, E. and {Omodei}, N. and {Orienti}, M. and {Orlando}, E. and {Paneque}, D. and {Pei}, Z. and {Perkins}, J.~S. and {Persic}, M. and {Pesce-Rollins}, M. and {Petrosian}, V. and {Pillera}, R. and {Poon}, H. and {Porter}, T.~A. and {Principe}, G. and {Rain{\`o}}, S. and {Rando}, R. and {Rani}, B. and {Razzano}, M. and {Razzaque}, S. and {Reimer}, A. and {Reimer}, O. and {Reposeur}, T. and {S{\'a}nchez-Conde}, M. and {Saz Parkinson}, P.~M. and {Scotton}, L. and {Serini}, D. and {Sgr{\`o}}, C. and {Siskind}, E.~J. and {Smith}, D.~A. and {Spandre}, G. and {Spinelli}, P. and {Sueoka}, K. and {Suson}, D.~J. and {Tajima}, H. and {Tak}, D. and {Thayer}, J.~B. and {Thompson}, D.~J. and {Torres}, D.~F. and {Troja}, E. and {Valverde}, J. and {Wood}, K. and {Zaharijas}, G.},
        title = "{Incremental Fermi Large Area Telescope Fourth Source Catalog}",
      journal = {\apjs},
         year = 2022,
        month = jun,
       volume = {260},
       number = {2},
          eid = {53},
        pages = {53},
          doi = {10.3847/1538-4365/ac6751},
archivePrefix = {arXiv},
       eprint = {2201.11184},
 primaryClass = {astro-ph.HE},
       adsurl = {https://ui.adsabs.harvard.edu/abs/2022ApJS..260...53A}
}

@INPROCEEDINGS{IceCubeAlerts,
       author = {{Blaufuss}, E. and {Kintscher}, T. and {Lu}, L. and {Tung}, C.~F.},
        title = "{The Next Generation of IceCube Real-time Neutrino Alerts}",
    booktitle = {36th International Cosmic Ray Conference (ICRC2019)},
         year = 2019,
       series = {International Cosmic Ray Conference},
       volume = {36},
        month = jul,
          eid = {1021},
        pages = {1021},
          doi = {10.22323/1.358.01021},
archivePrefix = {arXiv},
       eprint = {1908.04884},
 primaryClass = {astro-ph.HE},
       adsurl = {https://ui.adsabs.harvard.edu/abs/2019ICRC...36.1021B}
}

@ARTICLE{FermiLCrepo,
       author = {{Abdollahi}, S. and {Ajello}, M. and {Baldini}, L. and {Ballet}, J. and {Bastieri}, D. and {Becerra Gonzalez}, J. and {Bellazzini}, R. and {Berretta}, A. and {Bissaldi}, E. and {Bonino}, R. and {Brill}, A. and {Bruel}, P. and {Burns}, E. and {Buson}, S. and {Cameron}, R.~A. and {Caputo}, R. and {Caraveo}, P.~A. and {Cibrario}, N. and {Ciprini}, S. and {Cristarella Orestano}, P. and {Crnogorcevic}, M. and {Cutini}, S. and {D'Ammando}, F. and {De Gaetano}, S. and {Digel}, S.~W. and {Di Lalla}, N. and {Di Venere}, L. and {Dom{\'\i}nguez}, A. and {Ramazani}, V. Fallah and {Fegan}, S.~J. and {Ferrara}, E.~C. and {Fiori}, A. and {Fleischhack}, H. and {Franckowiak}, A. and {Fukazawa}, Y. and {Fusco}, P. and {Gammaldi}, V. and {Gargano}, F. and {Garrappa}, S. and {Gasbarra}, C. and {Gasparrini}, D. and {Giglietto}, N. and {Giordano}, F. and {Giroletti}, M. and {Green}, D. and {Grenier}, I.~A. and {Guiriec}, S. and {Gustafsson}, M. and {Hays}, E. and {Horan}, D. and {Hou}, X. and {J{\'o}hannesson}, G. and {Kerr}, M. and {Kocevski}, D. and {Kuss}, M. and {Latronico}, L. and {Li}, J. and {Liodakis}, I. and {Longo}, F. and {Loparco}, F. and {Lorusso}, L. and {Lott}, B. and {Lovellette}, M.~N. and {Lubrano}, P. and {Maldera}, S. and {Manfreda}, A. and {Mart{\'\i}-Devesa}, G. and {Mazziotta}, M.~N. and {Mereu}, I. and {Meyer}, M. and {Michelson}, P.~F. and {Mizuno}, T. and {Monzani}, M.~E. and {Morselli}, A. and {Moskalenko}, I.~V. and {Negro}, M. and {Omodei}, N. and {Orlando}, E. and {Ormes}, J.~F. and {Paneque}, D. and {Panzarini}, G. and {Perkins}, J.~S. and {Persic}, M. and {Pesce-Rollins}, M. and {Pillera}, R. and {Porter}, T.~A. and {Principe}, G. and {Racusin}, J.~L. and {Rain{\`o}}, S. and {Rando}, R. and {Rani}, B. and {Razzano}, M. and {Razzaque}, S. and {Reimer}, A. and {Reimer}, O. and {S{\'a}nchez-Conde}, M. and {Parkinson}, P.~M. Saz and {Scargle}, Jeff and {Scotton}, L. and {Serini}, D. and {Sgr{\`o}}, C. and {Siskind}, E.~J. and {Spandre}, G. and {Spinelli}, P. and {Suson}, D.~J. and {Tajima}, H. and {Thompson}, D.~J. and {Torres}, D.~F. and {Valverde}, J. and {Venters}, T. and {Wadiasingh}, Z. and {Wagner}, S. and {Wood}, K.},
        title = "{The Fermi-LAT Lightcurve Repository}",
      journal = {\apjs},
         year = 2023,
        month = apr,
       volume = {265},
       number = {2},
          eid = {31},
        pages = {31},
          doi = {10.3847/1538-4365/acbb6a},
archivePrefix = {arXiv},
       eprint = {2301.01607},
 primaryClass = {astro-ph.HE},
       adsurl = {https://ui.adsabs.harvard.edu/abs/2023ApJS..265...31A}
}

@ARTICLE{FermiIceCube24,
       author = {{Garrappa}, S. and {Buson}, S. and {Sinapius}, J. and {Franckowiak}, A. and {Liodakis}, I. and {Bartolini}, C. and {Giroletti}, M. and {Nanci}, C. and {Principe}, G. and {Venters}, T.~M.},
        title = "{Fermi-LAT follow-up observations in seven years of real-time high-energy neutrino alerts}",
      journal = {\aap},
         year = 2024,
        month = jul,
       volume = {687},
          eid = {A59},
        pages = {A59},
          doi = {10.1051/0004-6361/202449221},
archivePrefix = {arXiv},
       eprint = {2401.06666},
 primaryClass = {astro-ph.HE},
       adsurl = {https://ui.adsabs.harvard.edu/abs/2024A&A...687A..59G}
}

@ARTICLE{RFC,
       author = {{Petrov}, L.~Y. and {Kovalev}, Y.~Y.},
        title = "{The Radio Fundamental Catalog. I. Astrometry}",
      journal = {\apjs},
         year = 2025,
        month = feb,
       volume = {276},
       number = {2},
          eid = {38},
        pages = {38},
          doi = {10.3847/1538-4365/ad8c36},
archivePrefix = {arXiv},
       eprint = {2410.11794},
 primaryClass = {astro-ph.IM},
       adsurl = {https://ui.adsabs.harvard.edu/abs/2025ApJS..276...38P}
}

@ARTICLE{Murase2014,
       author = {{Murase}, Kohta and {Inoue}, Yoshiyuki and {Dermer}, Charles D.},
        title = "{Diffuse neutrino intensity from the inner jets of active galactic nuclei: Impacts of external photon fields and the blazar sequence}",
      journal = {\prd},
         year = 2014,
        month = jul,
       volume = {90},
       number = {2},
          eid = {023007},
        pages = {023007},
          doi = {10.1103/PhysRevD.90.023007},
archivePrefix = {arXiv},
       eprint = {1403.4089},
 primaryClass = {astro-ph.HE},
       adsurl = {https://ui.adsabs.harvard.edu/abs/2014PhRvD..90b3007M}
}

@ARTICLE{Redshift0215_first,
       author = {{Blades}, J.~C. and {Hunstead}, R.~W. and {Murdoch}, H.~S. and {Pettini}, M.},
        title = "{The near-ultraviolet spectrum of the high-redshift BL Lacertae object0215+015.}",
      journal = {\apj},
         year = 1985,
        month = jan,
       volume = {288},
        pages = {580-594},
          doi = {10.1086/162824},
       adsurl = {https://ui.adsabs.harvard.edu/abs/1985ApJ...288..580B}
}

@ARTICLE{Redshift0215,
       author = {{Foltz}, Craig B. and {Chaffee}, Jr., Frederic H.},
        title = "{Confirmation of the Emission-Line Redshift of the BL LAC Object PKS 0215+015}",
      journal = {\aj},
         year = 1987,
        month = mar,
       volume = {93},
        pages = {529},
          doi = {10.1086/114334},
       adsurl = {https://ui.adsabs.harvard.edu/abs/1987AJ.....93..529F}
}

@ARTICLE{Gaskell1982,
       author = {{Gaskell}, C.~M.},
        title = "{High redshift BL Lacertae object: PKS 0215+015.}",
      journal = {\apj},
         year = 1982,
        month = jan,
       volume = {252},
        pages = {447-454},
          doi = {10.1086/159572},
       adsurl = {https://ui.adsabs.harvard.edu/abs/1982ApJ...252..447G}
}

@ARTICLE{FermiAtel0215,
       author = {{Garrappa}, S. and {Buson}, S. and {Cheung}, C.~C. and {Sinapius}, J. and {Fermi-LAT Collaboration}},
        title = "{Fermi-LAT Gamma-ray Observations of IceCube-220225A}",
      journal = {GRB Coordinates Network},
         year = 2022,
        month = feb,
       volume = {31653},
        pages = {1},
       adsurl = {https://ui.adsabs.harvard.edu/abs/2022GCN.31653....1G}
}

@ARTICLE{Optical0215ATel,
       author = {{Nesci}, Roberto},
        title = "{PKS 0215+015 optically bright}",
      journal = {The Astronomer's Telegram},
         year = 2022,
        month = feb,
       volume = {15248},
        pages = {1},
       adsurl = {https://ui.adsabs.harvard.edu/abs/2022ATel15248....1N}
}

@ARTICLE{Nanci2022,
       author = {{Nanci}, C. and {Giroletti}, M. and {Orienti}, M. and {Migliori}, G. and {Mold{\'o}n}, J. and {Garrappa}, S. and {Kadler}, M. and {Ros}, E. and {Buson}, S. and {An}, T. and {P{\'e}rez-Torres}, M.~A. and {D'Ammando}, F. and {Mohan}, P. and {Agudo}, I. and {Sohn}, B.~W. and {Castro-Tirado}, A.~J. and {Zhang}, Y.},
        title = "{Observing the inner parsec-scale region of candidate neutrino-emitting blazars}",
      journal = {\aap},
         year = 2022,
        month = jul,
       volume = {663},
          eid = {A129},
        pages = {A129},
          doi = {10.1051/0004-6361/202142665},
archivePrefix = {arXiv},
       eprint = {2203.13268},
 primaryClass = {astro-ph.HE},
       adsurl = {https://ui.adsabs.harvard.edu/abs/2022A&A...663A.129N}
}

@ARTICLE{0446FermiAtel,
       author = {{Sinapius}, J. and {Garrappa}, S. and {Buson}, S. and {Bartolini}, C. and {Pfeiffer}, L. and {Fermi-LAT Collaboration}},
        title = "{Fermi-LAT detection of increased gamma-ray activity of blazar PKS 0446+11, located inside the IceCube-240105A error region}",
      journal = {GRB Coordinates Network},
         year = 2024,
        month = jan,
       volume = {35517},
        pages = {1},
       adsurl = {https://ui.adsabs.harvard.edu/abs/2024GCN.35517....1S}
}

@ARTICLE{Acharyya0735,
       author = {{Acharyya}, A. and {Adams}, C.~B. and {Archer}, A. and {Bangale}, P. and {Bartkoske}, J.~T. and {Batista}, P. and {Benbow}, W. and {Brill}, A. and {Buckley}, J.~H. and {Christiansen}, J.~L. and {Chromey}, A.~J. and {Errando}, M. and {Falcone}, A. and {Feng}, Q. and {Foote}, Juniper and {Fortson}, L. and {Furniss}, A. and {Gallagher}, G. and {Hanlon}, W. and {Hanna}, D. and {Hervet}, O. and {Hinrichs}, C.~E. and {Hoang}, J. and {Holder}, J. and {Humensky}, T.~B. and {Jin}, W. and {Kaaret}, P. and {Kertzman}, M. and {Kherlakian}, M. and {Kieda}, D. and {Kleiner}, T.~K. and {Korzoun}, N. and {Kumar}, S. and {Lang}, M.~J. and {Lundy}, M. and {Maier}, G. and {McGrath}, C.~E. and {Millard}, M.~J. and {Millis}, J. and {Mooney}, C.~L. and {Moriarty}, P. and {Mukherjee}, R. and {O'Brien}, S. and {Ong}, R.~A. and {Pohl}, M. and {Pueschel}, E. and {Quinn}, J. and {Ragan}, K. and {Reynolds}, P.~T. and {Ribeiro}, D. and {Roache}, E. and {Sadeh}, I. and {Sadun}, A.~C. and {Saha}, L. and {Santander}, M. and {Sembroski}, G.~H. and {Shang}, R. and {Splettstoesser}, M. and {Talluri}, A. Kaushik and {Tucci}, J.~V. and {Vassiliev}, V.~V. and {Weinstein}, A. and {Williams}, D.~A. and {Wong}, S.~L. and {Woo}, J. and {Aharonian}, F. and {Aschersleben}, J. and {Backes}, M. and {Martins}, V. Barbosa and {Batzofin}, R. and {Becherini}, Y. and {Berge}, D. and {Bernl{\"o}hr}, K. and {Bi}, B. and {B{\"o}ttcher}, M. and {Boisson}, C. and {Bolmont}, J. and {de Bony de Lavergne}, M. and {Borowska}, J. and {Bouyahiaoui}, M. and {Bradascio}, F. and {Breuhaus}, M. and {Brose}, R. and {Brun}, F. and {Bruno}, B. and {Bulik}, T. and {Burger-Scheidlin}, C. and {Caroff}, S. and {Casanova}, S. and {Cecil}, R. and {Celic}, J. and {Cerruti}, M. and {Chand}, T. and {Chandra}, S. and {Chen}, A. and {Chibueze}, J. and {Chibueze}, O. and {Cotter}, G. and {Dai}, S. and {Mbarubucyeye}, J. Damascene and {Djannati-Ata{\"\i}}, A. and {Dmytriiev}, A. and {Doroshenko}, V. and {Einecke}, S. and {Ernenwein}, J. -P. and {de Clairfontaine}, G. Fichet and {Filipovic}, M. and {Fontaine}, G. and {F{\"u}{\ss}ling}, M. and {Funk}, S. and {Gabici}, S. and {Ghafourizadeh}, S. and {Giavitto}, G. and {Glawion}, D. and {Glicenstein}, J.~F. and {Goswami}, P. and {Grolleron}, G. and {Haerer}, L. and {Hinton}, J.~A. and {Holch}, T.~L. and {Holler}, M. and {Horns}, D. and {Jamrozy}, M. and {Jankowsky}, F. and {Joshi}, V. and {Jung-Richardt}, I. and {Kasai}, E. and {Katarzy{\'n}ski}, K. and {Khatoon}, R. and {Kh{\'e}lifi}, B. and {Klepser}, S. and {Klu{\'z}niak}, W. and {Kosack}, K. and {Kostunin}, D. and {Lang}, R.~G. and {Le Stum}, S. and {Lemi{\`e}re}, A. and {Lenain}, J. -P. and {Leuschner}, F. and {Lohse}, T. and {Luashvili}, A. and {Lypova}, I. and {Mackey}, J. and {Malyshev}, D. and {Marandon}, V. and {Marchegiani}, P. and {Marcowith}, A. and {Mart{\'\i}-Devesa}, G. and {Marx}, R. and {Mitchell}, A. and {Moderski}, R. and {Mohrmann}, L. and {Montanari}, A. and {Moulin}, E. and {Murach}, T. and {Nakashima}, K. and {Niemiec}, J. and {Noel}, A. Priyana and {O'Brien}, P. and {Olivera-Nieto}, L. and {de Ona Wilhelmi}, E. and {Ostrowski}, M. and {Panny}, S. and {Panter}, M. and {Peron}, G. and {Prokhorov}, D.~A. and {P{\"u}hlhofer}, G. and {Punch}, M. and {Quirrenbach}, A. and {Reichherzer}, P. and {Reimer}, A. and {Reimer}, O. and {Ren}, H. and {Renaud}, M. and {Rieger}, F. and {Rudak}, B. and {Ruiz-Velasco}, E. and {Sahakian}, V. and {Santangelo}, A. and {Sasaki}, M. and {Sch{\"a}fer}, J. and {Sch{\"u}ssler}, F. and {Schutte}, H.~M. and {Schwanke}, U. and {Shapopi}, J.~N.~S. and {Specovius}, A. and {Spencer}, S. and {Stawarz}, {\L}. and {Steenkamp}, R. and {Steinmassl}, S. and {Sushch}, I. and {Suzuki}, H. and {Takahashi}, T. and {Tanaka}, T. and {Terrier}, R. and {van Eldik}, C. and {Vecchi}, M. and {Veh}, J. and {Venter}, C. and {Vink}, J.},
        title = "{Multiwavelength Observations of the Blazar PKS 0735+178 in Spatial and Temporal Coincidence with an Astrophysical Neutrino Candidate IceCube-211208A}",
      journal = {\apj},
         year = 2023,
        month = sep,
       volume = {954},
       number = {1},
          eid = {70},
        pages = {70},
          doi = {10.3847/1538-4357/ace327},
archivePrefix = {arXiv},
       eprint = {2306.17819},
 primaryClass = {astro-ph.HE},
       adsurl = {https://ui.adsabs.harvard.edu/abs/2023ApJ...954...70A}
}

@ARTICLE{Sahakyan2023_0735,
       author = {{Sahakyan}, N. and {Giommi}, P. and {Padovani}, P. and {Petropoulou}, M. and {B{\'e}gu{\'e}}, D. and {Boccardi}, B. and {Gasparyan}, S.},
        title = "{A multimessenger study of the blazar PKS 0735+178: a new major neutrino source candidate}",
      journal = {\mnras},
         year = 2023,
        month = feb,
       volume = {519},
       number = {1},
        pages = {1396-1408},
          doi = {10.1093/mnras/stac3607},
archivePrefix = {arXiv},
       eprint = {2204.05060},
 primaryClass = {astro-ph.HE},
       adsurl = {https://ui.adsabs.harvard.edu/abs/2023MNRAS.519.1396S}
}

@ARTICLE{0735ATelTelamon,
       author = {{Kadler}, Matthias and {Benke}, Petra and {Gokus}, Andrea and {Hessdoerfer}, Jonas and {Sinapius}, Jonas and {Amp} and {Weber}, Philip and {TELAMON Team} and {Tornikoski}, Merja and {Righini}, Simona and {Marchili}, Nicola and {Hovatta}, Talvikki and {Readhead}, Anthony C. and {Kiehlmann}, Sebastian and {Kovalev}, Yuri A. and {Popkov}, Alexander V. and {Kovalev}, Yuri Y.},
        title = "{TELAMON, Metsahovi, Medicina, OVRO and RATAN-600 programs find a long-term radio flare in PKS0735+17 coincident with IceCube-211208A}",
      journal = {The Astronomer's Telegram},
         year = 2021,
        month = dec,
       volume = {15105},
        pages = {1},
       adsurl = {https://ui.adsabs.harvard.edu/abs/2021ATel15105....1K}
}

@INPROCEEDINGS{0215Proceedings,
       author = {{Eppel}, F. and {Kadler}, M. and {Ros}, E. and {R{\"o}sch}, F. and {He{\ss}d{\"o}rfer}, J. and {Benke}, P. and {Edwards}, P.~G. and {Fromm}, C.~M. and {Giroletti}, M. and {Gokus}, A. and {G{\'o}mez}, J.~L. and {H{\"a}mmerich}, S. and {Kirchner}, D. and {Kovalev}, Y.~Y. and {Krichbaum}, T.~P. and {Lister}, M.~L. and {Nanci}, C. and {Ojha}, R. and {Paraschos}, G.~F. and {Plavin}, A. and {Readhead}, A.~C.~S. and {Stevens}, J. and {Weber}, P.},
        title = "{VLBI Scrutiny of a New Neutrino-Blazar Multiwavelength-Flare Coincidence}",
    booktitle = {The Multimessenger Chakra of Blazar Jets},
         year = 2023,
       editor = {{Liodakis}, Ioannis and {Aller}, Margo F. and {Krawczynski}, Henric and {L{\"a}hteenm{\"a}ki}, Anne and {Pearson}, Timothy J.},
       series = {IAU Symposium},
       volume = {375},
        month = jan,
        pages = {91-95},
          doi = {10.1017/S1743921323000893},
archivePrefix = {arXiv},
       eprint = {2301.13859},
 primaryClass = {astro-ph.HE},
       adsurl = {https://ui.adsabs.harvard.edu/abs/2023IAUS..375...91E}
}

@INPROCEEDINGS{Sanghyun2024_0735,
       author = {{Kim}, S. and {Lee}, S. -S. and {Cheong}, W.~Y. and {Jeong}, H. -W.},
        title = "{Interferometric Monitoring of a Potential Neutrino-Emitting Blazar PKS 0735+178: A Connection Between Neutrino Events and Radio Flares?}",
    booktitle = {Proceedings of the 16th EVN Symposium},
         year = 2024,
       editor = {{Ros}, E. and {Benke}, P. and {Dzib}, S.~A. and {Rottmann}, I. and {Zensus}, J.~A.},
        month = sep,
        pages = {123-126},
       adsurl = {https://ui.adsabs.harvard.edu/abs/2024evn..conf..123K}
}

@ARTICLE{Kim2025_0735,
       author = {{Kim}, Yu-Sik and {Kim}, Jae-Young},
        title = "{The dynamics of the parsec-scale jet in the neutrino blazar PKS 0735+178}",
      journal = {\aap},
         year = 2025,
        month = jul,
       volume = {699},
          eid = {A381},
        pages = {A381},
          doi = {10.1051/0004-6361/202452111},
archivePrefix = {arXiv},
       eprint = {2505.13876},
 primaryClass = {astro-ph.HE},
       adsurl = {https://ui.adsabs.harvard.edu/abs/2025A&A...699A.381K}
}

@ARTICLE{0446YuriATel,
       author = {{Kovalev}, Y.~Y. and {Plavin}, A.~V. and {Troitsky}, S.~V. and {Kovalev}, Yu. A. and {Popkov}, A.~V. and {Pushkarev}, A.~B.},
        title = "{Radio-flaring blazar PKS 0446+11 with bright parsec-scale core as a candidate for IceCube-240105A: RATAN-600 and MOJAVE VLBA observations}",
      journal = {The Astronomer's Telegram},
         year = 2024,
        month = jan,
       volume = {16409},
        pages = {1},
       adsurl = {https://ui.adsabs.harvard.edu/abs/2024ATel16409....1K}
}

@ARTICLE{KadlerATel0215,
       author = {{Kadler}, Matthias and {Stevens}, Jamie and {Ojha}, Roopesh and {Edwards}, Philip G.},
        title = "{ATCA finds a long-term radio flare of PKS0215+015 coincident with IceCube-220225A}",
      journal = {The Astronomer's Telegram},
         year = 2022,
        month = feb,
       volume = {15245},
        pages = {1},
       adsurl = {https://ui.adsabs.harvard.edu/abs/2022ATel15245....1K}
}

@ARTICLE{Telamon0446ATel,
       author = {{Eppel}, F. and {Kadler}, M. and {Debbrecht}, L. and {Eich}, J. and {Gokus}, A. and {Hessdoerfer}, J. and {Kim}, S. -H. and {Kirchner}, D. and {Schulga}, W.},
        title = "{TELAMON detection of an inverted spectrum and radio flare of PKS 0446+11 coincident with IceCube-240105A}",
      journal = {The Astronomer's Telegram},
         year = 2024,
        month = jan,
       volume = {16399},
        pages = {1},
       adsurl = {https://ui.adsabs.harvard.edu/abs/2024ATel16399....1E}
}

@ARTICLE{Plavin0215ATel,
       author = {{Plavin}, A.~V. and {Kovalev}, Y.~Y. and {Troitsky}, S.~V. and {Kovalev}, Yu. A. and {Popkov}, A.~V. and {Lister}, M.~L. and {Gomez}, J.~L. and {Hovatta}, T. and {Kiehlmann}, S. and {Liodakis}, I. and {Readhead}, A.~C.~S. and {Erkenov}, A.~K. and {Sotnikova}, Y.~V. and {Zhekanis}, G.~V.},
        title = "{The flaring radio-bright blazar PKS 0215+015 coincident with the high energy neutrino alert IceCube-220225A}",
      journal = {The Astronomer's Telegram},
         year = 2022,
        month = feb,
       volume = {15247},
        pages = {1},
       adsurl = {https://ui.adsabs.harvard.edu/abs/2022ATel15247....1P}
}

@ARTICLE{IceCubeCorrelation,
       author = {{Abbasi}, R. and {Ackermann}, M. and {Adams}, J. and {Agarwalla}, S.~K. and {Aguilar}, J.~A. and {Ahlers}, M. and {Alameddine}, J.~M. and {Amin}, N.~M. and {Andeen}, K. and {Anton}, G. and {Arg{\"u}elles}, C. and {Ashida}, Y. and {Athanasiadou}, S. and {Axani}, S.~N. and {Bai}, X. and {Balagopal}, V.~A. and {Baricevic}, M. and {Barwick}, S.~W. and {Basu}, V. and {Bay}, R. and {Beatty}, J.~J. and {Becker}, K. -H. and {Becker Tjus}, J. and {Beise}, J. and {Bellenghi}, C. and {Benning}, C. and {BenZvi}, S. and {Berley}, D. and {Bernardini}, E. and {Besson}, D.~Z. and {Binder}, G. and {Blaufuss}, E. and {Blot}, S. and {Bontempo}, F. and {Book}, J.~Y. and {Meneguolo}, C. Boscolo and {B{\"o}ser}, S. and {Botner}, O. and {B{\"o}ttcher}, J. and {Bourbeau}, E. and {Braun}, J. and {Brinson}, B. and {Brostean-Kaiser}, J. and {Burley}, R.~T. and {Busse}, R.~S. and {Butterfield}, D. and {Campana}, M.~A. and {Carloni}, K. and {Carnie-Bronca}, E.~G. and {Chattopadhyay}, S. and {Chau}, N. and {Chen}, C. and {Chen}, Z. and {Chirkin}, D. and {Choi}, S. and {Clark}, B.~A. and {Classen}, L. and {Coleman}, A. and {Collin}, G.~H. and {Connolly}, A. and {Conrad}, J.~M. and {Coppin}, P. and {Correa}, P. and {Countryman}, S. and {Cowen}, D.~F. and {Dave}, P. and {De Clercq}, C. and {DeLaunay}, J.~J. and {Delgado}, D. and {Dembinski}, H. and {Deng}, S. and {Deoskar}, K. and {Desai}, A. and {Desiati}, P. and {de Vries}, K.~D. and {de Wasseige}, G. and {DeYoung}, T. and {Diaz}, A. and {D{\'\i}az-V{\'e}lez}, J.~C. and {Dittmer}, M. and {Domi}, A. and {Dujmovic}, H. and {DuVernois}, M.~A. and {Ehrhardt}, T. and {Eller}, P. and {El Mentawi}, S. and {Engel}, R. and {Erpenbeck}, H. and {Evans}, J. and {Evenson}, P.~A. and {Fan}, K.~L. and {Fang}, K. and {Farrag}, K. and {Fazely}, A.~R. and {Fedynitch}, A. and {Feigl}, N. and {Fiedlschuster}, S. and {Finley}, C. and {Fischer}, L. and {Fox}, D. and {Franckowiak}, A. and {Friedman}, E. and {Fritz}, A. and {F{\"u}rst}, P. and {Gaisser}, T.~K. and {Gallagher}, J. and {Ganster}, E. and {Garcia}, A. and {Gerhardt}, L. and {Ghadimi}, A. and {Glaser}, C. and {Glauch}, T. and {Gl{\"u}senkamp}, T. and {Goehlke}, N. and {Gonzalez}, J.~G. and {Goswami}, S. and {Grant}, D. and {Gray}, S.~J. and {Gries}, O. and {Griffin}, S. and {Griswold}, S. and {G{\"u}nther}, C. and {Gutjahr}, P. and {Haack}, C. and {Hallgren}, A. and {Halliday}, R. and {Halve}, L. and {Halzen}, F. and {Hamdaoui}, H. and {Minh}, M. Ha and {Hanson}, K. and {Hardin}, J. and {Harnisch}, A.~A. and {Hatch}, P. and {Haungs}, A. and {Helbing}, K. and {Hellrung}, J. and {Henningsen}, F. and {Heuermann}, L. and {Heyer}, N. and {Hickford}, S. and {Hidvegi}, A. and {Hill}, C. and {Hill}, G.~C. and {Hoffman}, K.~D. and {Hori}, S. and {Hoshina}, K. and {Hou}, W. and {Huber}, T. and {Hultqvist}, K. and {H{\"u}nnefeld}, M. and {Hussain}, R. and {Hymon}, K. and {In}, S. and {Ishihara}, A. and {Jacquart}, M. and {Janik}, O. and {Jansson}, M. and {Japaridze}, G.~S. and {Jayakumar}, K. and {Jeong}, M. and {Jin}, M. and {Jones}, B.~J.~P. and {Kang}, D. and {Kang}, W. and {Kang}, X. and {Kappes}, A. and {Kappesser}, D. and {Kardum}, L. and {Karg}, T. and {Karl}, M. and {Karle}, A. and {Katz}, U. and {Kauer}, M. and {Kelley}, J.~L. and {Zathul}, A. Khatee and {Kheirandish}, A. and {Kiryluk}, J. and {Klein}, S.~R. and {Kochocki}, A. and {Koirala}, R. and {Kolanoski}, H. and {Kontrimas}, T. and {K{\"o}pke}, L. and {Kopper}, C. and {Koskinen}, D.~J. and {Koundal}, P. and {Kovacevich}, M. and {Kowalski}, M. and {Kozynets}, T. and {Kruiswijk}, K. and {Krupczak}, E. and {Kumar}, A. and {Kun}, E. and {Kurahashi}, N. and {Lad}, N. and {Lagunas Gualda}, C. and {Lamoureux}, M. and {Larson}, M.~J. and {Latseva}, S.},
        title = "{Search for Correlations of High-energy Neutrinos Detected in IceCube with Radio-bright AGN and Gamma-Ray Emission from Blazars}",
      journal = {\apj},
         year = 2023,
        month = sep,
       volume = {954},
       number = {1},
          eid = {75},
        pages = {75},
          doi = {10.3847/1538-4357/acdfcb},
archivePrefix = {arXiv},
       eprint = {2304.12675},
 primaryClass = {astro-ph.HE},
       adsurl = {https://ui.adsabs.harvard.edu/abs/2023ApJ...954...75A}
}

@ARTICLE{Bellenghi2023,
       author = {{Bellenghi}, Chiara and {Padovani}, Paolo and {Resconi}, Elisa and {Giommi}, Paolo},
        title = "{Correlating High-energy IceCube Neutrinos with 5BZCAT Blazars and RFC Sources}",
      journal = {\apjl},
         year = 2023,
        month = oct,
       volume = {955},
       number = {2},
          eid = {L32},
        pages = {L32},
          doi = {10.3847/2041-8213/acf711},
archivePrefix = {arXiv},
       eprint = {2309.03115},
 primaryClass = {astro-ph.HE},
       adsurl = {https://ui.adsabs.harvard.edu/abs/2023ApJ...955L..32B}
}

@misc{Lobanov05,
       author = {{Lobanov}, A.~P.},
      journal = {arXiv e-prints},
         year = 2005,
        month = mar,
          eid = {astro-ph/0503225},
        pages = {astro-ph/0503225},
          doi = {10.48550/arXiv.astro-ph/0503225},
archivePrefix = {arXiv},
       eprint = {astro-ph/0503225},
       primaryClass = {astro-ph},
       adsurl = {https://ui.adsabs.harvard.edu/abs/2005astro.ph..3225L},
    note = {arXiv e-prints, astro-ph/0503225}
}

\begin{appendix}

\section{Collimation profile fit}
\label{app:collimation}

\begin{figure}
    \centering
    \includegraphics[width=\linewidth]{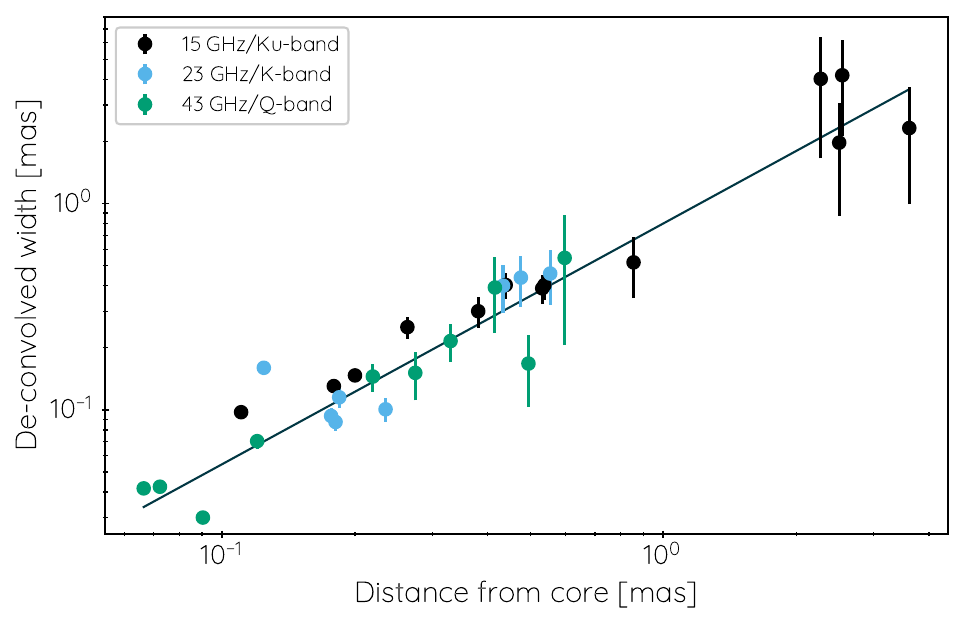}
    \caption{Collimation profile obtained from the combined VLBI data at 15\,GHz (Ku-band), 23\,GHz (K-band), and 43\,GHz (Q-band).}
    \label{fig:collimation}
\end{figure}

In order to probe the collimation profile of PKS\,0215+015, we have fitted a power law of the form 
\begin{equation}
    w(r)=w_0(r+r_0)^k,
\end{equation}
where $w_0$ is the jet width 1\,mas from the jet apex, $r$ is the distance of a component to the radio core, $r_0$ the distance from the radio core to the jet apex, and $k$ the power law/collimation index (see Fig.\,\ref{fig:collimation}). We excluded any unresolved components, following the definition by \cite{Lobanov05} and \cite{Kovalev05}, as well as the core components from the fit. The fit suggest a collimation index of $k=1.17\pm0.30$, which is consistent with a conical jet profile (i.e., k$\sim$1). Additionally, we find $w_0=(0.80\pm0.23)$\,mas, and $r_0=(0.045\pm0.039)$\,mas. In principle one has to consider the core shift effect due to synchrotron opacity \citep{Lobanov98,Kovalev2008_coreshift} before performing a combined fit of the collimation profile across frequencies, however, due to the compact nature of the source, as well as its large redshift, the core shift cannot be reliably estimated. Given the component cross identifications across frequencies (see App.\,\ref{app:modelfits}), no significant systematic shift across frequencies is evident from the optically thin components 1 and 2. Therefore, the core-shift is expected to be very small, which is consistent with the small value obtained for $r_0$.

\section{Component identification}
\label{app:modelfits}

\begin{figure*}
    \centering
    \includegraphics[width=0.735\linewidth]{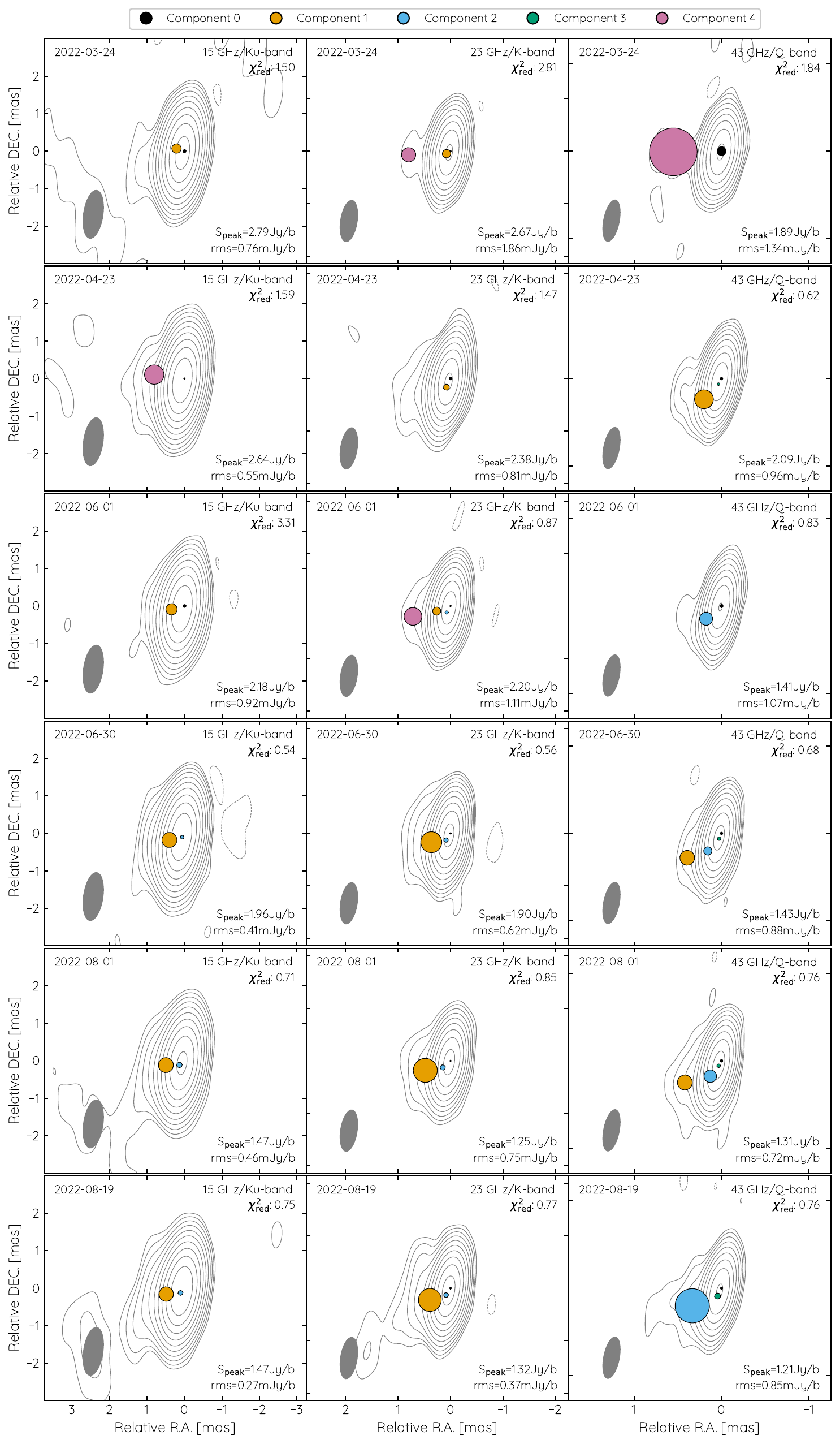}
    \caption{Total intensity contour plots, as shown in Fig.\,\ref{fig:all_pol_images}, contours start at four times the noise level, increasing by a factor of 2. The core component (component 0), and the modelfit components 1$-$4 are plotted on top of the contours, with their size corresponding to the \review{FWHM (full width at half maximum)} of the modelfit component. The reduced-$\chi^2$ value for the modelfit is shown in each panel. For unresolved components, the plotted component size corresponds to the resolution limit.}
    \label{fig:all_comps_plot}
\end{figure*}

In this section, we provide a detailed overview about the obtained modelfits and their identification across epochs and frequencies. The positions, sizes, flux densities, and S/N values for all components are listed in Tab.\,\ref{tab:components}. Figure\,\ref{fig:all_comps_plot} shows the position of the core component (component 0), as well as the jet components 1$-$4 for all epochs and frequencies, on top of the total intensity maps. The component identification for this dataset is not trivial since in several epochs, multiple components are located within the size of the beam. The presented modelfit represents the most consistent model that we obtained across all epochs and frequencies. 

In the first epoch, on March 24, 2022, at 15\,GHz and 43\,GHz, only two components are detected, while at 23\,GHz, three components are found. We decided to label the component furthest away from the core as component 4, while the component very close to the core is labeled as component 1. It is not clear, why component 1 cannot be fitted at 43\,GHz in this epoch, but we note the extended core size in this observation, thus, it might be \review{blended} with the core component. Similarly, component 4 could not be fitted at 15\,GHz, likely due to the limited resolution in this epoch, since the HN station of the VLBA had to be flagged for most of the observation.

In the second epoch, on April 23, 2022, at 15\,GHz, we can detect only the core component and one extended component, located at a distance of $\sim0.8$\,mas away from the core. This extended component is clearly not consistent with component 1 that was detected in the previous epoch and was thus identified as component 4, which was already found at 23\,GHz and 43\,GHz in the previous epoch at a similar core distance. At 23\,GHz, component 1 and the core are clearly detected, while the extended component 4 was not detected, likely due to limited S/N. At 43\,GHz a consistent component 1 was found, as well as a new component very close to the core, that we labeled as component 3.

In Epoch 3, on June 1, 2022, at 15\,GHz, again, the outermost component 4 is not detected, but a component consistent with the position of component 1 was found, in addition to the core component. At 23\,GHz we identified three components, in addition to the core component. The outermost component is consistent with the position of component 4, while the next component upstream must be labeled component 1, given its position at 15\,GHz. Additionally, this introduces a new component 2, located close to the 23\,GHz core. At 43\,GHz, only one jet component is detected in addition to the core. While this component could be consistent with component 1, we decided to assign it to component 2, which is more consistent with the 23\,GHz modelfit and the modelfits of the later 43\,GHz epochs (see also Fig.\,\ref{fig:kinematics}).

In Epoch 4, on June 30, 2022, a new component also appears at 15\,GHz, which is blended with the core component. This component was already visible in the previous epoch with better resolution at the higher frequencies and must therefore be consistent with component 2. The other detected jet component is clearly associated with component 1. A similar picture is seen at 23\,GHz, with component 2 located a bit further from the core than at 15\,GHz, but also component 1 being clearly detected. At 43\,GHz, another component shows up very close to the core which was detected before in Epoch 2 as component 3.

Epoch 5, observed on August 1, 2022 is largely consistent with Epoch 4, with only minor changes of the component positions between these two epochs at all frequencies.

At the final epoch, observed on 19 August 2022, there are again no major changes at 15\,GHz and 23\,GHz compared to the previous epoch. At 43\,GHz, the component labeled as component 2 is likely a combination of component 1 and component 2, and, thus, has a large positional uncertainty (see also Fig.\,\ref{fig:kinematics}). We decided to label it as component 2, since its position is more consistent with the position of component 2 detected at the other frequencies.

The strongest evidence for the fast motion of component 1 clearly comes from 15\,GHz and 23\,GHz, while the structure at 43\,GHz is less constrained by the modelfits. This might seem contradictory since 43\,GHz offers the best angular resolution, however, due to the limited S/N as compared to the lower frequencies, as well as the detection of an additional component close to the core, the cross-identification of components at 43\,GHz is especially challenging.  

\begin{table*}[h!]
\caption{Model fit component parameters \reviewtwo{for all epochs and frequencies.}}

\label{tab:components}
\centering
\resizebox{1.91\columnwidth}{!}{%
\begin{tabular}{c c c c c c c c}
\hline
ID & Epoch &
$X^a$ [mas] & $Y^a$ [mas] &
FWHM$^b$ [mas] &
Flux Density [mJy] &
S/N &
\review{$T_\mathrm{b}^c$ [K]} \\
\hline
\multicolumn{8}{c}{\textbf{15\,GHz (Ku-band)}} \\
\hline
0 & 2022-03-24 & \(-0.006 \pm 0.027\) & \(-0.003 \pm 0.017\) & \(0.0846 \pm 0.0025\) & \(2720 \pm 180\) & \(1176.3\) & \review{\( (5.33 \pm 0.41) \times 10^{12}\)} \\
0 & 2022-04-23 & \(-0.005 \pm 0.021\) & \(0.0014 \pm 0.0088\) & \(0.03579 \pm 0.00086\) & \(2590 \pm 160\) & \(1745.7\) & \review{\( (2.84 \pm 0.20) \times 10^{13}\)} \\
0 & 2022-06-01 & \(-0.016 \pm 0.024\) & \(0.026 \pm 0.027\) & \(0.0815 \pm 0.0029\) & \(1850 \pm 130\) & \(797.2\) & \review{\( (3.91 \pm 0.34) \times 10^{12}\)} \\
0 & 2022-06-30 & \(-0.029 \pm 0.027\) & \(0.044 \pm 0.028\) & \review{\(<0.040\)} & \(1102 \pm 77\) & \(829.5\) & \review{\(>9.53 \times 10^{12}\)} \\
0 & 2022-08-01 & \(-0.015 \pm 0.020\) & \(0.012 \pm 0.019\) & \review{\(<0.033\)} & \(1295 \pm 81\) & \(1400.3\) & \review{\(>1.62 \times 10^{13}\)} \\
0 & 2022-08-19 & \(-0.02 \pm 0.023\) & \(0.023 \pm 0.024\) & \review{\(<0.039\)} & \(1181 \pm 77\) & \(1132.8\) & \review{\(>1.09 \times 10^{13}\)} \\
1 & 2022-03-24 & \(0.21 \pm 0.13\) & \(0.07 \pm 0.12\) & \(0.252 \pm 0.031\) & \(168 \pm 29\) & \(69.0\) & \review{\( (3.74 \pm 0.91) \times 10^{10}\)} \\
1 & 2022-04-23 & \(0.27 \pm 0.12\) & \(-0.08 \pm 0.11\) & \review{\(<0.153\)} & \(97 \pm 18\) & \(64.2\) & \review{\(>5.84 \times 10^{10}\)} \\
1 & 2022-06-01 & \(0.33 \pm 0.14\) & \(-0.06 \pm 0.13\) & \(0.301 \pm 0.050\) & \(97 \pm 22\) & \(36.7\) & \review{\( (1.51 \pm 0.49) \times 10^{10}\)} \\
1 & 2022-06-30 & \(0.37 \pm 0.14\) & \(-0.13 \pm 0.13\) & \(0.403 \pm 0.060\) & \(79 \pm 15\) & \(45.9\) & \review{\( (6.9 \pm 2.0) \times 10^{9}\)} \\
1 & 2022-08-01 & \(0.48 \pm 0.15\) & \(-0.1 \pm 0.14\) & \(0.403 \pm 0.063\) & \(48.2 \pm 9.9\) & \(41.3\) & \review{\( (4.2 \pm 1.3) \times 10^{9}\)} \\
1 & 2022-08-19 & \(0.47 \pm 0.15\) & \(-0.13 \pm 0.15\) & \(0.388 \pm 0.063\) & \(50 \pm 11\) & \(39.3\) & \review{\( (4.7 \pm 1.5) \times 10^{9}\)} \\
2 & 2022-06-01 & \(0.072 \pm 0.063\) & \(-0.155 \pm 0.064\) & \review{\(<0.090\)} & \(347 \pm 44\) & \(150.8\) & \review{\(>6.05 \times 10^{11}\)} \\
2 & 2022-06-30 & \(0.034 \pm 0.031\) & \(-0.056 \pm 0.032\) & \(0.0974 \pm 0.0039\) & \(864 \pm 64\) & \(636.9\) & \review{\( (1.28 \pm 0.12) \times 10^{12}\)} \\
2 & 2022-08-01 & \(0.119 \pm 0.058\) & \(-0.099 \pm 0.057\) & \(0.147 \pm 0.010\) & \(199 \pm 22\) & \(206.0\) & \review{\( (1.29 \pm 0.19) \times 10^{11}\)} \\
2 & 2022-08-19 & \(0.087 \pm 0.050\) & \(-0.102 \pm 0.050\) & \(0.1304 \pm 0.0077\) & \(307 \pm 30\) & \(285.9\) & \review{\( (2.54 \pm 0.32) \times 10^{11}\)} \\
3 & 2022-03-24 & \(2.2 \pm 1.2\) & \(-1.0 \pm 1.2\) & \(2.0 \pm 1.1\) & \(21 \pm 12\) & \(4.0\) & \review{\( (7.5 \pm 7.4) \times 10^{7}\)} \\
3 & 2022-04-23 & \(3.6 \pm 1.4\) & \(-0.1 \pm 1.4\) & \(2.3 \pm 1.3\) & \(12.3 \pm 7.3\) & \(3.8\) & \review{\( (3.2 \pm 3.2) \times 10^{7}\)} \\
3 & 2022-08-01 & \(2.2 \pm 2.0\) & \(-1.1 \pm 1.8\) & \(4.2 \pm 2.0\) & \(31 \pm 15\) & \(5.0\) & \review{\( (2.5 \pm 2.1) \times 10^{7}\)} \\
3 & 2022-08-19 & \(2.1 \pm 2.4\) & \(-0.7 \pm 1.9\) & \(4.0 \pm 2.4\) & \(30 \pm 18\) & \(3.7\) & \review{\( (2.6 \pm 2.7) \times 10^{7}\)} \\
4 & 2022-04-23 & \(0.80 \pm 0.34\) & \(0.11 \pm 0.33\) & \(0.52 \pm 0.17\) & \(20.1 \pm 8.5\) & \(10.2\) & \review{\( (1.05 \pm 0.66) \times 10^{9}\)} \\
\hline
\multicolumn{8}{c}{\textbf{23\,GHz (K-band)}} \\
\hline
0 & 2022-03-24 & \(-0.008 \pm 0.024\) & \(0.006 \pm 0.019\) & \review{\(<0.035\)} & \(2330 \pm 160\) & \(830.2\) & \review{\(>1.09 \times 10^{13}\)} \\
0 & 2022-04-23 & \(-0.016 \pm 0.033\) & \(0.038 \pm 0.038\) & \review{\(<0.053\)} & \(1820 \pm 200\) & \(218.0\) & \review{\(>3.86 \times 10^{12}\)} \\
0 & 2022-06-01 & \(-0.009 \pm 0.015\) & \(0.014 \pm 0.017\) & \review{\(<0.025\)} & \(1920 \pm 130\) & \(852.4\) & \review{\(>1.74 \times 10^{13}\)} \\
0 & 2022-06-30 & \(-0.01 \pm 0.016\) & \(0.015 \pm 0.018\) & \review{\(<0.028\)} & \(1660 \pm 120\) & \(719.0\) & \review{\(>1.25 \times 10^{13}\)} \\
0 & 2022-08-01 & \(-0.004 \pm 0.013\) & \(0.004 \pm 0.011\) & \(0.02592 \pm 0.00086\) & \(1203 \pm 83\) & \(901.7\) & \review{\( (1.052 \pm 0.088) \times 10^{13}\)} \\
0 & 2022-08-19 & \(-0.014 \pm 0.019\) & \(0.022 \pm 0.021\) & \review{\(<0.036\)} & \(1084 \pm 86\) & \(530.0\) & \review{\(>5.02 \times 10^{12}\)} \\
1 & 2022-03-24 & \(0.067 \pm 0.064\) & \(-0.041 \pm 0.059\) & \(0.160 \pm 0.013\) & \(426 \pm 52\) & \(151.0\) & \review{\( (9.8 \pm 1.7) \times 10^{10}\)} \\
1 & 2022-04-23 & \(0.062 \pm 0.062\) & \(-0.125 \pm 0.064\) & \(0.115 \pm 0.013\) & \(710 \pm 110\) & \(82.7\) & \review{\( (3.15 \pm 0.71) \times 10^{11}\)} \\
1 & 2022-06-01 & \(0.25 \pm 0.11\) & \(-0.08 \pm 0.11\) & \review{\(<0.150\)} & \(53 \pm 15\) & \(25.1\) & \review{\(>1.37 \times 10^{10}\)} \\
1 & 2022-06-30 & \(0.36 \pm 0.17\) & \(-0.15 \pm 0.16\) & \(0.40 \pm 0.11\) & \(52 \pm 16\) & \(15.2\) & \review{\( (1.90 \pm 0.93) \times 10^{9}\)} \\
1 & 2022-08-01 & \(0.48 \pm 0.19\) & \(-0.18 \pm 0.19\) & \(0.46 \pm 0.13\) & \(34 \pm 11\) & \(12.5\) & \review{\( (9.6 \pm 5.1) \times 10^{8}\)} \\
1 & 2022-08-19 & \(0.38 \pm 0.18\) & \(-0.2 \pm 0.18\) & \(0.44 \pm 0.12\) & \(51 \pm 16\) & \(14.0\) & \review{\( (1.56 \pm 0.79) \times 10^{9}\)} \\
2 & 2022-06-01 & \(0.064 \pm 0.043\) & \(-0.109 \pm 0.044\) & \review{\(<0.063\)} & \(306 \pm 40\) & \(137.9\) & \review{\(>4.46 \times 10^{11}\)} \\
2 & 2022-06-30 & \(0.076 \pm 0.046\) & \(-0.112 \pm 0.047\) & \(0.0874 \pm 0.0080\) & \(285 \pm 39\) & \(119.0\) & \review{\( (2.19 \pm 0.41) \times 10^{11}\)} \\
2 & 2022-08-01 & \(0.144 \pm 0.062\) & \(-0.123 \pm 0.062\) & \(0.101 \pm 0.013\) & \(87 \pm 16\) & \(60.7\) & \review{\( (5.0 \pm 1.3) \times 10^{10}\)} \\
2 & 2022-08-19 & \(0.072 \pm 0.044\) & \(-0.11 \pm 0.044\) & \(0.0935 \pm 0.0082\) & \(283 \pm 37\) & \(129.9\) & \review{\( (1.91 \pm 0.34) \times 10^{11}\)} \\
3 & 2022-03-24 & \(0.79 \pm 0.32\) & \(-0.06 \pm 0.31\) & \review{\(<0.366\)} & \(15.0 \pm 8.1\) & \(8.3\) & \review{\(>6.58 \times 10^{8}\)} \\
3 & 2022-06-01 & \(0.71 \pm 0.28\) & \(-0.18 \pm 0.27\) & \review{\(<0.332\)} & \(12.9 \pm 7.6\) & \(5.5\) & \review{\(>6.87 \times 10^{8}\)} \\

\hline
\multicolumn{8}{c}{\textbf{43\,GHz (Q-band)}} \\
\hline
0 & 2022-03-24 & \(0.001 \pm 0.012\) & \(-0.002 \pm 0.024\) & \(0.1024 \pm 0.0066\) & \(2160 \pm 220\) & \(244.0\) & \review{\( (3.67 \pm 0.50) \times 10^{11}\)} \\
0 & 2022-04-23 & \(-0.013 \pm 0.014\) & \(0.024 \pm 0.015\) & \(0.0289 \pm 0.0013\) & \(1340 \pm 110\) & \(483.2\) & \review{\( (2.86 \pm 0.30) \times 10^{12}\)} \\
0 & 2022-06-01 & \(-0.001 \pm 0.013\) & \(0.0002 \pm 0.0041\) & \(0.0371 \pm 0.0015\) & \(1420 \pm 110\) & \(617.7\) & \review{\( (1.85 \pm 0.17) \times 10^{12}\)} \\
0 & 2022-06-30 & \(-0.017 \pm 0.016\) & \(0.037 \pm 0.016\) & \review{\(<0.025\)} & \(557 \pm 57\) & \(252.8\) & \review{\(>1.54 \times 10^{12}\)} \\
0 & 2022-08-01 & \(-0.023 \pm 0.017\) & \(0.038 \pm 0.017\) & \review{\(<0.026\)} & \(470 \pm 50\) & \(232.8\) & \review{\(>1.28 \times 10^{12}\)} \\
0 & 2022-08-19 & \(-0.011 \pm 0.015\) & \(0.023 \pm 0.016\) & \review{\(<0.027\)} & \(875 \pm 81\) & \(331.7\) & \review{\(>2.14 \times 10^{12}\)} \\
1 & 2022-04-23 & \(0.188 \pm 0.082\) & \(-0.213 \pm 0.082\) & \(0.216 \pm 0.045\) & \(88 \pm 23\) & \(23.6\) & \review{\( (3.4 \pm 1.3) \times 10^{9}\)} \\
1 & 2022-06-30 & \(0.38 \pm 0.11\) & \(-0.24 \pm 0.11\) & \review{\(<0.148\)} & \(13.9 \pm 7.2\) & \(8.0\) & \review{\(>1.14 \times 10^{9}\)} \\
1 & 2022-08-01 & \(0.40 \pm 0.12\) & \(-0.21 \pm 0.12\) & \(0.168 \pm 0.064\) & \(18.2 \pm 8.5\) & \(7.7\) & \review{\( (1.15 \pm 0.82) \times 10^{9}\)} \\
2 & 2022-06-01 & \(0.176 \pm 0.095\) & \(-0.146 \pm 0.094\) & \(0.151 \pm 0.040\) & \(42 \pm 15\) & \(15.2\) & \review{\( (3.3 \pm 1.7) \times 10^{9}\)} \\
2 & 2022-06-30 & \(0.140 \pm 0.056\) & \(-0.164 \pm 0.056\) & \review{\(<0.083\)} & \(54 \pm 16\) & \(24.3\) & \review{\(>1.41 \times 10^{10}\)} \\
2 & 2022-08-01 & \(0.106 \pm 0.047\) & \(-0.139 \pm 0.047\) & \(0.145 \pm 0.023\) & \(109 \pm 22\) & \(42.3\) & \review{\( (9.2 \pm 2.7) \times 10^{9}\)} \\
2 & 2022-08-19 & \(0.32 \pm 0.19\) & \(-0.18 \pm 0.18\) & \(0.39 \pm 0.16\) & \(56 \pm 24\) & \(7.2\) & \review{\( (6.6 \pm 4.6) \times 10^{8}\)} \\
3 & 2022-04-23 & \(0.022 \pm 0.019\) & \(-0.04 \pm 0.019\) & \(0.0301 \pm 0.0018\) & \(801 \pm 78\) & \(288.9\) & \review{\( (1.58 \pm 0.20) \times 10^{12}\)} \\
3 & 2022-06-30 & \(0.011 \pm 0.013\) & \(-0.025 \pm 0.013\) & \(0.0424 \pm 0.0022\) & \(910 \pm 79\) & \(388.6\) & \review{\( (9.0 \pm 1.0) \times 10^{11}\)} \\
3 & 2022-08-01 & \(0.011 \pm 0.012\) & \(-0.018 \pm 0.013\) & \(0.0417 \pm 0.0021\) & \(834 \pm 72\) & \(401.1\) & \review{\( (8.56 \pm 0.95) \times 10^{11}\)} \\
3 & 2022-08-19 & \(0.034 \pm 0.024\) & \(-0.066 \pm 0.025\) & \(0.0705 \pm 0.0059\) & \(412 \pm 51\) & \(144.6\) & \review{\( (1.47 \pm 0.25) \times 10^{11}\)} \\
4 & 2022-03-24 & \(0.55 \pm 0.42\) & \(-0.01 \pm 0.36\) & \(0.54 \pm 0.34\) & \(37 \pm 25\) & \(3.4\) & \review{\( (2.2 \pm 2.4) \times 10^{8}\)} \\
\hline
\hline
\multicolumn{8}{l}{\footnotesize \reviewtwo{$^a$ component X- and Y-position -- $^b$ component size (FWHM) shown for resolved components. For unresolved components an upper limit is provided,}}\\
\multicolumn{8}{l}{\footnotesize \reviewtwo{according to the resolution limit defined by \cite{Lobanov05} and \cite{Kovalev05} -- $^c$ brightness temperature, following \cite{Kovalev05}}}
\end{tabular}
}
\end{table*}

\end{appendix}

%
%
%

\end{document}